%% file: rc_lqt.tex
\documentclass[
aps,
pra,showpacs,showkeys,groupedaddress,nofootinbib,longbibliography]{revtex4-2}

\input{preamble}
\usepackage{orcidlink}
\usepackage{scalerel,stackengine}
\DeclareMathOperator*{\bigboxplus}{\scalerel*{\boxplus}{\bigotimes}}
\usepackage{bbm}
\usepackage{bbold}
\usepackage{xcolor}
\definecolor{darkgreen}{rgb}{0.0, 0.5, 0.0}
\definecolor{bleudefrance}{rgb}{0.19, 0.55, 0.91}
\definecolor{upsdellred}{rgb}{0.78, 0.03, 0.08}

\usepackage[T1]{fontenc}

\makeatletter
\let\save@mathaccent\mathaccent
\newcommand*\if@single[3]{%
	\setbox0\hbox{${\mathaccent"0362{#1}}^H$}%
	\setbox2\hbox{${\mathaccent"0362{\kern0pt#1}}^H$}%
	\ifdim\ht0=\ht2 #3\else #2\fi
}
\newcommand*\rel@kern[1]{\kern#1\dimexpr\macc@kerna}
\newcommand*\widebar[1]{\@ifnextchar^{{\wide@bar{#1}{0}}}{\wide@bar{#1}{1}}}
\newcommand*\wide@bar[2]{\if@single{#1}{\wide@bar@{#1}{#2}{1}}{\wide@bar@{#1}{#2}{2}}}
\newcommand*\wide@bar@[3]{%
	\begingroup
	\def\mathaccent##1##2{%
		\let\mathaccent\save@mathaccent
		\if#32 \let\macc@nucleus\first@char \fi
		\setbox\z@\hbox{$\macc@style{\macc@nucleus}_{}$}%
		\setbox\tw@\hbox{$\macc@style{\macc@nucleus}{}_{}$}%
		\dimen@\wd\tw@
		\advance\dimen@-\wd\z@
		\divide\dimen@ 3
		\@tempdima\wd\tw@
		\advance\@tempdima-\scriptspace
		\divide\@tempdima 10
		\advance\dimen@-\@tempdima
		\ifdim\dimen@>\z@ \dimen@0pt\fi
		\rel@kern{0.6}\kern-\dimen@
		\if#31
		\overline{\rel@kern{-0.6}\kern\dimen@\macc@nucleus\rel@kern{0.4}\kern\dimen@}%
		\advance\dimen@0.4\dimexpr\macc@kerna
		\let\final@kern#2%
		\ifdim\dimen@<\z@ \let\final@kern1\fi
		\if\final@kern1 \kern-\dimen@\fi
		\else
		\overline{\rel@kern{-0.6}\kern\dimen@#1}%
		\fi
	}%
	\macc@depth\@ne
	\let\math@bgroup\@empty \let\math@egroup\macc@set@skewchar
	\mathsurround\z@ \frozen@everymath{\mathgroup\macc@group\relax}%
	\macc@set@skewchar\relax
	\let\mathaccentV\macc@nested@a
	\if#31
	\macc@nested@a\relax111{#1}%
	\else
	\def\gobble@till@marker##1\endmarker{}%
	\futurelet\first@char\gobble@till@marker#1\endmarker
	\ifcat\noexpand\first@char A\else
	\def\first@char{}%
	\fi
	\macc@nested@a\relax111{\first@char}%
	\fi
	\endgroup
}
\makeatother

\stackMath
\newcommand\reallywidehat[1]{%
	\savestack{\tmpbox}{\stretchto{%
			\scaleto{%
				\scalerel*[\widthof{\ensuremath{#1}}]{\kern-.6pt\bigwedge\kern-.6pt}%
				{\rule[-\textheight/2]{1ex}{\textheight}}
			}{\textheight}%
		}{0.5ex}}%
	\stackon[1pt]{#1}{\tmpbox}%
}

\newcommand{\gategroupColor}[7]{\POS"#1,#2"."#3,#2"."#1,#4"."#3,#4"!C*+<#5>[#7]\frm{#6}}

\usepackage[normalem]{ulem}

\begin{document}
\title{The composition rule for quantum systems is not the only possible one}

\author{Marco Erba~\orcidlink{0000-0002-2172-592X}}
\thanks{Corresponding author, electronic mail: \href{mailto:recobrama@gmail.com}{\texttt{recobrama@gmail.com}}.}
\affiliation{International Centre for Theory of Quantum Technologies (ICTQT), University of Gdańsk, ul.~Prof.~Marii Janion 4, 80-309 Gdańsk, Poland}
\author{Paolo Perinotti~\orcidlink{0000-0003-4825-4264}}
\affiliation{Universit\`a degli Studi di Pavia, Dipartimento di Fisica, QUIT Group, and INFN Gruppo IV, Sezione di Pavia, via Bassi 6, 27100 Pavia, Italy}
\begin{abstract}
	Quantum theory provides a significant example of two intermingling hallmarks of science: the ability to consistently combine physical systems and study them compositely, and the power to extract predictions in the form of correlations. A striking consequence of this facet is the violation of Bell inequalities, which has been experimentally demonstrated via Bell tests, thus attesting a classical/quantum divide. The prediction of this phenomenon originates as quantum systems are prescribed to combine according to the \emph{composition postulate}, i.e.~the tensor-product rule. This rule has also an operationally salient formulation---rather than just
	a purely mathematical one---given in terms of discriminability of composite states via local measurements. However, both the theoretical and the empirical status of such a postulate have been repeatedly challenged, questioning its independence from other physical principles---most notably from quantum postulates pertaining solely to single systems. Is the composition postulate the only viable way to combine quantum systems into a consistent physical theory? Here, this long-standing problem is resolved by answering in the negative. This is achieved by adopting an operational approach to physical theories and exhibiting a family of theories that differ from standard quantum theory in their system-composition rule. These theories have the same predictions as standard quantum theory as far as Bell-like correlation scenarios are concerned. Quantum theory is thus established to embody genuinely more than quantum correlations. As a result, foundational programmes based
	on single-system principles only, or on mere Bell-like
	correlations, are operationally incomplete. On the experimental side, ascertaining the independence of postulates is a fundamental step to adjudicate between quantum theory and alternative physical theories: hence, the composition postulate calls for
    experimental scrutiny independently of the other features of quantum theory.
\end{abstract}
\maketitle

{
	\small
	\begin{quote}
		\guillemetleft It is striking that it has so far not been possible to find a logically consistent theory that is close to quantum mechanics, other than quantum mechanics itself\guillemetright\ --- Steven Weinberg (cited in Ref.~\cite{aaronson2004quantummechanicsislandtheoryspace})
\end{quote}
}

Quantum theory enabled scientific practice to provide a physical account of a wide and ever-growing domain of empirical phenomena. Testing the way in which multiple quantum systems combine into composite systems has led to striking experimental discoveries, such as the existence of entanglement, which makes tasks like quantum teleportation~\cite{PhysRevLett.70.1895} both theoretically and practically feasible---despite the impossibility of cloning unknown quantum states~\cite{Wootters:1982aa,DIEKS1982271,10.1119/1.5021356}. Yet, the bewildering nature of entanglement has been problematised since the early years of quantum mechanics~\cite{PhysRev.47.777,PhysRev.48.696,schrodinger1935discussion,PhysicsPhysiqueFizika.1.195}. The existence of entangled states is essentially a manifestation of how quantum systems are presumed to mathematically combine into joint systems, as already noted by Schr\"odinger: \guillemetleft I would not call that \emph{one} but rather \emph{the} characteristic trait of quantum mechanics\guillemetright~\cite{schrodinger1935discussion}. On the theoretical level, every quantum system comes with an associated complex Hilbert space. When two distinct quantum systems are considered jointly, the resulting composite quantum system is associated with the tensor product of the Hilbert spaces of the individual components. It has frequently been noted from various perspectives that the rule for composing quantum systems lacks compelling physical justification, relying instead on heuristic principles~\cite{Rosen:1960aa,Aerts:0aa,Kennedy_1995,PhysRevA.61.022117,fuchs2002quantummechanicsquantuminformation,aaronson2018complex,barrett2007information,PhysRevLett.104.140401,PhysRevLett.109.090403,Hardy2013,Barnum_2014,Krumm:2019aa,10.1007/978-3-030-26980-7_67,PhysRevLett.126.110402,PhysRevLett.128.140401,PhysRevA.106.062406,PhysRevLett.130.110202,centeno2024twirledworldssymmetryinducedfailures,lismer2025experimentaltestprincipletomographic}.\footnote{This issue arises even in modern textbooks of quantum mechanics, see, e.g.,~Refs.~\cite[\S3.5]{ballentine2014quantum} and~\cite[\S2.2.8]{Nielsen_Chuang_2010}.} As put by Fuchs~\cite{fuchs2002quantummechanicsquantuminformation}: «The technical translation of this question is, why do we combine systems according to the
tensor-product rule?».
An operational answer was provided by the operational-theoretic reformulation of quantum theory~\cite{PhysRevA.84.012311}, that features the composition rule in terms of \emph{local discriminability}---i.e.~the possibility of distinguishing any two different states of a composite system just by using the statistics of measurements on the components. To study the interrelations between quantum principles, it has been instrumental, since the earliest approaches to quantum foundations~\cite{PhysRev.47.777,PhysicsPhysiqueFizika.1.195}, to highlight what is distinctive about quantum theory by contrasting it with theories in which fundamental properties may fail (this strategy is also referred to as the methodology of \emph{foil theories}~\cite{ChiribellaSpekkens2016}). Constructing such theories is an essential step for proving the independence of a set of quantum postulates, as, e.g., done in the \emph{quantum reconstruction programmes}~\cite{ChiribellaSpekkens2016}. Quantum theory can thus be identified against a landscape of conceivable theories by means of postulates, thereby enabling the examination of their interdependence within a theory-independent background framework. Here we adopt the framework of operational theories, and within the latter we address the long-standing question as to whether the postulates that single out quantum theory can be simplified by removing the composition postulate and deriving it from the others. This problem is equivalent to the question as to whether the tensor-product rule is the only viable way to combine systems that in all other respects abide by quantum theory, obtaining a consistent theory within the wider framework. What we show is that it is in fact possible to consistently introduce alternative combination prescriptions for quantum systems such that the resulting quantum theories: (a) differ from the standard quantum one solely in the composition postulate---i.e.~they satisfy all the remaining axioms of quantum theory---and (b) do not have a diminished predictive power. This establishes that any argument showing that the tensor-product rule follows from single-system postulates must inevitably appeal to some substitute assumption, provably needed to replace the former. What is more, the quantum theories here presented possess, in arbitrary correlation scenarios~\cite{Fritz_2012,Renou:2021aa}, the same set of correlations as standard quantum theory (these are the ones allowing, e.g.,~for experimental demonstrations of the violation of Bell inequalities~\cite{PhysicsPhysiqueFizika.1.195,PhysRevLett.47.460,PhysRevLett.115.250402,PhysRevLett.115.250401,Hensen:2015aa,:2022aa,nobel2022}, attesting a classical/quantum divide). As a consequence, it is impossible to experimentally reconstruct quantum theory by solely probing Bell-like scenarios. These results also naturally elicit a number of intriguing consequences and questions, which are here inquired.

\section{Origin and motivation of the problem}\label{subsec:standard_quantum}

\textbf{Composition postulate and superposition principle in quantum mechanics. }Originally, to the purpose of describing composite quantum systems, the tensor product was introduced by the founding parents of the nascent quantum mechanics as a natural choice borrowed from matrix mechanics as well as by its suitable Hilbert-space representation.
Indeed, every quantum system $\sys{Q}$ has a
complex Hilbert space $\mathscr{H}_{\sys{Q}}$ associated to it, where the \emph{pure states} of $\sys{Q}$---i.e.~the normalised vectors of $\mathscr{H}_{\sys{Q}}$---``live in''. Every pair of quantum systems $\sys{Q}_{\text{I}}$ and $\sys{Q}_{\text{II}}$ defines a composite quantum system $\sys{Q}_{\text{I}}\sys{Q}_{\text{II}}$, whose associated Hilbert space is the tensor product of the components: $\mathscr{H}_{\sys{Q}_{\text{I}}\sys{Q}_{\text{II}}}=\mathscr{H}_{\sys{Q}_{\text{I}}}\otimes\mathscr{H}_{\sys{Q}_{\text{II}}}$~\cite{weyl1928theory,dirac1930,von1955mathematical}.
This is called \emph{the composition postulate}, or \emph{tensor-product rule}. Equivalently, the states of the composite system $\sys{Q}_{\text{I}}\sys{Q}_{\text{II}}$ are given by all conceivable \emph{superpositions} of product states of the components.\footnote{Dirac describes this choice as follows~\cite[III, \S20]{dirac1930}: \guillemetleft The multiplication here is of quite a different kind from any that occurs earlier in the theory. The ket vectors [\dots\!] are in two different vector spaces and their product is in a third vector space, which may be called the product of the two previous vector spaces\guillemetright\ and \guillemetleft A general ket vector of the product space [\dots\!] is a sum or integral of kets of this form\guillemetright. A seminal attempt to justify this choice via physical requirements is due to Weyl~\cite[II, \S10]{weyl1928theory}. Von Neumann, after noting that tensor products comply with the remaining axioms of the nascent theory, concludes: \guillemetleft We therefore postulate them generally. (This is the customary procedure in quantum mechanics.)\guillemetright~\cite[VI, \S2]{von1955mathematical}. See also, interestingly, Ref.~\cite[\S2.2.8]{Nielsen_Chuang_2010} discussing the heuristics of this aspect.} The consequences of this \emph{choice} were investigated soon after~\cite{PhysRev.47.777,schrodinger1935discussion}, with puzzling results. However, initially the question of whether the tensor-product rule may have a fundamental status was not addressed, this postulate being simply accepted \emph{prima facie}. As observed by Kennedy~\cite{Kennedy_1995}: \guillemotleft The intuition here is classical: the subsystems possess their own individual characteristics and combining the systems merely combines their characteristics\guillemetright; these physical assumptions \guillemetleft were immediately plausible only in an atmosphere still colored by the worldview of classical mechanics\guillemetright. In essence, the guiding intuition behind this choice can be equivalently framed as follows: first and foremost, subsystems of a composite system can be prepared \emph{independently}---resulting in any possible product of states---and, secondly, the superpositions of product states are \emph{exhaustive} for the joint states. Great emphasis has been given throughout the decades to quantum principles which purely pertain to single systems. Feynman famously described the double-slit experiment, stemming from the superposition principle, as a phenomenon \guillemotleft which has in it the heart of quantum mechanics\guillemetright, containing \guillemetleft the \emph{only} mystery\guillemotright~\cite{feynman2011feynman}. Here, we adopt a constructive operational approach to physical theories and show that there exist a wealth of theories with the same superposition principle, yet that are different from standard quantum theory, thus highlighting the crucial relevance of the behaviour of composite systems in characterising the latter (see also Ref.~\cite[\S2.2.8]{Nielsen_Chuang_2010} about this point). In particular, we introduce a family of quantum theories respecting the standard superposition principle while violating the composition postulate.

\textbf{Standard quantum theory. }Quantum theory (\qt, here also referred to as \emph{standard \qt}) can be succintly, but comprehensively, framed as follows. Each quantum system $\sys{Q}$ is associated with a corresponding complex Hilbert space $\mathscr{Q}$. The family of quantum systems is closed under an associative composition rule,
and the Hilbert space associated with each composite $\sys Q=\sys{Q}_{1}\sys{Q}_{2}\cdots\sys{Q}_{n}$ is the tensor product $\mathscr Q=\mathscr{Q}_{1}\otimes\mathscr{Q}_{2}\otimes\cdots\otimes\mathscr{Q}_{n}$ of the individual Hilbert spaces $\mathscr Q_i$ associated with the components $\sys Q_i$. The \emph{states} of a system $\sys{Q}$ are all the positive semidefinite linear operators on the space $\mathscr{Q}$ with trace in $\left[0,1\right]$, called the \emph{density operators on $\mathscr{Q}$}. The physical transformations from a system $\sys{Q}_1$ to a system $\sys{Q}_2$ are given by the \emph{quantum operations} $\sys{Q}_1\to\sys{Q}_2$, namely, the completely positive and trace non-increasing (CPTNI) linear maps from the density operators on $\mathscr{Q}_1$ to those on $\mathscr{Q}_2$. A collection of quantum operations describes the possible events in a test if its elements sum up to a \emph{quantum channel}, i.e.~a completely positive trace-preserving (CPTP) linear map. As special cases of quantum tests, demolitive measurements are described by POVMs (more details in Section~\ref{subsec:framework}). Finally, quantum operations also compose according to the tensor product $\otimes$ of linear maps.

The above-presented \emph{density-operator formulation} of \qt\ 
is a standard in 
quantum information-theoretic treatments, and allows one to treat nondemolitive measurements and mixed states~\cite{ballentine2014quantum,Nielsen_Chuang_2010,bookDCP2017}. However, it is worth remarking that this density-operator formulation also embodies the traditional \emph{Hilbert-space formulation} of \qt~\cite{Nielsen_Chuang_2010,Renou:2021aa}.\footnote{This is due to the fact that the composition of pure states is still pure. Remarkably, there exist theories violating this property~\cite{PhysRevA.102.052216}.} In particular, the Hilbert-space formulation of \qt\ encodes, on the one hand, the single-system postulates: (P1) pure states are in a one-to-one correspondence with the rank-one projectors $\ket{\Psi}\!\!\bra{\Psi}$ on a complex Hilbert space such that $\text{Tr}{\ket{\Psi}\!\!\bra{\Psi}}=1$, (P2) states of isolated systems evolve unitarily (Schr\"odinger equation), and (P3) measurements are represented by projection-valued measures (PVMs),
namely, collections of self-adjoint, mutually orthogonal projections $\lbrace\Pi_{\tilde{a}}\rbrace_{\tilde{a}}$ such that $\text{P}{\left(a\middle|\lbrace\Pi_{\tilde{a}}\rbrace_{\tilde{a}};\Psi\right)}=\text{Tr}{\left(\ket{\Psi}\!\!\bra{\Psi}\Pi_a\right)}$ (Born rule); on the other hand, the composition postulate asserts that: (P4) the Hilbert space of composite systems is the tensor product of those of the components and independent processes (i.e.~states, unitaries, PVMs) compose according to the tensor product. This is important, since we will present a family of theories of quantum systems satisfying all of the single-system postulates, while violating the composition postulate, in both their density-operator and Hilbert-space formulations.

\textbf{Reconstruction programmes and the local accessibility of quantum states. }Inspired by Hardy's seminal work~\cite{hardy2001quantumtheoryreasonableaxioms} and by a new wave of foundational questions~\cite{fuchs2002quantummechanicsquantuminformation}, principled research programmes collectively known as \emph{quantum reconstruction programmes} resulted in several derivations (or ``reconstructions'') of quantum theory from first principles (see, e.g.,~\cite{hardy2001quantumtheoryreasonableaxioms,dakic2009quantum,Masanes_2011} and references therein). Some of these approaches \cite{PhysRevA.84.012311,bookDCP2017,Selby_2021} significantly emphasise the notion of \emph{compositionality} in physics~\cite{coecke2005kindergartenquantummechanics,Hardy2013}, and the formalism used provides an account for combining physical systems in an operational manner and studying how they behave jointly 
in a \emph{theory-independent} fashion. Physical processes are then abstractly treated as black-boxes, without making assumptions on their internal modes of operation. This in turn provides a way to study physical theories ``from the outside''. Decoupling quantum axioms is in turn fundamental to understanding what is essential and distinctive about the theory.

\begin{figure}
	\begin{center}
		\includegraphics[scale=0.37]{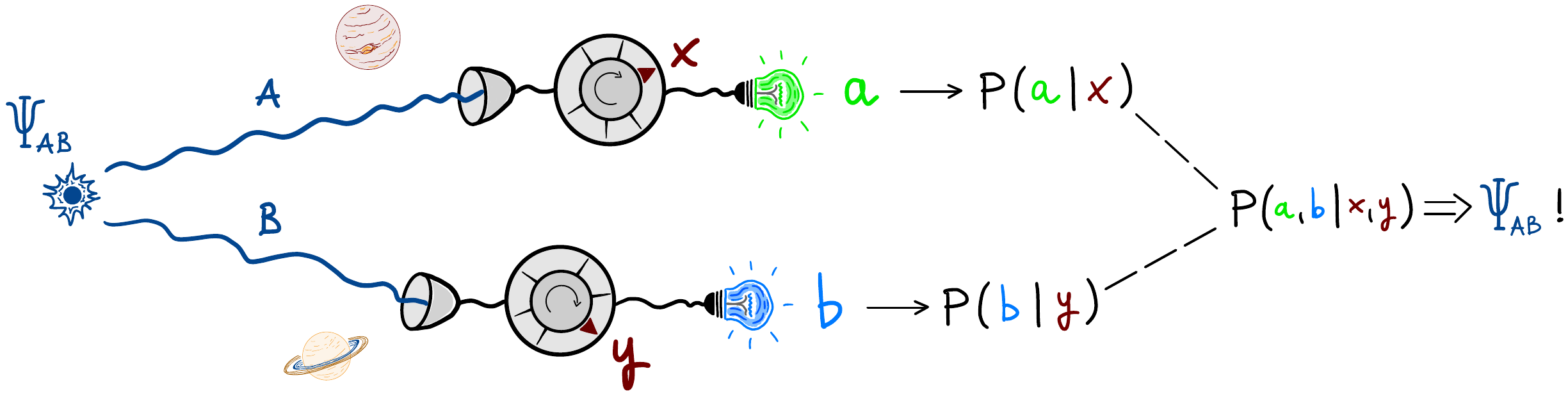}
	\end{center}
	\caption{
		Exemplification of the property of \emph{local tomography} satisfied by standard \qt. Two distant experimenters, Alice and Bob, gather, respectively, subsystem $\sys{A}$ and $\sys{B}$ of a standard quantum joint source $\sys{AB}$ emitting state $\Psi_{\sys{AB}}$. In their local laboratories, Alice and Bob independently choose, respectively, quantum measurements $x$ (with outcomes $a$) and $y$ (with outcomes $b$), thus observing local statistics $\text{P}{\left(a\middle|x\right)}$ and $\text{P}{\left(b\middle|y\right)}$. By communicating to each other the co-occurrences of their respective local outcomes via a line of classical communication (dashed lines), they are able to reconstruct the joint state of the standard quantum source: \guillemetleft In this sense quantum mechanics uses its information economically\guillemetright~\cite{wootters1990local}.
	}
	\label{fig:local_tomography}
\end{figure}

There have been several attempts at justifying the tensor-product rule from a variety of physical requirements by looking at quantum theory from the outside~\cite{hardy2001quantumtheoryreasonableaxioms,fuchs2002quantummechanicsquantuminformation,barrett2007information,dakic2009quantum,Masanes_2011,PhysRevLett.109.090403,bookDCP2017,10.1007/978-3-030-26980-7_67}. In their essence, these arguments appeal to a feature pinpointed by Wootters and dubbed ``local accessibility of quantum states''~\cite{wootters1990local}: \guillemetleft \emph{Any sets of measurements which are just sufficient for determining the states of the subsystems are, when performed jointly, also just sufficient for determining the state of the combined system}\guillemetright. This significant feature has been argued to reconcile quantum holism with scientific reductionism~\cite{bookDCP2017}, and has gained the status of a physical principle known as \emph{local tomography}~\cite{hardy2012limited} or \emph{local discriminability}~\cite{PhysRevA.81.062348,bookDCP2017}. Indeed, in the operational probabilistic framework~\cite{PhysRevA.81.062348,PhysRevA.84.012311,bookDCP2017}, local tomography allows one to derive the tensor-product rule for quantum systems in a theory-independent manner.\footnote{Notably, a related argument was already put forward by Aerts and Daubechies in the context of quantum logic~\cite{Aerts:0aa} (see also Ref.~\cite{centeno2024twirledworldssymmetryinducedfailures} and references therein for relevant works prior to the reconstruction programmes).}
The operational meaning and relevance of this principle are illustrated by Figure~\ref{fig:local_tomography}.
Besides, in a locally tomographic
scenario, arbitrary processes have the additional desirable property of being completely experimentally characterised by using input-states and output-measurements without the need for any ancillary systems (\emph{local process tomography})~\cite{bookDCP2017}. Accordingly, every quantum process $\T{T}$ can be characterised as the unique linear map $\hat{\T{T}}$ expressing its local action $\hat{\T{T}}{\left(\rho\right)}=\T{T}\rho$ on states $\rho$. These facts have interesting consequences for the alternative theories of quantum systems that we shall now present. In particular, these quantum theories \emph{must} violate local tomography, otherwise they would fall back into the tensor-product rule.

\section{Results}\label{subsec:framework}


\textbf{Setting the framework. }The superposition principle---being a single-system principle---must hold for both elementary and composite systems; accordingly, the tensor-product rule is implied by the conjunction of \emph{compositional compatibility of systems}---the products of \emph{all} local states can be formed independently---and \emph{exhaustivity of superpositions of product states}---the tensor-product space of the components \emph{entirely} describes the composite system. Accordingly, in order to modify the tensor-product rule it is necessary to drop either property. However, the price to pay, if states (and measurements) could not be prepared (nor performed) in a \emph{statistically independent} manner, would be too high. In particular, it would preclude an agent the experimental freedom of performing local experiments just because there exists another agent controlling a remote system.
This means that \emph{at least} the tensor product of quantum systems shall always be included in their joint system. For further arguments in defense of \emph{compositional compatibility of systems}, see, e.g.,~Refs.~\cite{PhysRevLett.126.110402,Renou:2021aa}, where the status of the principle in algebraic quantum field theory is also discussed. Therefore, our choice shall be to drop \emph{exhaustivity of superpositions of product states}.
Note that standard \qt\ is thus not restricted in any way, rather, it is extended.

We are now in the position to present the family of operational theories of quantum systems with an alternative composition postulate. These will be collectively called \emph{latent quantum theories} (\lqt s in brief). We resort to the framework of
\emph{operational probabilistic theories}~\cite{bookDCP2017}. This allows us to give an account of \lqt s consistent with both of the standard formulations of \qt\ recalled in Section~\ref{subsec:standard_quantum}. The systems of \lqt s are defined to be regular quantum systems, and their states, which we refer to as \emph{latent quantum states}, are given by the usual set of quantum states. For every latent quantum system $\sys{Q}$, its states correspond to density matrices $\rho$ on the Hilbert space of the system, and will be conveniently denoted by $\rket{\rho}_{\sys{Q}}$. The transformations of the theory have a type $\sys{Q}\to\sys{H}$ identifying their input and output systems, and their mathematical representation will be discussed later. Transformations are denoted by the calligraphic notation $\T{G}$, and every pair $\T{G}\colon\sys{Q}\to\sys{H}$ and $\T{F}\colon\sys{H}\to\sys{F}$ is, as in standard \qt, \emph{sequentially composable}, namely, $\T{F}\T{G}\colon\sys{Q}\to\sys{F}$ is still a transformation of the theory. Sequential composition is
associative. As a special case of transformations, the measurements (effects) of a latent quantum system $\sys{Q}$ will be denoted by $\rbra{a}_{\sys{Q}}$, while their pairing with states by $\rbraket{a}{\rho}_{\sys{Q}}\in\left[0,1\right]$ (i.e.~the Born rule). The \emph{latent quantum effects} of a system $\sys{Q}$ are, as in the standard density-operator formulation of \qt, elements of positive operator-valued measures (POVMs), the latter being collections of functionals $\lbrace\rbra{{\tilde{a}}}_{\sys{Q}}\coloneqq\text{Tr}{\left(\left.\cdot\right.\Pi_{\tilde{a}}\right)}\rbrace_{\tilde{a}}$ where every $\Pi_a$ is an arbitrary operator on $\mathscr{Q}$ such that $\mathbb{0}_{\mathscr{Q}}\leq\Pi_a\leq\mathbb{1}_{\mathscr{Q}}$ and $\sum_{\tilde{a}}\Pi_{\tilde{a}}=\mathbb{1}_{\mathscr{Q}}$. Note that POVMs contain PVMs---the measurements in the standard Hilbert-space formulation of \qt---as a strict subset. The composition rule (\emph{parallel composition}), denoted by $\boxtimes$, is required to be \emph{associative}---just like $\otimes$ is---because the theoretical partition of a physical system into subsystems shall not affect the predictions of the theory~\cite{hardy2012limited,Hardy2013,Masanes:2019aa,PhysRevA.102.052216}. Importantly, the operations of sequential and parallel compositions are required to satisfy a salient property, called the \emph{interchange law}, as it happens in standard \qt\ for tensor products:
\begin{equation}\label{eq:interchange_law}
	\left(\T{B}\boxtimes\T{D}\right)\left(\T{A}\boxtimes\T{C}\right)
	=
	\left(\T{BA}\right)\boxtimes\left(\T{DC}\right)
	.
\end{equation}
This means that sequential and parallel compositions are compatible, and, accordingly, they define reasonable notions of \emph{local and statistically independent operations}, encapsulating the property of \emph{compositional compatibility of systems}. Moreover, the existence of \emph{identity processes} (``doing nothing'' on systems) is required; these are the neutral elements of sequential composition and satisfy the property that when two identities are composed in parallel, the resulting process is still an identity (``doing nothing locally amounts to doing nothing globally''). Such processes, via the interchange law~\eqref{eq:interchange_law}, are pivotal in defining how agents can act locally within their laboratories, thereby ensuring the possibility of performing local experiments without interacting with remote systems. Finally, the ability to consistently swap systems between local laboratories is also demanded. More details about these operational \emph{desiderata} can be found in Appendix~\ref{app:quantum_theory_operational}.

Now, there will be essentially two main variations, on a formal level, with respect to standard \qt. First, the composition rule $\boxtimes$, that shall be presented below, \emph{is not the usual tensor product} $\otimes$. In addition, in light of how the specific alternative product rule is defined, the family of theories that we are going to introduce also violates local process tomography. However, a universal minimal ancilla---the same for all processes in the theories---is sufficient to perform process tomography. This constrains the degree of holism required to experimentally characterise the physical transformations, as it also happens in physical theories with superselection rules, such as fermionic quantum theory or ``real quantum theory''.\footnote{By ``real quantum theory'' we precisely mean the density-operator formulation of \qt\ based on \emph{real} (as opposed to \emph{complex}) Hilbert spaces. For relevant examples of theories violating local tomography see Refs.~\cite{PhysRevA.102.052216,Renou:2021aa,centeno2024twirledworldssymmetryinducedfailures} and references therein, or Appendix~\ref{app:transformations}.} As a consequence, in order to characterise a \emph{latent quantum operation} $\T{G}\colon\sys{Q}\to\sys{H}$ it is necessary to probe its action in the presence of an (arbitrary, but non-trivial) ancillary system $\sys{E}$, i.e.~$\left(\T{G}\boxtimes\T{I}_{\sys{E}}\right)\rket{\rho}_{\sys{Q}\sys{E}}=\rket{\widehat{\T{G}\boxtimes\T{I}_{\sys{E}}}{\left(\rho\right)}}_{\sys{H}\sys{E}}$, where $\T{I}_{\sys{E}}$ denotes the identity process (``doing nothing'') on system $\sys{E}$.
More details on this feature can be found in Appendix~\ref{app:transformations}.

\textbf{The quantum realm is compatible with alternative composition rules. }The general infinite family of \lqt s is constructed in Appendix~\ref{app:lqts}. Here, to fix ideas, we illustrate the simplest example of a latent quantum theory. Let us consider the \emph{latent quantum composition} of $n$ elementary quantum systems $\{\sys{Q}_{1},\sys{Q}_{2},\ldots,\sys{Q}_{n}\}$. This is the rule assigning a Hilbert space to the resulting latent quantum composite $\sys{Q}_{1}\sys{Q}_{2}\cdots\sys{Q}_{n}$, reading:
\begin{equation}\label{eq:composition}
	\mathscr{Q}_{1}\boxtimes\mathscr{Q}_{2}\boxtimes\cdots\boxtimes\mathscr{Q}_{n}
	=
	\underbrace{\mathscr{L}^{\otimes{n\choose2}}}_{\coloneqq\mathscr{L}_n}\otimes\left(\mathscr{Q}_{1}\otimes\mathscr{Q}_{2}\otimes\cdots\otimes\mathscr{Q}_{n}\right)
	,
\end{equation}
where $\mathscr{L}$ is an arbitrary Hilbert space called the \emph{latent factor}. Note that $n\choose2$ is the number of unordered pairs of $n$ elements, accordingly each copy of the latent factor is thought to be a degree of freedom expressing a ``link'' between each pair of subsystems. The states of latent quantum composites $\sys{Q}_{1}\sys{Q}_{2}\cdots\sys{Q}_{n}$ are just the density operators on $\mathscr{Q}_{1}\boxtimes\mathscr{Q}_{2}\boxtimes\cdots\boxtimes\mathscr{Q}_{n}$ as per~\eqref{eq:composition}. An important consequence of rule~\eqref{eq:composition} is that
\begin{equation*}
	\mathscr{Q}_{1}\otimes\mathscr{Q}_{2}\otimes\cdots\otimes\mathscr{Q}_{n}\subseteq\mathscr{Q}_{1}\boxtimes\mathscr{Q}_{2}\boxtimes\cdots\boxtimes\mathscr{Q}_{n}
	,
\end{equation*}
namely, the standard quantum products are always included in the latent quantum composites (as a manifestation of the above-discussed compositional compatibility of systems). Note that by choosing $\mathscr{L}\cong\mathbb{C}$ one recovers standard \qt. Locally, latent quantum systems (both elementary and composite) look exactly like standard quantum ones. This alternative composition postulate, though, is still not enough to specify the full theory. Indeed, one also needs to postulate how processes compose in parallel, as this is now not granted as in the case of $\otimes$. Importantly, this choice is decisive also for assessing the associativity of $\boxtimes$ and whether the interchange law~\eqref{eq:interchange_law} is satisfied. States are prescribed to compose as follows:
\begin{equation*}
	\rket{\rho_1}_{\sys{Q}_{1}}\boxtimes\rket{\rho_2}_{\sys{Q}_{2}}\boxtimes\cdots\boxtimes\rket{\rho_n}_{\sys{Q}_{n}}
	=
	\rket{\xi}_{\sys{L}}^{\otimes{n\choose2}}\otimes\rket{\rho_1}_{\sys{Q}_{1}}\otimes\rket{\rho_2}_{\sys{Q}_{2}}\otimes\cdots\otimes\rket{\rho_n}_{\sys{Q}_{n}}
	,
\end{equation*}
where $\rket{\xi}_{\sys{L}}$ is a fixed pure state of the latent factor, expressing that every pair of subsystems is bound by a product relation. Note that pure states compose to pure states, just as in the usual \qt. Furthermore, latent quantum effects, being POVM elements, are prescribed to compose as follows:
\begin{equation}\label{eq:effect_composition}
	\rbra{a_1}_{\sys{Q}_1}\boxtimes\rbra{a_2}_{\sys{Q}_2}\boxtimes\cdots\boxtimes\rbra{a_n}_{\sys{Q}_n}
	=
	\text{Tr}{\left\{\left.\cdot\right.\left(\mathbb{1}_{\mathscr{L}_{n}}\otimes\Pi_{a_1}\otimes\Pi_{a_2}\otimes\cdots\otimes\Pi_{a_n}\right)\right\}}
	,
\end{equation}
Note that not only POVMs compose to POVMs, but also PVMs compose to PVMs, as 
in 
standard \qt.

We now illustrate how arbitrary processes are defined and how they compose.
First, let us define arbitrary processes $\sys{S}\to\sys{S}$ for every elementary latent quantum system $\sys{S}$: these are characterised as pairs $\T{T}^{(i)}=\left(\hat{\T{T}},\hat{\T{Z}}^{(i)}\right)$, where $\hat{\T{T}}$ is the linear map associated with a standard quantum operation $\sys{S}\to\sys{S}$, while $i\in\lbrace0,1\rbrace$ with $\hat{\T{Z}}^{(0)}=\T{I}_{\sys{L}}$ (identity process) and $\hat{\T{Z}}^{(1)}=\rket{\xi}_{\sys{L}}\rbra{\text{Tr}}_{\sys{L}}$ (both standard quantum operations). The general rationale behind this specific form can be found in Appendix~\ref{app:transformations}. Let us now consider two elementary latent quantum systems $\sys{Q}$ and $\sys{E}$; we can then define the composition of processes on the latent quantum composite $\sys{Q}\sys{E}$:
\begin{equation}\label{eq:process_composition}
	\left(\T{F}^{(i)}\boxtimes\T{G}^{(j)}\right)\rket{\Sigma}_{\sys{Q}\sys{E}}
	\coloneqq
	\left[\left(\hat{\T{Z}}^{(j)}\hat{\T{Z}}^{(i)}\right)\otimes\hat{\T{F}}\otimes\hat{\T{G}}\right]\rket{\Sigma}_{\sys{Q}\sys{E}}
	.
\end{equation}
Via the composition rule~\eqref{eq:process_composition}, the interchange law~\eqref{eq:interchange_law} is easily verified for this particular scenario.

\begin{figure}
	\begin{center}
		\includegraphics[scale=0.37]{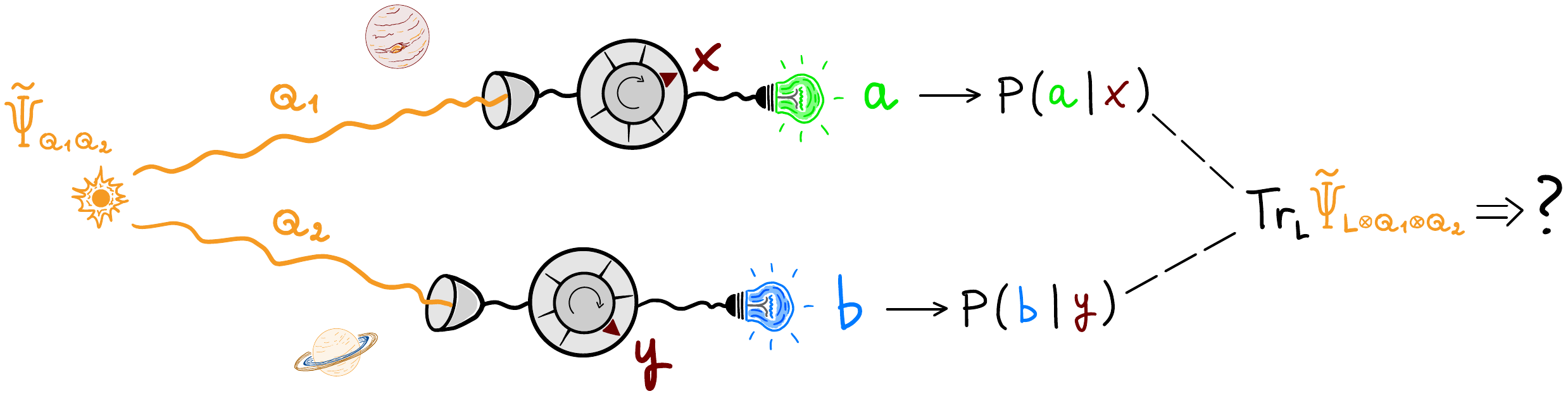}
	\end{center}
	\caption{
		Example of \emph{violation of local tomography} in \lqt s. The scenario is the same as the one in Figure~\ref{fig:local_tomography}, except that Alice and Bob here have access to their respective subsystem of a joint \lqt~source $\sys{Q}_1\sys{Q}_2$ emitting state $\tilde{\Psi}_{\sys{Q}_1\sys{Q}_2}$. 
		The two parties are now able to experimentally reconstruct just $\text{Tr}_{\sys{L}}{\tilde{\Psi}_{\sys{Q}_1\sys{Q}_2}}$, namely, the standard quantum counterpart of the full state after discarding the latent factor $\sys{L}$; yet, they cannot reconstruct the full state $\tilde{\Psi}_{\sys{Q}_1\sys{Q}_2}$ by just performing independent local measurements. Then, the latent factor can be considered a \emph{locally hidden}
		degree of freedom popping up \emph{just} when subsystems are considered within their composite. It can be \emph{locally manipulated} (with suitable restrictions) but cannot be \emph{measured} unless the parties cooperate \emph{by performing global measurements}. This is opposed to, e.g.,~the model studied in Ref.~\cite{PhysRevA.87.052106}, where the extra factor corresponds to a system that can be
	manipulated and measured by any observer.}
	\label{fig:nonlocal_tomography}
\end{figure}

In the Appendices, the above examples are generalised, also accounting for an arbitrary number of (generally composite) latent quantum systems, by using a convenient circuital representation. In Appendix~\ref{app:check}, it is verified that every \lqt\ defines an operational probabilistic theory of quantum systems: (i) \lqt s' systems are quantum systems, (ii) \lqt s comply with the operational \emph{desiderata}, and therefore (iii) \emph{\lqt s are theories differing from \qt\ just in the composition postulate}. Moreover, as anticipated, \lqt s exhibit a violation of local tomography, that is exemplified in Figure~\ref{fig:nonlocal_tomography}.
It is manifest that no reference is made to the dimensionalities of the systems, since the construction of these theories is grounded solely on the mathematical existence of the usual tensor product of Hilbert spaces. In fact, \emph{\lqt s are defined in any dimension}, finite or infinite.

A final result regards a paramount quantum feature marking a departure from a classical worldview: the violation of Bell inequalities~\cite{PhysicsPhysiqueFizika.1.195,nobel2022}, which is experimentally demonstrated by \emph{Bell tests}~\cite{Hensen:2015aa,Renou:2021aa,nobel2022}. These consist in experiments probing \emph{$n$-party scenarios}, where each agent can independently choose some local measurements $\left(x_1,x_2,\ldots,x_n\right)$, with corresponding outcomes $\left(a_1,a_2,\ldots,a_n\right)$, to be performed on a shared setting $\Sigma$ (a $n$-partite state); the sets of Bell-like correlations are then given by all of the probability tables of the form $\text{P}{\left(a_1,a_2,\ldots,a_n\middle|x_1,x_2,\ldots,x_n;\Sigma\right)}$ provided by a theory.
Remarkably, the set of latent quantum Bell-like correlations coincides with the quantum ones. By resorting to rules~\eqref{eq:composition} and~\eqref{eq:effect_composition}, this can be easily proven for scenarios of elementary systems: $\text{P}_{\lqt}{\left(a_1,a_2,\ldots,a_n\middle|x_1,x_2,\ldots,x_n;\Sigma\right)}\equiv\left(\rbra{a_1^{(x_1)}}_{\sys{Q}_{1}}\boxtimes\rbra{a_2^{(x_2)}}_{\sys{Q}_{2}}\boxtimes\cdots\boxtimes\rbra{a_n^{(x_n)}}_{\sys{Q}_{n}}\right)\rket{\Sigma}=\left(\rbra{a_1^{(x_1)}}_{\sys{Q}_{1}}\otimes\rbra{a_2^{(x_2)}}_{\sys{Q}_{2}}\otimes\cdots\otimes\rbra{a_n^{(x_n)}}_{\sys{Q}_{n}}\right)\rket{\text{Tr}_{\mathscr{L}_n}{\Sigma}}\equiv\text{P}_{\qt}{\left(a_1,a_2,\ldots,a_n\middle|x_1,x_2,\ldots,x_n;\text{Tr}_{\mathscr{L}_n}\Sigma\right)}$.
The proof of the general case is given in Appendix~\ref{app:correlations}, where it is also proven that the correspondence between latent and standard quantum scenarios
preserves the causal relations between the operations,
i.e.~\emph{the product (space-like) structure of both states and measurements is retained} (on the relevance of arbitrary prepare-and-measure networks see, e.g.,~Refs.~\cite{Fritz_2012,Renou:2021aa}).
Finally, the construction introduced here (in Appendices~\ref{app:preliminaries} and~\ref{app:lqts}) represents an advancement in the formulation of new theories, being applicable to every arbitrary (possibly yet unknown) parent theory.

\section{Discussion}
{
	\small
	\begin{quote}
		\guillemetleft Of every would be describer of the universe one has
		a right to ask immediately two general questions.
		The first is: ``What are the materials of your universe's composition?'' And the second: ``In what
		manner or manners do you represent them to be
		connected?''\guillemetright\ --- William James (cited in Ref.~\cite{RevModPhys.85.1693})
\end{quote}}

This work is not intended to show that the tensor-product rule is theoretically or empirically inadequate. On the contrary, the results of the paper highlight its importance in the reconstruction of standard quantum theory. 
The \emph{latent quantum theories} presented here are, in fact, a family of \emph{generalisations of standard quantum theory} that violate the tensor-product rule. These quantum theories provide a reasonable \emph{extension} of standard quantum theory, subsuming it as a special case.

\textbf{Fundamental consequences. }Our results are complementary to those of Ref.~\cite{Navascues:2015aa}: they show that it is possible to modify the operational structure of quantum theory without running into consequences for quantum correlations. In principle, it is by no means obvious that alternative composition postulates would even be logically, or empirically, consistent with a theory of quantum systems.\footnote{In fact, it can be shown that the quantum generalisation of the composition rule presented in Ref.~\cite{PhysRevA.102.052216} leads to an obstruction in meeting the operational \emph{desiderata}:
accordingly, these requirements are, in this respect, {genuinely selective}.} On the one hand, had we found that an operational theory of quantum systems necessarily satisfied the standard composition rule, this instance of the postulate would have been redundant within the framework. Conversely, our result effectively decouples the tensor-product rule from other physical principles, notably from single-system quantum postulates---while at the same time \emph{extending} quantum theory, as a further desirable feature. This contributes to a structural mapping of the interrelations between the quantum principles, highlighting what is distinctive about quantum composition by contrasting it with fully quantum theories where the fundamental property of interest---that is, the tensor-product rule---concretely fails. Accordingly, such a valid counter-model also directly informs prospective examinations of quantum postulates. Indeed, we showed that a theory of quantum systems can violate the tensor-product rule while preserving every other single-system feature of quantum states and transformations.
The theories presented here are linear and exhibit
neither superluminal signalling nor supra-quantum correlations~\cite{Popescu:1994aa,Navascues:2015aa}. Crucially, they respect all proposed principles to capture quantum correlations~\cite{Navascues:2015aa} (such as \emph{information causality}~\cite{Pawowski:2009ab}), showing that modifying the composition postulate need not entail features that are widely regarded as problematic. Furthermore, generic \emph{correlation scenarios}~\cite{Fritz_2012}---despite being sufficient for certifying a \emph{classical/quantum divide}, or even a \emph{real/complex} one within the quantum domain~\cite{Renou:2021aa}---are here established to be insufficient for singling out standard quantum theory, even when the preservation of the causal (space-like) structure is imposed---as was done, e.g.,~in Ref.~\cite{Renou:2021aa}. This is the demonstration that there are aspects of the theory that cannot be inferred by correlation scenarios, namely, that even multipartite principles characterising quantum correlations alone~\cite{PhysRevLett.107.210403,Fritz:2013aa} would be insufficient to recover the full theory. Quantum theory is thus established to embody genuinely more than just quantum correlations.

Our work has several significant foundational consequences. There exists a widespread belief that the quantum composition rule may be redundant, as it could be implied by other quantum principles~\cite{PhysRevLett.126.110402}. In this regard, it is worth noting that endeavours to derive or study other quantum postulates ``from the outside''---like the Born rule~\cite{Masanes:2019aa,fiorentino2025gleasonstheoremqubitcomposite}, collapse~\cite{fiorentino2023quantumtheorynoncollapsingmeasurements}, Schr\"odinger equation or general dynamics~\cite{wilson2023originlinearityunitarityquantum}, and even complex
Hilbert spaces~\cite{Renou:2021aa}---typically assume and employ the composition postulate.\footnote{See also Ref.~\cite{Kennedy_1995} about the circularity of non-signalling proofs.}
On the one hand, in recent years interesting proposals have been put forward for characterising single-system quantum theory~\cite{Barnum_2014} or recovering the tensor-product rule for pure states~\cite{PhysRevLett.126.110402}: our work demonstrates that imposing a compositional assumption---such as local discriminability or an analogous principle---is unavoidable should one aim to single out the tensor-product rule within a landscape of conceivable theories of quantum systems. On the other hand, ambitious and intriguing reconstruction programmes aim at constraining quantum theory on the sole basis of the rules for calculating probabilities of events, which the predictive power of a theory is hinged upon. For instance, this is the case for endeavours like \emph{QBism}~\cite{RevModPhys.85.1693}---a well-known interpretation of quantum theory---or the \emph{device-independent programme}~\cite{Navascues:2015aa}---providing a framework for the most robust experimental demonstrations of the non-classicality of quantum theory via Bell tests.
However, such approaches are now prompted to confront the existence of conceivable alternatives to standard quantum theory demonstrating that the composition postulate is not operationally redundant. What is more, the existing statistics-based approaches are thus demonstrated to be operationally incomplete---in the sense that they cannot single out quantum theory \emph{qua physical theory}---unless they incorporate further hypotheses, e.g.~on the nature of subsystems or of the correlation scenarios considered. Accordingly, their discriminating power is inherently limited: for instance, Bell-like scenarios need to be surpassed in order to achieve a thorough reconstruction of quantum theory, or even just to experimentally probe the composition postulate. Remarkably, this limitation is also deeply connected to Tsirelson's problem~\cite{scholz2008tsirelsonsproblem}, to the general non-factorisability of multi-party Hilbert spaces in algebraic formulations of quantum field theory~\cite{Fewster:2016aa}, and, finally, to the very axiomatisation of the tensor product of Hilbert spaces~\cite{doi:10.1073/pnas.2117024119}. Therefore, the present results also represent a step in pursuing a reasonable notion of space-like separation for local laboratories within relativistic quantum arenas. For example, operational probabilistic theories of quantum systems naturally incorporate (via the interchange law~\eqref{eq:interchange_law}, yet relaxing the tensor-product rule) commutator-models~\cite{scholz2008tsirelsonsproblem} for space-like separated measurements. Therefore, modelling infinite-dimensional quantum theories (where the tensor-model and the commutator-model are known to be inequivalent~\cite{cabello2025possibleconsequencesphysicsnegative}) as fully-fledged operational theories, can result in singling out specific models, for space-like separation, based on operational grounds.

\textbf{Experimental pathways. }Beyond these foundational consequences, the independence of the composition postulate also opens up new perspectives across different areas of physics and technology.
Within the modern sciences, the experimental study of systems behaving as complex networks is manifest across physics---for instance, in the interaction between celestial bodies like coalescing black holes, in the coupling of information carriers in computation, and even in the interplay between gravitational and quantum degrees of freedom~\cite{Galley_2023}. Moreover, this aspect is tightly related to the study of causal structures, which is a fundamental part of the scientific method.
On the technological side, this is a crucial aspect for the
\emph{quantum internet}~\cite{doi:10.1126/science.aam9288} which is currently under way. Therefore, interestingly, our results may find an application to the semi-device-independent framework~\cite{VanHimbeeck2017semidevice} and to the phenomenon of dimension leakage in quantum computers~\cite{rybotycki2024deviceindependentdimensionleakagenull}: in particular, our work shows that adopting overarching assumptions on the dimensions of systems is highly restrictive even in a purely quantum domain. Indeed, the simplest experimental setting where latent quantum theories predict observable departures from standard quantum theory is the one depicted by Fig.~\ref{fig:nonlocal_tomography}, modified to allow for global measurements, thus becoming a \emph{time-like test}~\cite{PhysRevLett.128.140401,PhysRevA.106.062406}. Such an experiment would seek for evidence that the ``operational size''---or ``dimension'', which can be experimentally constrained---$D_{\sys{AB}}$ of a composite system $\sys{AB}$ would exceed the product $D_{\sys{A}}D_{\sys{B}}$ of the individual ``sizes'' of the components $\sys{A}$ and $\sys{B}$. This could be done by, e.g.,~finding that the maximum number of jointly perfectly discriminable states of the composite system is greater than the product of those of the presumed components. Moreover, the theories of quantum systems formulated here exhibit, by construction, a wealth of novel phenomena that would be interesting to investigate,
such as other testable unusual behaviours in time-like scenarios~\cite{PhysRevLett.119.020401,PhysRevLett.128.140401,PhysRevA.106.062406,PhysRevLett.130.110202} and in \emph{higher-order dynamics} (a highly active area of research whose paradigmatic example is~\emph{indefinite causal order}~\cite{Oreshkov_2012,Chiribella_2013}). The latter aspects are deeply related to our understanding and formulation of space-time theories---including, but not limited to, those where gravity is described as an emergent phenomenon.

Some authors have argued that local tomography is not falsifiable, and that its violations can be explained away by appealing to technological or fundamental limitations (some of these arguments are reviewed, e.g.,~in Refs.~\cite{PhysRevLett.126.110402,Renou:2021aa}). Although it is true that this can be consistently done in a variety of physically relevant cases~\cite{centeno2024twirledworldssymmetryinducedfailures}, it is not obvious whether this stance is in principle always tenable: on the contrary, our results strongly challenge this view,\footnote{In fact, there is evidence that a large class of the theories here presented does not admit of an \emph{ontological model}~\cite{Schmid_2024,soltani2025noncontextualontologicalmodelsoperational} in terms of standard quantum theory and even of any theory which abides by local tomography. In this sense, they could not be considered subtheories of standard quantum theory, featuring a genuine violation of local tomography that cannot be explained away by an ontological model. This feature contrasts with the fact that every prepare-and-measure correlation scenario, in the theories introduced, admits of a standard-quantum \emph{hidden-variable model}---the latter being, unlike ontological models, insensitive to composition rules---thus highlighting another remarkable aspect of compositional properties in physical theories (see also Ref.~\cite{soltani2025decouplinglocalclassicalityclassical}).} which is not grounded in any experimental observation. \emph{Latent quantum theories} allow for testable predictions departing from standard quantum theory, including beyond-standard-quantum phenomena such as \emph{hyperdense coding}~\cite{Massar_2015}, \emph{superadditivity of classical capacities}~\cite{Massar_2015}, and \emph{hypersignalling}~\cite{PhysRevLett.119.020401}. This means that local tomography might be relaxed at a fundamental level, and even experimentally violated, \emph{without necessarily implying a breakdown of the quantum nature of the systems probed as single systems}---but rather, possibly, of the tensor-product rule alone. Furthermore, there exist recent experimental proposals for adjudicating between different composition rules for quantum theory, see~\cite{PhysRevLett.128.140401,PhysRevA.106.062406,PhysRevLett.130.110202,lismer2025experimentaltestprincipletomographic} and references therein. However, none of the aforementioned experimental scenarios are \emph{device-independent}: the effects of the above-discussed dimensional mismatches are in \emph{principle standard-quantum embeddable} (unlike other, dimensionally robust violations, such as that of Tsirelson's bound); accordingly, they would not provide \emph{per se} conclusive tests of quantum composition in the same compelling way as that offered by the violation of Bell inequalities for testing hidden-variable models. This facet might be viewed as a drawback of the above time-like scenarios, yet at the same time providing an interesting and stimulating perspective. Indeed, devising such a bona fide test is one of the main experimental challenges posed by our results. An intriguing possibility for experimentally testing the composition postulate---informed by device-independent principles yet surpassing Bell-like scenarios---would be to consider \emph{dynamical causal structures} (i.e.~higher-order dynamics), a line of research that is left for future work.

\textbf{Outlooks. }In a sense made precise in this work, Schr\"odinger truly captured an essential attribute of quantum theory: the composition postulate is «not \emph{one} but rather \emph{the} characteristic trait of quantum mechanics»~\cite{schrodinger1935discussion}---even amongst the conceivable operational theories of quantum systems. This work identifies a new piece in the foundational programme of reconstructing quantum theory from physical principles and its search for an experimental disproof of the alternatives. Quantum composition yields holistic, non-classical phenomena~\cite{PhysicsPhysiqueFizika.1.195,PhysRevLett.47.460,PhysRevLett.115.250402,PhysRevLett.115.250401,Hensen:2015aa,:2022aa,nobel2022} whose prediction does not single out a unique composition postulate. In the context of space-time theories, especially general relativity, the problem of \emph{underdetermination}~\cite{sep-scientific-underdetermination} has been frequently raised,
being relevant also in the present context. Although quantum theory and its modifications presented here share the same set of Bell-like correlations, they are inequivalent, quite different in fact, as fully-fledged theories.
Is quantum theory underdetermined by our current experimental methods~\cite{PhysRevLett.125.060406}? Under what physical conditions might quantum systems exhibit correlations beyond the standard ones~\cite{PhysRevLett.104.140401}? The problem of experimentally testing the composition postulate poses several interesting and independent challenges in its own right. The present work paves the way for overcoming these challenges. First, our work shows that fixed prepare-and-measure networks are fundamentally insufficient for detecting deviations from the standard quantum composition \emph{in the absence of dimensional assumptions}. In light of this, besides considering dynamical causal structures as suggested above, the question of thoroughly surveying correlation scenarios with arbitrary causal structures~\cite{Fritz_2016} remains an open problem. In particular, this entails searching for deviations in scenarios where arbitrary networks (i.e.~arbitrary transformations) are taken into account, thus highlighting the importance of general processes, and the compositionality thereof, in these experimental contexts. More generally, the theories presented offer a concrete test bench to inspire the formulation of new compositional principles and guide their incorporation into experimental scenarios, ultimately aiming either to single out standard quantum theory within a landscape of possible theories or, conversely, to supersede standard quantum composition---yet, crucially, without renouncing the quantum nature of the systems.

The theories of quantum systems presented here show that it is actually \guillemetleft possible to find a logically consistent theory that is close to quantum mechanics, other than quantum mechanics itself\guillemetright.\footnote{Steven Weinberg (cited in Ref.~\cite{aaronson2004quantummechanicsislandtheoryspace}), as many others (see, e.g.,~Ref.~\cite{Navascues:2015aa}), considered the difficulty in doing so a \guillemetleft striking\guillemetright\ fact.} Quantum theory---at least among the logically conceivable theories of quantum systems---is not ``an island in theoryspace''~\cite{aaronson2004quantummechanicsislandtheoryspace}, as the principle of local tomography can be provably relaxed. Our work effectively decouples the quantum composition postulate from the other (single-system) quantum principles---\emph{both from a theoretical and from an empirical standpoint}; this calls for a reconsideration of the current theoretical relevance and empirical status of this postulate \emph{independently of the other quantum principles}. Indeed, there have been several proposals that local tomography, or quantum compositionality \emph{tout court}, is challenged or subjected to experimental scrutiny~\cite{Rosen:1960aa,Kennedy_1995,PhysRevLett.109.090403,PhysRevLett.128.140401,PhysRevA.106.062406,PhysRevLett.130.110202,centeno2024twirledworldssymmetryinducedfailures,lismer2025experimentaltestprincipletomographic}. Our results make the pursuit of an experimental justification for quantum composition fundamentally compelling.


\section*{Acknowledgements}
All diagrams in the Appendices are typeset using the package \href{https://ctan.org/pkg/qcircuit}{\texttt{qcircuit}} and
the editor \href{https://tikzit.github.io/}{\texttt{TikZiT}}. P.P.~acknowledges financial support from European Union - Next Generation EU through the PNNR MUR Project No.~PE0000023-NQSTI. M.E.~acknowledges financial support from the National Science Centre, Poland (Opus Project, Categorical Foundations of the Non-Classicality of Nature, Project No.~2021/41/B/ST2/03149 and Grant Sonata 16 No.~2020/39/D/ST2/01234).

\section*{Author contributions, Competing interests, Data availability}
{All authors contributed equally to this work. The authors declare no competing interests. There are no data to be shared.}


\bibliography{opt_bib}

\appendix

\section{Preliminaries}\label{app:preliminaries}

\subsection{Quantum theory and operational theories}\label{app:quantum_theory_operational}

Quantum theory (\qt, also referred to as \emph{standard} \qt) can be framed, broadly speaking, as an \emph{operational probabilistic theory} of quantum systems~\cite{PhysRevA.81.062348,chiribella2016quantum,bookDCP2017}. As discussed in Sections~\ref{subsec:standard_quantum} and~\ref{subsec:framework} of the main text, the operational formalism allows one to study the properties of general physical theories ``from outside'', investigating the interdependence of relevant axioms~\cite{hardy1999disentanglingnonlocalityteleportation,PhysRevA.102.052216,D_Ariano_2020,PhysRevA.109.022239,rolino2024minimaloperationaltheoriesclassical}. The primitives of the operational probabilistic framework are \emph{systems}, \emph{processes}, and their modes of \emph{compositions}. We first present standard \qt\ in terms of these primitives, and then we will relate it to a more general circuital representation that we will be using in the following.

Each quantum system $\sys{Q}$ is associated with a complex Hilbert space $\mathscr{Q}$, and the family of quantum systems is closed under an associative composition rule, denoted by $\otimes$ and called \emph{parallel composition}. Indeed, the Hilbert spaces associated with composites $\sys{Q}_{1}\sys{Q}_{2}\cdots\sys{Q}_{n}$ are the tensor products $\mathscr{Q}_{1}\otimes\mathscr{Q}_{2}\otimes\cdots\otimes\mathscr{Q}_{n}$ of the associated individual Hilbert spaces (\emph{composition postulate}). The family of standard quantum systems will be denoted by $\Sys{\qt}=\lbrace\sys{Q}_1,\sys{Q}_2,\ldots\rbrace$. There exists a system---that is, the \emph{trivial system} $\sys{I}$---whose Hilbert-space dimension is $d=1$ and such that the relations $\sys{I}\sys{Q}=\sys{Q}=\sys{Q}\sys{I}$ hold true for all systems $\sys{Q}$; this means the space $\mathscr{I}\cong\mathbb{C}$ is the unit of the tensor product of Hilbert spaces. The \emph{states} of a system $\sys{Q}$ are all the positive semidefinite linear operators on $\mathscr{Q}$ with trace in $\left[0,1\right]$, called \emph{the density operators on $\mathscr{Q}$}. The physical quantum transformations from a system $\sys{Q}_1$ to a system $\sys{Q}_2$ are given by the \emph{quantum operations} $\sys{Q}_1\to\sys{Q}_2$, namely, the completely positive and trace non-increasing (CPTNI) linear maps from the density operators on $\mathscr{Q}_1$ to those on $\mathscr{Q}_2$. Every pair of quantum operations $\T{A}\colon\sys{A}\to\sys{B}$ and $\T{B}\colon\sys{B}\to\sys{C}$ is \emph{sequentially composable} via map composition, namely, $\T{B}\T{A}\colon\sys{A}\to\sys{C}$ is a quantum operation. Finally, quantum operations compose in parallel according to the tensor product $\otimes$ of linear maps.

The above definitional ingredients of \qt\ can be distilled in a generic fashion by resorting to a suitable operational framework. We will make use of the circuital representation for operational probabilistic theories~\cite{PhysRevA.84.012311,bookDCP2017,Coecke_Kissinger_2017,Selby_2021}. Technically, operational theories are modelled as \emph{symmetric strict monoidal categories}~\cite{maclane,heunen2019categories}. Processes $\T{T}$ are diagrammatically represented as boxes carrying a pair of wires, which in turn correspond to the relevant input/output systems, say $\sys{A}$ and $\sys{B}$:
\begin{align*}
	\begin{aligned}
		\Qcircuit @C=1.5em @R=1.5em
		{
			&\s{A}&\gate{\T{T}}&\s{B}\qw&
		}
	\end{aligned}
	.
\end{align*}
Just as in the conventional circuital representation of quantum processes, which is ubiquitously employed in quantum information, the horizontal (left-to-right) pictorial connection of wires will represent sequential composition, while the vertical (top-to-bottom) juxtaposition of boxes will represent their parallel composition;
moreover, both sequential and parallel composition are required to be associative:
\begin{align}\label{eq:associativity_sequential}
	\begin{aligned}
		\Qcircuit @C=1.5em @R=1.5em
		{
			&\s{A}&\qw&\gate{\T{A}}&\s{B}\qw&\gate{\T{B}}&\s{C}\qw&\gate{\T{C}}&\qw&\s{D}\qw&
			\relax\gategroupColor{1}{4}{1}{8}{2.75em}{--}{bleudefrance}
			\relax\gategroupColor{1}{4}{1}{6}{1.5em}{--}{bleudefrance}
		}
	\end{aligned}
	&=
	\begin{aligned}
		\Qcircuit @C=1.5em @R=1.5em
		{
			&\s{A}&\qw&\gate{\T{A}}&\s{B}\qw&\gate{\T{B}}&\s{C}\qw&\gate{\T{C}}&\qw&\s{D}\qw&
			\relax\gategroupColor{1}{4}{1}{8}{2.75em}{--}{bleudefrance}
			\relax\gategroupColor{1}{6}{1}{8}{1.5em}{--}{bleudefrance}
		}
	\end{aligned}
\end{align}
and
\begin{align}\label{eq:associativity_parallel}
	\begin{aligned}
		\Qcircuit @C=1.5em @R=1.5em
		{
			&\s{A}&\qw&\gate{\T{A}}&\qw&\s{B}\qw&\\
			&\s{C}&\qw&\gate{\T{C}}&\qw&\s{D}\qw&\\
			&\s{E}&\qw&\gate{\T{E}}&\qw&\s{F}\qw&
			\relax\gategroupColor{1}{4}{3}{4}{2.75em}{--}{upsdellred}
			\relax\gategroupColor{1}{4}{2}{4}{1.5em}{--}{upsdellred}
		}
	\end{aligned}
	&=
	\begin{aligned}
		\Qcircuit @C=1.5em @R=1.5em
		{
			&\s{A}&\qw&\gate{\T{A}}&\qw&\s{B}\qw&\\
			&\s{C}&\qw&\gate{\T{C}}&\qw&\s{D}\qw&\\
			&\s{E}&\qw&\gate{\T{E}}&\qw&\s{F}\qw&
			\relax\gategroupColor{1}{4}{3}{4}{2.75em}{--}{upsdellred}
			\relax\gategroupColor{2}{4}{3}{4}{1.5em}{--}{upsdellred}
		}
	\end{aligned}
\end{align}
must hold. For every system $\sys{S}$ there exists an \emph{identity process}
\begin{align}\label{eq:identity_op}
	\begin{aligned}
		\Qcircuit @C=1.5em @R=1.5em
		{
			&\s{S}&\gate{\T{I}}&\s{S}\qw&
		}
	\end{aligned}
	=
	\begin{aligned}
		\Qcircuit @C=1.5em @R=1.5em
		{
			&&\qw&\s{S}\qw&\qw&\qw&
		}
	\end{aligned}
\end{align}
satisfying, for every process $\T{T}$ from system $\sys{A}$ to system $\sys{B}$, the relations
\begin{equation}\label{eq:identity_property}
	\T{I}_{\sys{B}}\T{T} = \T{T} = \T{T}\T{I}_{\sys{A}}
	.
\end{equation}
Parallel composition of processes will be denoted by $\boxtimes$ (where, for standard \qt, obviously $\boxtimes=\otimes$). Every pair of identities is required to parallel compose to the identity of the composite of the individual systems:
\begin{equation}\label{eq:identity_parallel}
	\T{I}_{\sys{A}}\boxtimes\T{I}_{\sys{B}} = \T{I}_{\sys{AB}}
	.
\end{equation}
There exists a system $\sys{I}$, called the \emph{trivial system}, being the unit of parallel composition of systems, i.e.~$\sys{AI}=\sys A=\sys{IA}$. Accordingly, an explicit diagrammatic representation of the trivial system can be omitted:
\begin{align}\label{eq:unit}
	\begin{aligned}
		\Qcircuit @C=1.5em @R=1.5em
		{
			&\s{I}&\gate{\rho}&\s{A}\qw&
		}
	\end{aligned}
	=
	\begin{aligned}
		\Qcircuit @C=1.4em @R=1em {
			&\prepareC{\rho}&\s{A}\qw&
		}
	\end{aligned}
	,\quad
	\begin{aligned}
		\Qcircuit @C=1.5em @R=1.5em
		{
			&\s{A}&\gate{a}&\s{I}\qw&
		}
	\end{aligned}
	=
	\begin{aligned}
		\Qcircuit @C=1.4em @R=1em {
			&\s{A}&\measureD{a}&
		}
	\end{aligned}
	,
\end{align}
where the above represent, respectively, the states and the effects of the theory; also, the identity of $\sys{I}$ satisfies the following:
\begin{equation}\label{eq:identity_trivial}
	\T{T}\boxtimes\T{I}_{\sys{I}} = \T{T} = \T{I}_{\sys{I}}\boxtimes\T{T}
	.
\end{equation}
Processes $\sys{I}\to\sys{I}$ just represent probabilities. The above is clearly satisfied by \qt. In order to provide reasonable notions of \emph{local} and \emph{statistically independent} operations, sequential and parallel compositions are required to be compatible, namely, to satisfy the following \emph{interchange law}:
\begin{align}\label{eq:interchange_law_diagrammatic}
	\begin{aligned}
		\Qcircuit @C=1.5em @R=2.5em
		{
			&\s{A}&\qw&\gate{\T{A}}&\s{B}\qw&\gate{\T{B}}&\qw&\s{C}\qw&
			\\
			&\s{D}&\qw&\gate{\T{D}}&\s{E}\qw&\gate{\T{E}}&\qw&\s{F}\qw&
			\relax\gategroupColor{1}{4}{1}{6}{1.6em}{--}{bleudefrance}
			\relax\gategroupColor{2}{4}{2}{6}{1.6em}{--}{bleudefrance}
			\relax\gategroupColor{1}{4}{2}{6}{3em}{--}{upsdellred}
		}
	\end{aligned}
	=
	\begin{aligned}
		\Qcircuit @C=1.5em @R=2.5em
		{
			&\s{A}&\qw&\gate{\T{A}}&\s{B}\qw&\gate{\T{B}}&\qw&\s{C}\qw&
			\\
			&\s{D}&\qw&\gate{\T{D}}&\s{E}\qw&\gate{\T{E}}&\qw&\s{F}\qw&
			\relax\gategroupColor{1}{4}{2}{4}{1.6em}{--}{upsdellred}
			\relax\gategroupColor{1}{6}{2}{6}{1.6em}{--}{upsdellred}
			\relax\gategroupColor{1}{4}{2}{6}{3em}{--}{bleudefrance}
		}
	\end{aligned}.
\end{align}
The above relation is satisfied by composition of linear maps along with the choice $\boxtimes=\otimes$ in \qt. Finally, a \emph{swap process} $\T{S}$---\emph{involutive} (and, in particular, \emph{reversible})---which is diagrammatically represented as 
\begin{align}\label{eq:swap_op}
	\begin{aligned}
		\Qcircuit @C=1em @R=2em
		{
			&\s{A}&\multigate{1}{\T{S}}&\s{B}\qw&
			\\
			&\s{B}&\ghost{\T{S}}&\s{A}\qw&
		}
	\end{aligned}
	\coloneqq
	\input{braiding.tikz}
	,
\end{align}
is required to exist for every pair of systems $\sys{A}$ and $\sys{B}$. The following relations are imposed to be satisfied by the swap:
\begin{align}
	&\input{swap_narrow_test_2.tikz}=\input{swap_narrow_test_1.tikz}
	,\label{eq:swap_naturality}
	\\
	&\input{braiding_hexagon1.tikz}
	=
	\input{braiding_threewires_intwo1.tikz}
	.\label{eq:swap_hexagon}
\end{align}
Relation~\eqref{eq:swap_naturality} means that $\T{S}$ correctly swaps local experiments, while relation~\eqref{eq:swap_hexagon} guarantees that swaps on different systems compose just as one would expect, that is, consistently with the action of the symmetric group on the wires~\cite{maclane}.

The above operational \emph{desiderata} are the essential features that are required to be satisfied by the alternative theories of quantum system presented in Appendix~\ref{app:lqts}. Satisfying these constraints is nontrivial, since they can lead to obstructions for given choices of $\boxtimes$~\cite{PhysRevA.101.042118,PhysRevA.102.052216}. Nevertheless, we will define an alternative composition postulate for quantum systems verifying all of the operational requirements. Accordingly, we will show that quantum systems can define operational theories with alternative composition postulates.

\subsection{Preparatory notation}\label{app:preparatory_notation}

Let us first posit some preparatory notation, which will allow us to provide a convenient illustration of the alternative composites of quantum systems. For a start, let $\mathtt{C},\mathtt{N},\mathtt{M},\ldots$ stand for (finite) strings of respectively $c,n,m,\ldots$ characters of an arbitrary alphabet. Every such string $\mathtt{C}$ can be then denoted as a concatenation
of totally ordered labelled integers as follows:
\begin{equation*}
	\mathtt{C}
	=
	1_{\mathtt{C}}\boxplus2_{\mathtt{C}}\boxplus\cdots\boxplus c_{\mathtt{C}}
	,
\end{equation*}
where the (associative) operation of concatenation $\boxplus$ of strings
has been introduced.
A concatenation of strings $\mathtt{N}\boxplus\mathtt{M}$ has clearly length $n+m$. Moreover, we will be considering ordered
pairs of characters of the form $\left(i_{\mathtt{N}},j_{\mathtt{M}}\right)$;
this allows one to define two binary operations $\times$ and $\odot$, each mapping pairs of strings $\left(\mathtt{N},\mathtt{M}\right)$ to strings of composite characters of the form $\left(i_{\mathtt{N}},j_{\mathtt{M}}\right)$. The operation $\times$ is a total operation defined as follows:
\begin{equation*}
	\mathtt{N}\times\mathtt{M}
	\coloneqq
	\bigboxplus_{i=1}^{n}\left(\bigboxplus_{j=1}^{m}\left(i_{\mathtt{N}},j_{\mathtt{M}}\right)\right)
	.
\end{equation*}
The string $\mathtt{N}\times\mathtt{M}$ has clearly length $n\cdot m$. Finally, the operation $\odot$ is a partial operation defined as follows:
\begin{equation*}
\mathtt{N}\odot\mathtt{N}
	\coloneqq
	\bigboxplus_{i=1}^{n}\left(\bigboxplus_{j=1}^{i-1}\left(i_{\mathtt{N}},j_{\mathtt{N}}\right)\right)
	.
\end{equation*}
Note that the number of characters forming the string $\mathtt{N}\odot\mathtt{N}$ is 
triangular
therefore the string $\mathtt{N}\odot\mathtt{N}$ has clearly length $n\choose 2$. Observe that both in $\mathtt{N}\times\mathtt{M}$ and in $\mathtt{N}\odot\mathtt{N}$ the composite characters (ordered pairs) are concatenated in the lexicographic order. 

The following elementary relations hold:
\begin{align}
	\label{eq:odot_decomposition}
	&(\mathtt{N}\boxplus\mathtt{E})^{\odot 2}
	\cong
	\left(\mathtt{E}\odot\mathtt{E}\right)\boxplus\left(\mathtt{E}\times\mathtt{N}\right)\boxplus\left(\mathtt{N}\odot\mathtt{N}\right),
	\\
	\label{eq:boxplus_symmetric}
	&\mathtt{E}\times\left(\mathtt{N}_1\boxplus\mathtt{N}_2\right)
	\cong
	\mathtt{E}\times\left(\mathtt{N}_2\boxplus\mathtt{N}_1\right),
	\\
	\label{eq:times_symmetric}
	&\mathtt{N}_1\times\mathtt{N}_2
	\cong
	\mathtt{N}_2\times\mathtt{N}_1
	,
\end{align}
where the isomorphisms in relations~\eqref{eq:odot_decomposition}
and~\eqref{eq:boxplus_symmetric} are just permutations of characters, while the one in relation~\eqref{eq:times_symmetric} is a permutation of characters precomposed with the mapping $\left(i_{\mathtt{N}_1},j_{\mathtt{N}_2}\right)\mapsto\left(j_{\mathtt{N}_2},i_{\mathtt{N}_1}\right)$. The above relations will come in handy in the following. 

\section{Latent quantum theory}\label{app:lqts}

In the present section, we will define a family of theories collectively called \emph{latent quantum theories} (\lqt s). We will build \lqt s upon standard \qt\ (Appendix~\ref{app:quantum_theory_operational}). We emphasise that we are not assuming finite-dimensional \qt, therefore our results hold for arbitrary dimensionalities. In particular, this formalism
is suited to encompass algebraic quantum field theories (see, e.g.,~also Ref.~\cite[Suppl.~Info.~I]{Renou:2021aa} for a discussion of the status of the composition postulate in the framework of algebraic quantum field theories).

\subsection{Systems}\label{app:systems}


Given the family of \emph{standard quantum systems} $\Sys{\qt}=\lbrace\sys{Q}_1,\sys{Q}_2,\ldots\rbrace$, each latent quantum system $\sys{Q}_{\mathtt{N}}$ is posited to be characterised as a (finite) formal string
\begin{equation*}
	\sys{Q}_{\mathtt{N}}\coloneqq
	\sys{Q}_{1_\mathtt{N}}\sys{Q}_{2_\mathtt{N}}\cdots\sys{Q}_{n_\mathtt{N}}
\end{equation*}
on the alphabet $\Sys{\qt}$,
where
strings differing just by the presence of the character $\sys{I}$ in some positions are identified.\footnote{Such a characterisation of systems amounts to the property of \emph{existence and uniqueness of system-decomposition}~\cite[§6, Def.~14]{rolino2024minimaloperationaltheoriesclassical}, technically making an operational probabilistic theory satisfying it a \emph{coloured PROP}~\cite{maclane1965categorical,carette2022propificationscalablecomonad}.} This characterisation implies that, for standard quantum systems $\sys{Q}_{i_\mathtt{N}},\sys{Q}_{j_\mathtt{M}}\neq\sys{I}$ $\forall i_\mathtt{N}\in\mathtt{N},j_\mathtt{M}\in\mathtt{M}$, the following holds:
\begin{equation}\label{eq:indentity_criterion2}
	\sys{Q}_{1_\mathtt{N}}\sys{Q}_{2_\mathtt{N}}\cdots\sys{Q}_{n_\mathtt{N}}=\sys{Q}_{1_\mathtt{M}}\sys{Q}_{2_\mathtt{M}}\cdots\sys{Q}_{m_\mathtt{M}}\quad \Longleftrightarrow\quad \mathtt{N}=\mathtt{M}
	.
\end{equation}
Finally, a map $\mathfrak{q}$ for associating a standard quantum system (and, in particular, a Hilbert space) to each latent quantum system is posited as follows:
\begin{equation}\label{eq:latent quantum_composition}
	\mathfrak{q}{\left(\sys{Q}_{\mathtt{N}}\right)}
	=
	\mathfrak{q}{\left(\sys{Q}_{1_\mathtt{N}}\sys{Q}_{2_\mathtt{N}}\cdots\sys{Q}_{n_\mathtt{N}}\right)}
	\coloneqq
	\sys{L}_{\mathtt{N}\odot\mathtt{N}}\otimes\sys{O}_{\mathtt{N}}.	
\end{equation}
The usual tensor product
\begin{equation*}
	\sys{O}_{\mathtt{N}} \coloneqq\sys{Q}_{1_\mathtt{N}}\otimes\sys{Q}_{2_\mathtt{N}}\otimes\cdots\otimes\sys{Q}_{n_\mathtt{N}}=\bigotimes_{i=1}^n\sys{Q}_{i_\mathtt{N}},
\end{equation*}
is called the \emph{operational space} of the latent quantum system $\sys{Q}_{\mathtt{N}}$, while $\sys{L}_{\mathtt{N}\odot\mathtt{N}}$ is called the \emph{latent space} of the latent quantum system $\sys{Q}_{\mathtt{N}}$, and it is also defined in terms of the usual tensor product of standard quantum systems via the following relation:
\begin{equation*}
	\sys{L}_{\mathtt{C}} \coloneqq \bigotimes_{\left(i,j\right)\in\mathtt{C}}
	\sys{L}_{ij}
	,\ \ 
	\text{for $\mathtt{C}=\mathtt{N}\odot\mathtt{N}$ or $\mathtt{C}=\mathtt{N}\times\mathtt{M}$}
	.
\end{equation*}
Specifically, in relation~\eqref{eq:latent quantum_composition} we posit
\begin{align}
	&\sys{L}_{\mathtt{N}\odot\mathtt{N}} \coloneqq \bigotimes_{(i_{\mathtt{N}},j_{\mathtt{N}})\in{\mathtt{N}\odot\mathtt{N}}} \sys{L}_{i_{\mathtt{N}}j_{\mathtt{N}}},\label{eq:latent_spaces}\\
	&(\sys{Q}_{i_{\mathtt{N}}}=\sys{I}\vee \sys{Q}_{j_{\mathtt{N}}}=\sys{I})
	\ \Longrightarrow\  
	\sys{L}_{i_{\mathtt{N}}j_{\mathtt{N}}}=\sys{I}
	,\qquad	\sys{L}_{i_{\mathtt{N}}j_{\mathtt{M}}}=\sys{L}_{j_{\mathtt{M}}i_{\mathtt{N}}}\qquad \forall \mathtt{N},\mathtt{M},i,j,
	\label{eq:latent_spaces_2}
\end{align}
and the members of the collection of standard quantum systems {$\lambda\coloneqq\left\{\sys{L}_{i_{\mathtt{N}}j_{\mathtt{M}}} \mid \sys{Q}_{i_\mathtt{N}},\sys{Q}_{j_\mathtt{M}}\in\Sys{\qt}\right\}$}, called the \emph{latent factors}, are fixed (finite- or infinite-dimensional) standard quantum systems. 
Note that second relation in~\eqref{eq:latent_spaces_2} means that the latent factors depend only on \emph{unordered} pairs of indices. On the one hand, by setting $\sys{L}_{i_{\mathtt{N}}j_{\mathtt{N}}}\coloneqq\sys{I}$ for all $\mathtt{N}$ and all $\left(i_{\mathtt{N}},j_{\mathtt{N}}\right)$, standard \qt\ is recovered. On the other hand, every choice of collection $\lambda$ complying with~\eqref{eq:latent_spaces} and~\eqref{eq:latent_spaces_2} defines a distinct family of \lqt s---the remaining class of free parameters, that has to be fixed in order to pick out a given \lqt, will be specified in the following.
The map $\mathfrak{q}$ posited in~\eqref{eq:latent quantum_composition} is well-defined in light of relations~\eqref{eq:indentity_criterion2} and the first condition in~\eqref{eq:latent_spaces_2}. The full collection of latent quantum systems will be denoted by $\Sys{\lqt}$. Accordingly, we have posited that:
\begin{equation*}
	\forall
	\sys{Q}_{\mathtt{N}}\in\Sys{\lqt}
	,\quad
	\mathfrak{q}{\left(\sys{Q}_{\mathtt{N}}\right)}
	= 
	\sys{L}_{\mathtt{N}\odot\mathtt{N}}\otimes\sys{O}_{\mathtt{N}}
	\in\Sys{\qt}
	.
\end{equation*}
Let us then introduce the composite-system rule for arbitrary latent quantum systems as follows:
\begin{equation*}
	\sys{Q}_{\mathtt{N}}\boxtimes\sys{Q}_{\mathtt{E}}\coloneqq
	\sys{Q}_{\mathtt{N}}\sys{Q}_{\mathtt{E}}\equiv
	\sys{Q}_{\mathtt{N}\boxplus\mathtt{E}}
	\in\Sys{\lqt}
	.
\end{equation*}
The operation of parallel composition 
for latent quantum systems is well-defined in light of the identity criterion~\eqref{eq:indentity_criterion2} and relations~\eqref{eq:latent_spaces} and~\eqref{eq:latent_spaces_2}, it is
associative since $\boxplus$ is, and it also admits of a trivial system, given by the trivial standard quantum system $\sys{I}$.

\subsection{Transformations}\label{app:transformations}

We have set a family of quantum systems closed under an associative internal operation $\boxtimes$ whose unit is the trivial system $\sys{I}$. Now, we need to specify the family of latent quantum admissible (``physical'') transformations. We will make use of the circuital representation presented in Appendix~\ref{app:quantum_theory_operational}. In general, when \emph{local tomography}~\cite{hardy2012limited,bookDCP2017} holds, every process is completely identified by the single linear map expressing its local action (see also Section~\ref{subsec:standard_quantum} in the main text). However, as is known, in the absence of local tomography~\cite{hardy2012limited}, a single process 
identifies a \emph{collection of linear maps}~\cite{Chiribella_2014,PhysRevA.102.052216}. The reason of this fact is that, in general, the action of two processes having the same local actions may yet differ when they operate on the input system extended by considering the presence of ancillae. This happens for instance within \emph{fermionic quantum theory}~\cite{D_Ariano_2014,doi:10.1142/S0217751X14300257}, \emph{real quantum theory}~\cite{stueckelberg1960quantum,Renou:2021aa}, or other operational theories~\cite{Barnum_2015,PhysRevA.101.042118,Barnum2020composites,PhysRevA.102.052216,rolino2024minimaloperationaltheoriesclassical,centeno2024twirledworldssymmetryinducedfailures,soltani2025decouplinglocalclassicalityclassical,baldijão2026tomographicallynonlocalentanglement}. Operationally, this feature precludes the ability to perform \emph{local process tomography}~\cite{bookDCP2017,sym13111985}. As a consequence, transformations will 
read:
\begin{equation*}
	\input{transformation2.tikz}
	\quad=\quad\{\widehat{\mathcal G\boxtimes \mathcal I_{\sys Q_\mathtt E}}\}_\mathtt E
	,
\end{equation*}
where each map
\begin{equation}\label{eq:mapping}
\widehat{\mathcal G\boxtimes \mathcal I_{\sys Q_\mathtt E}}
	\ \colon\ 	\mathfrak{q}{\left(\sys{Q}_{\mathtt{N}}\sys{Q}_{\mathtt{E}}\right)} = \sys{L}_{\left(\mathtt{N}\boxplus\mathtt{E}\right)^{\odot 2}}\otimes\sys{O}_{\mathtt{N}\boxplus\mathtt{E}} \to 
	\mathfrak{q}{\left(\sys{Q}_{\mathtt{M}}\sys{Q}_{\mathtt{E}}\right)} = \sys{L}_{\left(\mathtt{M}\boxplus\mathtt{E}\right)^{\odot 2}}\otimes\sys{O}_{\mathtt{M}\boxplus\mathtt{E}}
\end{equation}
is a standard quantum operation expressing the action of process $\T{G}$ when it is extended with identity processes on ancillary systems $\sys{Q}_{\mathtt{E}}$. 
Let us briefly inspect~\eqref{eq:mapping}. In light of relation~\eqref{eq:unit}, note that if $\sys{Q}_{\mathtt{N}}=\sys{I}$, the input-system wire can be omitted, and the quantum operations represent a \emph{state} of system $\sys{Q}_{\mathtt{M}}$ and ancillary extensions thereof; if $\sys{Q}_{\mathtt{M}}=\sys{I}$, the output-system wire can be omitted, and the quantum operations represent an \emph{effect} of system $\sys{Q}_{\mathtt{N}}$ and ancillary extensions thereof; finally, $\mathtt{E}=\varepsilon$ (the empty string) means that there is no ancillary system, and in that case the associated quantum operation just represents the local action $\hat {\T G}$ of process $\T{G}$ on system $\mathfrak{q}{\left(\sys{Q}_{\mathtt{N}}\right)}$, or its dual action on system $\mathfrak{q}{\left(\sys{Q}_{\mathtt{M}}\right)}$, as clear from the context.

In the following, in light of definition~\eqref{eq:mapping}, we will be able to make use of the standard quantum circuital formalism. More specifically, we will provide a representation of each quantum operation~\eqref{eq:mapping}, defining a latent quantum process, as a CPTNI linear map on the relevant Hilbert space of the input composite system. Therefore, for the sake of avoiding weighing down the notational clarity, in the quantum circuits below we will always write $\alpha$ instead of $\sys{L}_{\alpha}$, whether $\alpha$ denotes an arbitrary string of pairs or a single pair (see Section~\ref{app:systems}). Moreover, for the sake of clarity, we will distinguish latent quantum systems from standard quantum ones by representing the latter ones as thick dashed and dotted wires.

We are now in position of defining transformations for \lqt s. A latent quantum transformation $\sys{Q}_{\mathtt{N}}\to\sys{Q}_{\mathtt{M}}$ is characterised by a pair $(\T{T},\lbrace\T{S}_{k,k'}^{\mathtt{C}}\rbrace_\mathtt C)$
as per following definitions. The first entry $\T T$
represents an arbitrary standard quantum operation:
\begin{equation*}
	\T{T}\ \colon\ 
	\sys{L}_{\mathtt{N}\odot\mathtt{N}}\otimes\sys{O}_{\mathtt{N}} \to \sys{L}_{\mathtt{M}\odot\mathtt{M}}\otimes\sys{O}_{\mathtt{M}}
	;
\end{equation*}
the second entry $\T{S}_{k,k'}^{\mathtt{C}}$ is a family of (standard quantum) ``noisy permutations'':
\begin{equation}\label{eq:noisy_permutation}
\begin{split}
	\T{S}_{k,k'}^{\mathtt{C}}&\coloneqq
	\bigotimes_{j=1}^{c} \T{S}_{k,k'}^{j_\mathtt{C}},
	\\
	\T{S}_{k,k'}^{j_\mathtt{C}}\ &\coloneqq\  \scalebox{0.9}{\input{noisy_permutation.tikz}}
	\ \ \ ,
\end{split}
\end{equation}
where
\begin{equation*}
	\xi_{j_{\mathtt{C}}\times\mathtt{K}}
	\coloneqq
	\bigotimes_{i=1}^{k}\xi_{\sys{L}_{j_{\mathtt{C}}i_{\mathtt{K}}}}
	,
\end{equation*}
and the members of the collection
\begin{equation*}
{
	\xi\coloneqq\left\{\xi_{\sys{L}_{i_{\mathtt{N}}j_{\mathtt{M}}}} \mid \sys{Q}_{i_\mathtt{N}},\sys{Q}_{j_\mathtt{M}}\in\Sys{\qt} \right\}
	,
	}
\end{equation*}
that are called the \emph{latent states}, are fixed normalised states of the associated latent factors, while $\T{S}_{j_{\mathtt{C}}}$ is an arbitrary permutation
of the input wires; finally, $\text{Tr}$ indicates the trace functional on the corresponding factors. The ``noisy permutations'' $\T{S}_{k,k'}^{j_\mathtt{C}}$ are assumed to be in reduced form, 
so that the standard quantum channels $\T{S}_{k,k'}^{j_\mathtt{C}}$ (and $\T{S}_{k,k'}^{\mathtt{C}}$ thereon) are uniquely specified by a pair $(k,k')$ and a permutation $\T{S}$ of $k+n$ wires such that:
\begin{equation*}
	\forall i=1,2,\ldots,k,
	\ \ 
	\T{S}{\left(i\right)}>k'
	.
\end{equation*}
Let us also observe that, via relation~\eqref{eq:odot_decomposition}, for any given partition $\mathtt{C}=\mathtt{N}\boxplus\mathtt{E}$ into substrings, the associated latent space has the following tensor sectors:
\begin{equation}\label{eq:iso1}
	\sys{L}_{\mathtt{C}\odot\mathtt{C}} = \sys{L}_{\left(\mathtt{N}\boxplus\mathtt{E}\right)^{\odot 2}}
	\cong
	\sys{L}_{\mathtt{E}\odot\mathtt{E}}\otimes\sys{L}_{\mathtt{E}\times\mathtt{N}} \otimes \sys{L}_{\mathtt{N}\odot\mathtt{N}}
\end{equation}
(up to permutations of factors). Similarly, in light of relations~\eqref{eq:boxplus_symmetric}  and~\eqref{eq:times_symmetric} (the latter combined with relation~\eqref{eq:latent_spaces_2}),
one has:
\begin{align}
	&\sys{L}_{\mathtt{E}\times\left(\mathtt{N}_1\boxplus\mathtt{N}_2\right)}
	\cong
	\sys{L}_{\mathtt{E}\times\left(\mathtt{N}_2\boxplus\mathtt{N}_1\right)}
	,\label{eq:iso2}
	\\
	&\sys{L}_{\mathtt{N}_1\times\mathtt{N}_2}
	\cong
	\sys{L}_{\mathtt{N}_2\times\mathtt{N}_1}
	.\label{eq:iso3}
\end{align}
Accordingly, 
in the following symbols such as $\sigma,\gamma,\tau$ will be extensively used to label boxes where the appropriate permutation of quantum wires (uniquely deducible from the context) is intended. In conclusion, we observe that the collections of latent quantum transformations $\sys{Q}_{\mathtt{N}}\to\sys{Q}_{\mathtt{M}}$ are clearly well-defined.

Every pair $\left(\lambda,\xi\right)$ of collections of latent factors and states
defines a distinct $\left(\lambda,\xi\right)$-\lqt. The simplest \lqt\ is the one where all the nontrivial latent factors are equal to a fixed standard quantum system $\sys{L}$ and the only latent state is pure; this is the one illustrated in the main text
(Section~\ref{subsec:framework}). Latent quantum transformations, for every given $\left(\lambda,\xi\right)$-\lqt, are then defined as:
\begin{equation}\label{eq:quantum_circuit}
\begin{split}
	&\input{transformation.tikz}
	\quad=\quad
	\left\{\reallywidehat{\T{G}_{\T{T},\{\T{S}^{\mathtt{C}}_{k,k'}\}_{\mathtt{C}}}\boxtimes\T{I}_{\sys{Q}_{\mathtt{E}}}}\right\}_{\mathtt{E}}
	\quad,
    \\
	&\reallywidehat{\T{G}_{\T{T},\{\T{S}^{\mathtt{C}}_{k,k'}\}_{\mathtt{C}}}\boxtimes\T{I}_{\sys{Q}_{\mathtt{E}}}}
	\qquad\coloneqq\qquad
	\input{quantum_circuit.tikz}
	\quad,
\end{split}
\end{equation}
where the sector highlighted in light-teal (bottom) represents the standard quantum action of the transformation, while the one highlighted in light-periwinkle (top) represents its genuinely latent quantum action. Note that relation~\eqref{eq:iso1} guarantees that the operators $\sigma_i$ in definition~\eqref{eq:quantum_circuit} are indeed well-defined permutations. The function that maps pairs $(\T{T},\lbrace\T{S}_{k,k'}^{\mathtt{C}}\rbrace_{\mathtt{C}})$ to the families of quantum operations in~\eqref{eq:quantum_circuit} is well-defined. These contain a transformation which will be verified to be the latent quantum \emph{identity process}~\eqref{eq:identity_op}, that is:
\begin{equation}\label{eq:identity}
	\input{identity.tikz}
	\quad\coloneqq\quad
	\input{transformation_identity.tikz}
	\quad,
\end{equation}
where $\T{I}_{\sys{\mathtt{N}^{\odot 2}}\otimes\sys{O}_{\mathtt{N}}}$ and $\T{I}_{\mathtt{C}\times\mathtt{N}}$ are standard quantum identities on the relevant spaces. Furthermore, the latent quantum \emph{swap process}~\eqref{eq:swap_op} $\sys{Q}_{\mathtt{N}_1}\boxtimes\sys{Q}_{\mathtt{N}_2}=\sys{Q}_{\mathtt{N}_1}\sys{Q}_{\mathtt{N}_2} \to \sys{Q}_{\mathtt{N}_2}\boxtimes\sys{Q}_{\mathtt{N}_1}=\sys{Q}_{\mathtt{N}_2}\sys{Q}_{\mathtt{N}_1}$ is defined by the following family of permutations of the factors:
\begin{equation}\label{eq:swap}
	\reallywidehat{\T{S}_{\left(\sys{Q}_{\mathtt{N}_1}\sys{Q}_{\mathtt{N}_2}\right)\left(\sys{Q}_{\mathtt{N}_2}\sys{Q}_{\mathtt{N}_1}\right)}\boxtimes\T{I}_{\sys{Q}_\mathtt{E}}}
	\ \coloneqq\ 
	\scalebox{0.75}
	{
		\input{quantum_circuit_swap.tikz}
	}
	\quad
	,
\end{equation}
where
\begin{equation*}
	\begin{split}
		\T{S}_{\mathtt{N_1}\mathtt{N_2}}^{\mathtt{E}}\coloneqq
		\bigotimes_{j=1}^{e}\T{S}_{\mathtt{N_1}\mathtt{N_2}}^{j_\mathtt{E}}
	\end{split}
\end{equation*}
and $\T{S}_{\mathtt{N_1}\mathtt{N_2}}^{j_\mathtt{E}}$ just indicates the swap $\sys{L}_{j_\mathtt{E}\times\left(\mathtt{N_1}\boxplus\mathtt{N_2}\right)}\to\sys{L}_{j_\mathtt{E}\times\left(\mathtt{N_2}\boxplus\mathtt{N_1}\right)}$ (which is well defined by relation~\eqref{eq:iso2}). It is then straightforward to verify that every CPTNI map associated with each~\eqref{eq:swap} is indeed of the form~\eqref{eq:quantum_circuit}. Note that relations~\eqref{eq:iso1} and~\eqref{eq:iso3} guarantee that all of the operators $\sigma_i$ in definition~\eqref{eq:swap} are indeed well-defined permutations. It will also be verified that the latent quantum process defined by~\eqref{eq:swap} indeed behaves as a good swap for $\left(\lambda,\xi\right)$-\lqt s.

We are now ready to complete our presentation of \lqt s by endowing these theories with rules for both sequential and parallel compositions of their transformations.

\subsection{Composing transformations}

Let us start by defining:
\begin{equation}\label{eq:sequential_composition}
	\input{sequential1.tikz}
	\quad=\quad
	\input{sequential_composition.tikz}
	\quad\coloneqq\quad
    \left\{\left(\widehat{\T{G}\boxtimes\T{I}_{\sys{Q}_{\mathtt{E}}}}\right)\left(\widehat{\T{F}\boxtimes\T{I}_{\sys{Q}_{\mathtt{E}}}}\right)\right\}_{\mathtt{E}}
    \quad,
\end{equation}
which is a well-defined expression---the right-hand side being given by the compositions of the corresponding quantum operations.
The transformation $\T{G}\T{F}$ is defined to be the \emph{latent quantum sequential composition} of $\T{F}$ and $\T{G}$. Explicitly, sequential composition reads:
\begin{equation}\label{eq:sequential_composition_explicit}
	\T{G}_{\tilde{\T{T}},\lbrace\tilde{\T{S}}_{z,z'}^{\mathtt{C}}\rbrace_{\mathtt{C}}}\T{G}_{\T{T},\lbrace\T{S}_{k,k'}^{\mathtt{C}}\rbrace_{\mathtt{C}}}
    =
    \T{G}_{\tilde{\T{T}}\T{T},\lbrace\tilde{\T{S}}_{z,z'}^{\mathtt{C}}\T{S}_{k,k'}^{\mathtt{C}}\rbrace_{\mathtt{C}}}
    .
\end{equation}
Note that $\tilde{\T{S}}_{z,z'}^{\mathtt{C}}\T{S}_{k,k'}^{\mathtt{C}}$ is, by construction, always of the form~\eqref{eq:noisy_permutation}; explicitly, it reads:
\begin{equation*}
	\tilde{\T{S}}_{z,z'}^{\mathtt{C}}\T{S}_{k,k'}^{\mathtt{C}}
	=
	\bigotimes_{j_\mathtt{C}=1}^{c}\left(\tilde{\T{S}}_{z,z'}^{j_\mathtt{C}}\T{S}_{k,k'}^{j_\mathtt{C}}\right)
	.
\end{equation*}
Sequential composition is clearly associative---namely, it satisfies relation~\eqref{eq:associativity_sequential}---as inherited from associativity of quantum operations. Furthermore, the transformation $\T{G}_{\T{I}_{\sys{\mathtt{N}^{\odot 2}}\otimes\sys{O}_{\mathtt{N}}},\lbrace\T{I}_{\mathtt{C}\times\mathtt{N}}\rbrace_{\mathtt{C}}}$ defined in~\eqref{eq:identity} clearly satisfies~\eqref{eq:identity_property} for latent quantum sequential composition. Finally, as a direct application of definitions~\eqref{eq:swap},~(\ref{eq:sequential_composition}--\ref{eq:sequential_composition_explicit}), and~\eqref{eq:identity}, it follows that the swap is involutive (and, in particular, reversible).

We can now proceed by positing the last operational feature of \lqt s, that is, \emph{parallel composition of latent quantum transformations}:
\begin{equation}\label{eq:parallel_composition}
	\begin{split}
	&\input{parallel_composition.tikz}
	\quad\coloneqq\quad
    \left\{\reallywidehat{\left(\T{G}_{\T{T},\lbrace\T{S}_{k,k'}^{\mathtt{C}}\rbrace_{\mathtt{C}}}\boxtimes\T{G}_{\tilde{\T{T}},\lbrace\tilde{\T{S}}_{z,z'}^{\mathtt{C}}\rbrace_{\mathtt{C}}}\right)\boxtimes\T{I}_{\sys{Q}_{\mathtt{E}}}}\right\}_{\mathtt{E}}
    \ \ ,
	\\
    &\reallywidehat{\left(\T{G}_{\T{T},\lbrace\T{S}_{k,k'}^{\mathtt{C}}\rbrace_{\mathtt{C}}}\boxtimes\T{G}_{\tilde{\T{T}},\lbrace\tilde{\T{S}}_{z,z'}^{\mathtt{C}}\rbrace_{\mathtt{C}}}\right)\boxtimes\T{I}_{\sys{Q}_{\mathtt{E}}}}
	\coloneqq\\ 
	&\coloneqq
	\scalebox{0.87}
	{
		\input{quantum_circuit_parallel_composition_alternative.tikz}
	}
	\ \ ,
	\end{split}
\end{equation}
where 
\begin{equation}\label{eq:star_product}
	\input{asterisk_product.tikz}
	\ \ \coloneqq\ \ 
	\input{asterisk_product_a.tikz}
	\ \ ,
\end{equation}
where the permutations $\tau$ and $\tilde\tau$ permute the systems as to exchange indices 
in the pairs and reordering the resulting pairs in lexicographic order. The parallel composition operation is clearly well-defined.
Parallel compositions~\eqref{eq:parallel_composition} are still of the form~\eqref{eq:quantum_circuit}, namely, they are latent quantum transformations.
One can see that the action of~\eqref{eq:star_product} on local input states $\rho_{\sys{L}_{{\mathtt{N}_2}\times{\mathtt{N}_1}}}$ is:
\begin{equation*}
\begin{split}
	\rho_{\sys{L}_{{\mathtt{N}_2}\times{\mathtt{N}_1}}}
	\mapsto
	P(\text{Tr}_\mathtt{J}[\rho_{\sys{L}_{{\mathtt{N}_2}\times{\mathtt{N}_1}}}]\otimes\xi_{\mathtt J'})P^\dag,
	\quad
	&
	\mathtt{J}
	\coloneqq
	\lbrace\left(j_{\mathtt{N}_2},i_{\mathtt{N}_1}\right)\mid
	{\T{S}{\left(i\right)}\leq k'}
	\vee
	{\tilde{\T{S}}{\left(j\right)}\leq z'}\rbrace
	\\ 
	&\mathtt{J}'\coloneqq
    \lbrace\left(j_{\mathtt{M}_2}',i_{\mathtt{M}_1}'\right)\mid
    {({\T{S}^{-1}{\left(i'\right)}\leq k} \land {\tilde{\T{S}}^{-1}{\left(j'\right)}> z} )
    \vee
    {({\T{S}^{-1}{\left(i'\right)}>k} \land {\tilde{\T{S}}^{-1}{\left(j'\right)}\leq z})}
    \rbrace
    ,
    }
\end{split}
\end{equation*}
while $P$ is a suitable permutation of the wires. By the characterisation of the action of channels~\eqref{eq:star_product} just provided, it is now straightforward to verify the following intertwining property:
\begin{equation}\label{eq:intertwining}
	\scalebox{0.85}{\input{asterisk_product_a.tikz}}
	\ \ =\ \ 
	\scalebox{0.85}{\input{asterisk_product_b.tikz}},
\end{equation}
where $\gamma$ and $\tilde\gamma$ serve the same purpose as $\tau$ and $\tilde\tau$.
Note that the identity $\T{G}_{\T{I}_{\sys{I}},\left\{\T{I}_{\sys{I}}^{\mathtt{C}}\right\}_{\mathtt{C}}}$ of the trivial system $\sys{I}$ also satisfies~\eqref{eq:identity_trivial}, as it can verified by using definitions~\eqref{eq:latent_spaces_2} and~\eqref{eq:parallel_composition}. Via~\eqref{eq:parallel_composition}, one can also see that the family of identity processes defined in~\eqref{eq:identity} also satisfies the expected property~\eqref{eq:identity_parallel} for the parallel composition of two identities:
\begin{equation*}
	\input{identity_parallel2.tikz}
	\quad =\quad
	\input{identity_parallel.tikz}
	\quad.
\end{equation*}

\subsection{Latent quantum theories are operational quantum theories}\label{app:check} 

A few checks from Appendix~\ref{app:quantum_theory_operational} have been already made, as immediate consequences of the definitions, in the previous section. The remaining ones are the following:~\eqref{eq:associativity_parallel} associativity of parallel composition of processes;~\eqref{eq:interchange_law_diagrammatic} the interchange law;~\eqref{eq:swap_naturality}--\eqref{eq:swap_hexagon} the properties of the swap.

Let us start from the swap. In light of the definitions of sequential composition~(\ref{eq:sequential_composition}--\ref{eq:sequential_composition_explicit}) and identity process~\eqref{eq:identity}, the latent quantum swap process~\eqref{eq:swap} is involutive (therefore, in particular, reversible). Furthermore, by using the definitions of swap~\eqref{eq:swap}, parallel composition~\eqref{eq:parallel_composition}, and sequential composition~(\ref{eq:sequential_composition}--\ref{eq:sequential_composition_explicit}), combined with relation~\eqref{eq:intertwining}, one easily verifies the swapping of local laboratories~\eqref{eq:swap_naturality}. The symmetric-group property~\eqref{eq:swap_hexagon} is, in turn, just a consequence of the definitions of swap~\eqref{eq:swap}, identity process~\eqref{eq:identity}, parallel composition~\eqref{eq:parallel_composition}, and sequential composition~\eqref{eq:sequential_composition}: on the one hand, each quantum operation~\eqref{eq:quantum_circuit}, that is associated with the left-hand side of relation~\eqref{eq:swap_hexagon}, is a permutations of given subsystems; on the other hand, this is also the case for each quantum operation~\eqref{eq:quantum_circuit} associated with the right-hand side of relation~\eqref{eq:swap_hexagon}; this means that these quantum operations indeed pairwise coincide, since in \qt\ two permutations of given subsystems coincide if and only if their output systems coincide (as in this case).

Let us proceed with verifying the interchange law~\eqref{eq:interchange_law_diagrammatic}. This property can be actually split into three relations which---provided that associativity of sequential composition~\eqref{eq:associativity_sequential} and swapping of local experiments~\eqref{eq:swap_naturality} both hold---are equivalent to~\eqref{eq:interchange_law_diagrammatic}. These are given by:
\begin{align}
	&\begin{aligned}\label{eq:bifunct_paralell}
		\Qcircuit @C=1.5em @R=2.5em
		{
			&\s{A}&\qw&\qw&\s{A}\qw&\gate{\T{A}}&\qw&\s{B}\qw&
			\\
			&\s{D}&\qw&\gate{\T{D}}&\s{E}\qw&\qw&\qw&\s{E}\qw&
			\relax\gategroupColor{1}{4}{2}{4}{1.6em}{--}{gray}
			\relax\gategroupColor{1}{6}{2}{6}{1.6em}{--}{gray}
			\relax\gategroupColor{1}{4}{2}{6}{3em}{--}{gray}
		}
	\end{aligned}
	=
	\begin{aligned}
		\Qcircuit @C=1.5em @R=2.5em
		{
			&\s{A}&\gate{\T{A}}&\s{B}\qw&
			\\
			&\s{D}&\gate{\T{D}}&\s{E}\qw&
		}
	\end{aligned}
	=
	\begin{aligned}
		\Qcircuit @C=1.5em @R=2.5em
		{
			&\s{A}&\qw&\gate{\T{A}}&\s{B}\qw&\qw&\qw&\s{B}\qw&
			\\
			&\s{D}&\qw&\qw&\s{D}\qw&\gate{\T{D}}&\qw&\s{E}\qw&
			\relax\gategroupColor{1}{4}{2}{4}{1.6em}{--}{gray}
			\relax\gategroupColor{1}{6}{2}{6}{1.6em}{--}{gray}
			\relax\gategroupColor{1}{4}{2}{6}{3em}{--}{gray}
		}
	\end{aligned}
	,\\
	&\begin{aligned}\label{eq:bifunct_sequential}
		\Qcircuit @C=1.5em @R=2.5em
		{
			&\s{A}&\qw&\gate{\T{A}}&\s{B}\qw&\gate{\T{B}}&\qw&\s{C}\qw&
			\\
			&\s{D}&\qw&\qw&\qw&\qw&\qw&\s{D}\qw&
			\relax\gategroupColor{1}{4}{1}{6}{1.6em}{--}{gray}
			\relax\gategroupColor{2}{4}{2}{6}{1.6em}{--}{gray}
			\relax\gategroupColor{1}{4}{2}{6}{3em}{--}{gray}
		}
	\end{aligned}
	=
	\begin{aligned}
		\Qcircuit @C=1.5em @R=2.5em
		{
			&\s{A}&\qw&\gate{\T{A}}&\s{B}\qw&\gate{\T{B}}&\qw&\s{C}\qw&
			\\
			&\s{D}&\qw&\qw&\s{D}\qw&\qw&\qw&\s{D}\qw&
			\relax\gategroupColor{1}{4}{2}{4}{1.6em}{--}{gray}
			\relax\gategroupColor{1}{6}{2}{6}{1.6em}{--}{gray}
			\relax\gategroupColor{1}{4}{2}{6}{3em}{--}{gray}
		}
	\end{aligned}.
\end{align}
Relations~\eqref{eq:bifunct_paralell} and~\eqref{eq:bifunct_sequential} are easier to verify than the full interchange law~\eqref{eq:interchange_law_diagrammatic}. Indeed, by using the definitions of parallel~\eqref{eq:parallel_composition} and sequential~(\ref{eq:sequential_composition}--\ref{eq:sequential_composition_explicit}) compositions, combined with relation~\eqref{eq:intertwining}, properties~\eqref{eq:bifunct_paralell} and~\eqref{eq:bifunct_sequential} are easily verified; one can thus use these to reconstruct the interchange law~\eqref{eq:interchange_law_diagrammatic} by using associativity of sequential composition~\eqref{eq:associativity_sequential} and swapping of local experiments~\eqref{eq:swap_naturality}, that have been already checked.

Finally, we need to check associativity of $\boxtimes$~\eqref{eq:associativity_parallel} for processes. In the presence of all of the above other properties, that have been already verified, it is enough to prove the following particular case of full associativity~\eqref{eq:associativity_parallel}:
\begin{equation}\label{eq:preassociativity_diagrammatic}
		\input{pre-associativity_blank2.tikz}
		\quad=\quad
		\input{pre-associativity_blank.tikz}
		\quad .
\end{equation}
More explicitly, by resorting to the notation used so far, one needs to verify:
\begin{equation}\label{eq:preassociativity}
	\input{pre-associativity2.tikz}
	\quad=\quad
	\input{pre-associativity.tikz}
	\quad.
\end{equation}
Recalling the latent quantum identity process~\eqref{eq:identity} and using parallel composition~\eqref{eq:parallel_composition} twice for the two different associations, the quantum operations~\eqref{eq:quantum_circuit} associated with the left-hand and right-hand sides of~\eqref{eq:preassociativity} read, respectively:
\begin{align}\label{eq:quantum_circuit_check}
	\scalebox{0.87}
	{
	\input{quantum_circuit_check.tikz}
    }
    \quad
    ,
\end{align}
and
\begin{align}\label{eq:quantum_circuit_check2}
	\scalebox{0.87}
	{
		\input{quantum_circuit_check2.tikz}
	}
	\quad
	.
\end{align}
Checking that the standard quantum operations~\eqref{eq:quantum_circuit_check} and~\eqref{eq:quantum_circuit_check2} indeed coincide can be now done by direct inspection of the two expressions, and it suffices to note that, by definition~\eqref{eq:identity}, $\bigotimes_{j_{\mathtt{E}}=1}^{e}\left(\T{S}_{k,k'}^{j_\mathtt{E}}\otimes\T{I}_{\mathtt{N}\boxplus\mathtt{M}}^{j_\mathtt{E}}\right) = \bigotimes_{j_{\mathtt{E}}=1}^{e}\left[\T{S}_{k,k'}^{j_\mathtt{E}}\otimes\left(\T{I}_{\mathtt{N}}^{j_\mathtt{E}}\otimes\T{I}_{\mathtt{M}}^{j_\mathtt{E}}\right)\right] =  \bigotimes_{j_{\mathtt{E}}=1}^{e}\left[\left(\T{S}_{k,k'}^{j_\mathtt{E}}\otimes\T{I}_{\mathtt{N}}^{j_\mathtt{E}}\right)\otimes\T{I}_{\mathtt{M}}^{j_\mathtt{E}}\right]$ and, by definition~\eqref{eq:noisy_permutation}, $\T{S}_{k,k'}^{\mathtt{N}\boxplus\mathtt{M}} = \T{S}_{k,k'}^{\mathtt{N}}\otimes\left(\bigotimes_{j_{\mathtt{M}}=1}^{m}\T{S}_{k,k'}^{j_\mathtt{M}}\right)$.
Accordingly, identity~\eqref{eq:preassociativity} is verified, and it can be used combined with all of the other already verified operational relations, easily deducing its counterparts:
\begin{align*}
	&\input{pre-associativity_blank_var_1.tikz}
	\quad=\quad
	\input{pre-associativity_blank2_var_1.tikz}
	\quad
	,\\
	&\input{pre-associativity_blank_var_2.tikz}
	\quad=\quad
	\input{pre-associativity_blank2_var_2.tikz}
	\quad
	,
\end{align*}
which one can in turn exploit, combined with~\eqref{eq:preassociativity_diagrammatic}, to finally reconstruct associativity of $\boxtimes$~\eqref{eq:associativity_parallel}.

\subsection{Latent quantum Bell-like correlations coincide with standard quantum ones}\label{app:correlations}

The set of Bell-like correlations of an operational theory $\Theta$ are defined by so-called \emph{scenarios} of $n$ parties where each agent can independently choose some local measurements $\left(x_1,x_2,\ldots,x_n\right)$ to be performed on a shared setting $\nu$ (a $n$-partite state); the corresponding outcomes will be denoted by strings $\left(a_1,a_2,\ldots,a_n\right)$. Accordingly, the set of Bell-like correlations are given by the collection of all of the probability tables of the form:
\begin{equation*}
	\text{P}_{\Theta}{\left(a_1,a_2,\ldots,a_n\middle|x_1,x_2,\ldots,x_n;\nu\right)}
	\coloneqq
	\left(\rbra{a_1^{(x_1)}}\boxtimes_{\Theta}\rbra{a_2^{(x_2)}}\boxtimes_{\Theta}\cdots\boxtimes_{\Theta}\rbra{a_n^{(x_n)}}\right)\rket{\nu}
	,
\end{equation*}
where $\boxtimes_{\Theta}$ denotes the composition rule of the operational theory $\Theta$.

We can first easily prove that standard quantum Bell-like correlations are a subset of latent quantum ones:
\begin{equation}\label{eq:quantum_coincide_lq}
	\begin{split}
		\text{P}_{\qt}{\left(a_1,a_2,\ldots,a_n\middle|x_1,x_2,\ldots,x_n;\rho\right)}
		&\equiv
		\left(\rbra{a_1^{(x_1)}}_{\sys{Q}_{1}}\otimes\rbra{a_2^{(x_2)}}_{\sys{Q}_{2}}\otimes\cdots\otimes\rbra{a_n^{(x_n)}}_{\sys{Q}_{n}}\right)\rket{\rho}_{\sys{Q}_{1}\otimes\sys{Q}_{2}\otimes\cdots\otimes\sys{Q}_{n}}
		\\
		&=
		\text{Tr}{\left\{\rho\left(\Pi_{a_1}^{(x_1)}\otimes\Pi_{a_2}^{(x_2)}\otimes\cdots\otimes\Pi_{a_n}^{(x_n)}\right)\right\}}
		\\
		&=
		\text{Tr}{\left\{\text{Tr}_{\tilde{\sys{L}}}{\left[\left(\xi_{\tilde{\sys{L}}}\otimes\rho\right)\left(\mathbb{1}_{\tilde{\sys{L}}}\otimes\Pi_{a_1}^{(x_1)}\otimes\Pi_{a_2}^{(x_2)}\otimes\cdots\otimes\Pi_{a_n}^{(x_n)}\right)\right]}\right\}}
		\\
		&=
		\text{Tr}{\left\{\xi_{\tilde{\sys{L}}}\otimes\rho\left(\mathbb{1}_{\tilde{\sys{L}}}\otimes\Pi_{a_1}^{(x_1)}\otimes\Pi_{a_2}^{(x_2)}\otimes\cdots\otimes\Pi_{a_n}^{(x_n)}\right)\right\}}
		\\
		&=
		\left(\rbra{a_1^{(x_1)}}_{\sys{Q}_{1}}\boxtimes\rbra{a_2^{(x_2)}}_{\sys{Q}_{2}}\boxtimes\cdots\boxtimes\rbra{a_n^{(x_n)}}_{\sys{Q}_{n}}\right)\rket{\xi_{\tilde{\sys{L}}}\otimes\rho}_{\sys{Q}_{1}\boxtimes\sys{Q}_{2}\boxtimes\cdots\boxtimes\sys{Q}_{n}}
		\\
		&\equiv
		\text{P}_{\lqt}{\left(a_1,a_2,\ldots,a_n\middle|x_1,x_2,\ldots,x_n;\xi_{\tilde{\sys{L}}}\otimes\rho\right)}
		,
	\end{split}
\end{equation}
where $\xi_{\tilde{\sys{L}}}$ are the latent states associated with the parallel composition of the elementary latent quantum systems $\sys{Q}_i$.

Conversely, we can now prove that latent quantum Bell-like correlations are a subset of the standard quantum ones. By using the latent quantum composition rule~\eqref{eq:parallel_composition} $n$ times for the simple case of effects, one can then compute:
\begin{equation*}
	\begin{split}
		\text{P}_{\lqt}{\left(a_1,a_2,\ldots,a_n\middle|x_1,x_2,\ldots,x_n;\Sigma\right)}
		&\equiv
		\left(\rbra{a_1^{(x_1)}}_{\sys{Q}_{\mathtt{M}_1}}\boxtimes\rbra{a_2^{(x_2)}}_{\sys{Q}_{\mathtt{M}_2}}\boxtimes\cdots\boxtimes\rbra{a_n^{(x_n)}}_{\sys{Q}_{\mathtt{M}_n}}\right)\rket{\Sigma}_{\sys{Q}_{\mathtt{M}_1}\boxtimes\sys{Q}_{\mathtt{M}_2}\boxtimes\cdots\boxtimes\sys{Q}_{\mathtt{M}_n}}
		\\
		&=
		\text{Tr}{\left\{\tilde{\T{S}}{\left(\Sigma\right)}\left(\mathbb{1}_{\tilde{\sys{L}}}\otimes\Pi_{a_1}^{(x_1)}\otimes\Pi_{a_2}^{(x_2)}\otimes\cdots\otimes\Pi_{a_n}^{(x_n)}\right)\right\}}
		\\
		&=
		\text{Tr}{\left\{\text{Tr}_{\tilde{\sys{L}}}\tilde{\T{S}}{\left(\Sigma\right)}\left(\Pi_{a_1}^{(x_1)}\otimes\Pi_{a_2}^{(x_2)}\otimes\cdots\otimes\Pi_{a_n}^{(x_n)}\right)\right\}}
		\\
		&=
		\left(\rbra{a_1^{(x_1)}}_{\sys{Q}_{\mathtt{M}_1}}\otimes\rbra{a_2^{(x_2)}}_{\sys{Q}_{\mathtt{M}_2}}\otimes\cdots\otimes\rbra{a_n^{(x_n)}}_{\sys{Q}_{\mathtt{M}_n}}\right)\rket{\text{Tr}_{\tilde{\sys{L}}}\tilde{\T{S}}{\left(\Sigma\right)}}_{\sys{Q}_{\mathtt{M}_1}\otimes\sys{Q}_{\mathtt{M}_2}\otimes\cdots\otimes\sys{Q}_{\mathtt{M}_n}}
		\\
		&\equiv
		\text{P}_{\qt}{\left(a_1,a_2,\ldots,a_n\middle|x_1,x_2,\ldots,x_n;\text{Tr}_{\tilde{\sys{L}}}\tilde{\T{S}}{\left(\Sigma\right)}\right)}
		,
	\end{split}
\end{equation*}
where $\tilde{\sys{L}}$ denotes the Hilbert space associated with those latent factors that are not already counted within the (generally composite) subsystems $\sys{Q}_{\mathtt{M}_i}$, while $\tilde{\T{S}}$ is the convenient permutation of wires accounting for the suitable reordering of systems.

Therefore, we have shown that
\begin{equation*}
	\begin{split}
	&\forall
	\text{P}_{\lqt}{\left(a_1,a_2,\ldots,a_n\middle|x_1,x_2,\ldots,x_n;\Sigma\right)}
	\quad\exists
	\rho_{\Sigma}
	\colon
	\\
	&\text{P}_{\lqt}{\left(a_1,a_2,\ldots,a_n\middle|x_1,x_2,\ldots,x_n;\Sigma\right)}=
	\text{P}_{\qt}{\left(a_1,a_2,\ldots,a_n\middle|x_1,x_2,\ldots,x_n;\rho_{\Sigma}\right)}
	,
	\end{split}
\end{equation*}
and, vice versa:
\begin{equation*}
	\begin{split}
	&\forall
	\text{P}_{\qt}{\left(a_1,a_2,\ldots,a_n\middle|x_1,x_2,\ldots,x_n;\rho\right)}
	\quad\exists
	\Sigma_{\rho}
	\colon
	\\
	&\text{P}_{\qt}{\left(a_1,a_2,\ldots,a_n\middle|x_1,x_2,\ldots,x_n;\rho\right)}=
	\text{P}_{\lqt}{\left(a_1,a_2,\ldots,a_n\middle|x_1,x_2,\ldots,x_n;\Sigma_{\rho}\right)}
	.
\end{split}
\end{equation*}
Note that states $\rho_{\Sigma}$ and $\Sigma_{\rho}$ above generally depend on the particular partition into subsystems which is given by the product-structure of measurements, but not on their choices nor on their outcomes. However, this product-structure is fixed each time by the particular scenario under consideration. Incidentally, we observe that latent quantum correlations are \emph{closed under wirings} (or ``classical operations'')~\cite{PhysRevA.80.062107,Navascues:2015aa}: this simply follows from the fact that standard quantum correlations---which we just proved to coincide with latent quantum ones---are closed under wirings. This demonstrates that such a principle is also insufficient to single out a \emph{unique} quantum theory.

Finally, we can also show that, for each fixed scenario as above, the space-like structure of states can be also retained. This boils down to the following identity:
\begin{equation}\label{eq:scenarios_coincide}
	\begin{split}
	\scalebox{0.96}{\input{prepare_and_measure_lqt.tikz}}
	\quad &=\quad 
	\scalebox{0.96}{\input{prepare_and_measure_qt.tikz}}
	\\
	&\coloneqq\quad
	\scalebox{0.96}{\input{prepare_and_measure_qt_final.tikz}}
	\quad,
	\end{split}
\end{equation}
which is derived by using the latent quantum composition rule~\eqref{eq:parallel_composition} $m$ and $n$ times for the simple cases of, respectively, states and effects. On the one hand, $\mathtt{S}_1\mathtt{S}_2\cdots\mathtt{S}_m$ and $\mathtt{S}_1'\mathtt{S}_2'\cdots\mathtt{S}_n'$ correspond, up to the permutation $\mathcal{S}$, to generally different decompositions of the same composite system, while systems $\tilde{\sys{L}}'$ and $\tilde{\sys{L}}$ are those latent subsystems reflecting the product-structure of, respectively, states and effects appearing in the first side; on the other hand, the subsystem-structures of the $\tilde{\sys{Q}}_{\mathtt{S}_i}$ (respectively, of the $\tilde{\sys{Q}}_{\mathtt{S}_j'}$) are defined solely in terms of the product-structure of the effect $\rbra{a_1}_{\sys{Q}_{\mathtt{S}_1'}}\boxtimes\rbra{a_2}_{\sys{Q}_{\mathtt{S}_2'}}\boxtimes\cdots\boxtimes\rbra{a_n}_{\sys{Q}_{\mathtt{S}_n'}}$ (state $\rket{\Sigma_1}_{\sys{Q}_{\mathtt{S}_1}}\boxtimes\rket{\Sigma_2}_{\sys{Q}_{\mathtt{S}_2}}\boxtimes\cdots\boxtimes\rket{\Sigma_m}_{\sys{Q}_{\mathtt{S}_m}}$) of the original latent quantum scenario. Finally, note that the network of measurements in the first side of identity~\eqref{eq:scenarios_coincide} coincide with the one in its third side: this can be always achieved just by suitably choosing the permutation $\tilde{\T{S}}$ in terms of $\T{S}$ in the second side of identity~\eqref{eq:scenarios_coincide}---which is done by arbitrarily assigning every latent factor $i_{\mathtt{S}_{\tilde{i}}}j_{\mathtt{S}_{\tilde{j}}}$, by swapping it as a first-neighbour in the parallel composition, to the operational factor associated with $i_{\mathtt{S}_{\tilde{i}}}$ or $j_{\mathtt{S}_{\tilde{j}}}$ (or to some latent factor that was previously assigned in the same way). This means that, for every latent quantum scenario, there exists a standard quantum one with the same states-effects connectivity. Conversely, by retracing identity~\eqref{eq:scenarios_coincide} backward starting from the third side (as done in relation~\eqref{eq:quantum_coincide_lq}), it is now easy to see that, in turn, for every standard quantum scenario, there exists a latent quantum one with the same states-effects connectivity. Importantly, this result is robust under the implementation of shared randomness in both the preparational and the observational parts of such scenarios, since identity~\eqref{eq:scenarios_coincide} is linear in both states and effects. Such a possibility was considered e.g.~in Ref.~\cite{Renou:2021aa}. Remarkably, there is strong evidence that a large class of latent quantum theories, in spite having the same set of correlation scenarios~\cite{Fritz_2012} of standard quantum theory, does not admit of an \emph{ontological model}~\cite{Schmid_2024,soltani2025noncontextualontologicalmodelsoperational,soltani2025decouplinglocalclassicalityclassical} in terms of standard quantum theory and even of any theory which abides by local tomography.

\end{document}

%% file: arxiv_lqt-v4/figures/braiding.tikz
\begin{tikzpicture}
	\begin{pgfonlayer}{nodelayer}
		\node [style={system_label}] (0) at (-1.75, 1.5) {$\scriptstyle \sys  A$};
		\node [style=none] (1) at (1.25, 1) {};
		\node [style=none] (2) at (1.25, -1) {};
		\node [style={system_label}] (3) at (1.75, 1.5) {$\scriptstyle \sys  B$};
		\node [style={system_label}] (4) at (1.75, -0.5) {$\scriptstyle \sys  A$};
		\node [style=none] (5) at (-1.75, 1) {};
		\node [style=none] (6) at (-1.75, -1) {};
		\node [style=none] (7) at (-1.25, 1) {};
		\node [style=none] (8) at (-1.25, -1) {};
		\node [style={system_label}] (9) at (-1.75, -0.5) {$\scriptstyle \sys  B$};
		\node [style=none] (10) at (1.75, 1) {};
		\node [style=none] (11) at (1.75, -1) {};
	\end{pgfonlayer}
	\begin{pgfonlayer}{edgelayer}
		\draw (5.center) to (7.center);
		\draw (6.center) to (8.center);
		\draw (1.center) to (10.center);
		\draw (2.center) to (11.center);
		\draw [in=-180, out=0] (8.center) to (1.center);
		\draw [in=180, out=0] (7.center) to (2.center);
	\end{pgfonlayer}
\end{tikzpicture}

%% file: arxiv_lqt-v4/figures/swap_narrow_test_2.tikz
\begin{tikzpicture}
	\begin{pgfonlayer}{nodelayer}
		\node [style=none] (0) at (0.5, -1) {};
		\node [style=none] (1) at (0.5, 1) {};
		\node [style=none] (2) at (3, 1) {};
		\node [style=none] (3) at (3, -1) {};
		\node [style=none] (29) at (3, 1) {};
		\node [style=none] (30) at (3, -1) {};
		\node [style=none] (38) at (0.5, 1) {};
		\node [style=none] (39) at (0.5, -1) {};
		\node [style=none] (44) at (3, 1) {};
		\node [style=none] (45) at (3, -1) {};
		\node [style=none] (46) at (3.5, 1) {};
		\node [style=none] (47) at (3.5, -1) {};
		\node [style={system_label}] (48) at (-3, 1.5) {$\scriptstyle \sys  A$};
		\node [style={system_label}] (49) at (-3, -0.5) {$\scriptstyle \sys  C$};
		\node [style=none] (50) at (0.5, -1) {};
		\node [style=none] (51) at (0.5, 1) {};
		\node [style=small square] (52) at (-1.5, 1) {$\T{A}$};
		\node [style=small square] (53) at (-1.5, -1) {$\T{C}$};
		\node [style={system_label}] (54) at (0, -0.5) {$\scriptstyle \sys  D$};
		\node [style={system_label}] (55) at (0, 1.5) {$\scriptstyle \sys  B$};
		\node [style={system_label}] (56) at (3.5, 1.5) {$\scriptstyle \sys  D$};
		\node [style={system_label}] (57) at (3.5, -0.5) {$\scriptstyle \sys  B$};
		\node [style=none] (58) at (-3, 1) {};
		\node [style=none] (59) at (-3, -1) {};
	\end{pgfonlayer}
	\begin{pgfonlayer}{edgelayer}
		\draw [in=-180, out=0] (1.center) to (3.center);
		\draw (44.center) to (46.center);
		\draw (45.center) to (47.center);
		\draw (52) to (51.center);
		\draw (53) to (50.center);
		\draw (58.center) to (52);
		\draw (59.center) to (53);
		\draw [in=-180, out=0] (50.center) to (44.center);
	\end{pgfonlayer}
\end{tikzpicture}

%% file: arxiv_lqt-v4/figures/swap_narrow_test_1.tikz
\begin{tikzpicture}
	\begin{pgfonlayer}{nodelayer}
		\node [style=none] (0) at (-3, -1) {};
		\node [style=none] (1) at (-3, 1) {};
		\node [style=none] (2) at (-0.5, 1) {};
		\node [style=none] (3) at (-0.5, -1) {};
		\node [style=none] (25) at (3, -1) {};
		\node [style=none] (26) at (3, 1) {};
		\node [style=small square] (27) at (1.5, 1) {$\T{C}$};
		\node [style=small square] (28) at (1.5, -1) {$\T{A}$};
		\node [style=none] (29) at (-0.5, 1) {};
		\node [style=none] (30) at (-0.5, -1) {};
		\node [style={system_label}] (31) at (3, 1.5) {$\scriptstyle \sys  D$};
		\node [style={system_label}] (32) at (3, -0.5) {$\scriptstyle \sys  B$};
		\node [style={system_label}] (33) at (0, 1.5) {$\scriptstyle \sys  C$};
		\node [style={system_label}] (34) at (0, -0.5) {$\scriptstyle \sys  A$};
		\node [style=none] (36) at (-3.5, 1) {};
		\node [style=none] (37) at (-3.5, -1) {};
		\node [style=none] (38) at (-3, 1) {};
		\node [style=none] (39) at (-3, -1) {};
		\node [style={system_label}] (40) at (-3.5, 1.5) {$\scriptstyle \sys  A$};
		\node [style={system_label}] (41) at (-3.5, -0.5) {$\scriptstyle \sys  C$};
	\end{pgfonlayer}
	\begin{pgfonlayer}{edgelayer}
		\draw [in=-180, out=0] (1.center) to (3.center);
		\draw [in=180, out=0] (29.center) to (27);
		\draw (27) to (26.center);
		\draw (30.center) to (28);
		\draw (28) to (25.center);
		\draw (36.center) to (38.center);
		\draw (37.center) to (39.center);
		\draw [in=-180, out=0] (39.center) to (29.center);
	\end{pgfonlayer}
\end{tikzpicture}

%% file: arxiv_lqt-v4/figures/braiding_hexagon1.tikz
\begin{tikzpicture}
	\begin{pgfonlayer}{nodelayer}
		\node [style=none] (0) at (-3, 0) {};
		\node [style=none] (1) at (-3, 2) {};
		\node [style=none] (2) at (-0.5, 2) {};
		\node [style=none] (3) at (-0.5, 0) {};
		\node [style=none] (6) at (-0.5, 2) {};
		\node [style=none] (7) at (-0.5, 0) {};
		\node [style={system_label}] (8) at (1.75, 2.5) {$\scriptstyle \sys  B$};
		\node [style={system_label}] (9) at (0, 0.5) {$\scriptstyle \sys  A$};
		\node [style=none] (10) at (-3.5, 2) {};
		\node [style=none] (11) at (-3.5, 0) {};
		\node [style=none] (12) at (-3, 2) {};
		\node [style=none] (13) at (-3, 0) {};
		\node [style={system_label}] (14) at (-3.5, 2.5) {$\scriptstyle \sys  A$};
		\node [style={system_label}] (15) at (-3.5, 0.5) {$\scriptstyle \sys  B$};
		\node [style=none] (16) at (3.5, 2) {};
		\node [style=none] (17) at (0, 0) {};
		\node [style=none] (18) at (0.5, -2) {};
		\node [style=none] (19) at (0.5, 0) {};
		\node [style=none] (20) at (3, 0) {};
		\node [style=none] (21) at (3, -2) {};
		\node [style=none] (24) at (3, 0) {};
		\node [style=none] (25) at (3, -2) {};
		\node [style={system_label}] (26) at (3.5, 0.5) {$\scriptstyle \sys  C$};
		\node [style={system_label}] (27) at (3.5, -1.5) {$\scriptstyle \sys  A$};
		\node [style=none] (28) at (0, 0) {};
		\node [style=none] (29) at (-3.5, -2) {};
		\node [style=none] (30) at (0.5, 0) {};
		\node [style=none] (31) at (0.5, -2) {};
		\node [style={system_label}] (33) at (-1.75, -1.5) {$\scriptstyle \sys  C$};
		\node [style=none] (34) at (3.5, 0) {};
		\node [style=none] (35) at (3.5, -2) {};
		\node [style=none] (36) at (0, 3) {};
		\node [style=none] (37) at (0, -2.5) {};
		\node [style=none] (38) at (-4, 0) {};
		\node [style=none] (39) at (4, 0) {};
	\end{pgfonlayer}
	\begin{pgfonlayer}{edgelayer}
		\draw [in=-180, out=0, looseness=0.75] (1.center) to (3.center);
		\draw (10.center) to (12.center);
		\draw (11.center) to (13.center);
		\draw (6.center) to (16.center);
		\draw (7.center) to (17.center);
		\draw [in=-180, out=0, looseness=0.75] (19.center) to (21.center);
		\draw (28.center) to (30.center);
		\draw (29.center) to (31.center);
		\draw (24.center) to (34.center);
		\draw (25.center) to (35.center);
		\draw [in=180, out=0] (13.center) to (6.center);
		\draw [in=-180, out=0] (31.center) to (24.center);
	\end{pgfonlayer}
\end{tikzpicture}

%% file: arxiv_lqt-v4/figures/braiding_threewires_intwo1.tikz
\begin{tikzpicture}
	\begin{pgfonlayer}{nodelayer}
		\node [style={system_label}] (7) at (-3.5, 2.5) {$\scriptstyle \sys  A$};
		\node [style={system_label}] (8) at (-3.5, -1.5) {$\scriptstyle \sys{BC}$};
		\node [style=none] (10) at (3, 2) {};
		\node [style=none] (12) at (3, 2) {};
		\node [style={system_label}] (14) at (3.5, 2.5) {$\scriptstyle \sys{BC}$};
		\node [style={system_label}] (15) at (3.5, -1.5) {$\scriptstyle \sys  A$};
		\node [style=none] (16) at (-3.5, -2) {};
		\node [style=none] (17) at (-3, -2) {};
		\node [style=none] (18) at (3.5, 2) {};
		\node [style=none] (24) at (0, 3) {};
		\node [style=none] (25) at (0, -2.75) {};
		\node [style=none] (26) at (-4, 0) {};
		\node [style=none] (27) at (4, 0) {};
		\node [style=none] (28) at (-3, 2) {};
		\node [style=none] (30) at (-3.5, 2) {};
		\node [style=none] (31) at (-3, 2) {};
		\node [style=none] (39) at (3.5, -2) {};
		\node [style=none] (40) at (3, -2) {};
		\node [style=none] (41) at (3.5, -2) {};
		\node [style=none] (42) at (3, -2) {};
		\node [style=none] (45) at (0, 0) {};
	\end{pgfonlayer}
	\begin{pgfonlayer}{edgelayer}
		\draw (16.center) to (17.center);
		\draw (12.center) to (18.center);
		\draw (30.center) to (31.center);
		\draw (40.center) to (41.center);
		\draw [in=120, out=0] (31.center) to (45.center);
		\draw [in=180, out=-60] (45.center) to (42.center);
		\draw [in=-180, out=60] (45.center) to (12.center);
		\draw [in=-120, out=0] (17.center) to (45.center);
	\end{pgfonlayer}
\end{tikzpicture}

%% file: arxiv_lqt-v4/figures/transformation2.tikz
\begin{tikzpicture}
	\begin{pgfonlayer}{nodelayer}
		\node [style=small square] (0) at (0, 0) {$\T{G}$};
		\node [style=none] (1) at (1.25, 0.5) {$\scriptstyle \sys{Q}_{\mathtt{M}}$};
		\node [style=none] (2) at (-1.25, 0.5) {$\scriptstyle \sys{Q}_{\mathtt{N}}$};
		\node [style=none] (3) at (-1.75, 0) {};
		\node [style=none] (4) at (1.75, 0) {};
	\end{pgfonlayer}
	\begin{pgfonlayer}{edgelayer}
		\draw (3.center) to (0);
		\draw (4.center) to (0);
	\end{pgfonlayer}
\end{tikzpicture}

%% file: arxiv_lqt-v4/figures/noisy_permutation.tikz
\begin{tikzpicture}
	\begin{pgfonlayer}{nodelayer}
		\node [style=none] (1) at (4.5, 2.75) {};
		\node [style=none] (2) at (7.25, 1.25) {};
		\node [style=none] (3) at (-3, -0.5) {\vdots};
		\node [style=none] (6) at (3, -0.5) {\vdots};
		\node [style=none] (7) at (7.25, -2.25) {};
		\node [style=none] (9) at (7.25, 0) {};
		\node [style=none] (11) at (-4.5, 2.75) {};
		\node [style=none] (12) at (-7.25, 1.25) {};
		\node [style=none] (13) at (-7.25, 0) {};
		\node [style=none] (14) at (-7.25, -2.25) {};
		\node [style=tall rectangle] (19) at (0, 0.25) {$\T{S}_{j_{\mathtt{C}}}$};
		\node [style=none] (20) at (6, 3.25) {};
		\node [style=none] (21) at (6.5, 2.75) {};
		\node [style=none] (22) at (6.5, 2.75) {};
		\node [style=none] (23) at (6, 2.25) {};
		\node [style=none] (24) at (4.5, 2.25) {};
		\node [style=none] (25) at (4.5, 3.25) {};
		\node [style=none] (26) at (5.5, 2.75) {$\text{Tr}$};
		\node [style=none] (31) at (3, 3.25) {$\scriptstyle j_{\mathtt{C}}\times\mathtt{K}'$};
		\node [style=none] (32) at (-5.75, 3.25) {};
		\node [style=none] (33) at (-6.25, 2.75) {};
		\node [style=none] (34) at (-6.25, 2.75) {};
		\node [style=none] (35) at (-5.75, 2.25) {};
		\node [style=none] (36) at (-4.5, 2.25) {};
		\node [style=none] (37) at (-4.5, 3.25) {};
		\node [style=none] (38) at (-5.25, 2.75) {$\xi$};
		\node [style=none] (43) at (-3, 3.25) {$\scriptstyle j_{\mathtt{C}}\times\mathtt{K}$};
		\node [style=none] (44) at (3, 1.75) {$\scriptstyle j_{\mathtt{C}}1_\mathtt{M}$};
		\node [style=none] (45) at (-3, 1.75) {$\scriptstyle j_{\mathtt{C}}1_\mathtt{N}$};
		\node [style=none] (46) at (3, 0.5) {$\scriptstyle j_{\mathtt{C}}2_\mathtt{M}$};
		\node [style=none] (47) at (-3, 0.5) {$\scriptstyle j_{\mathtt{C}}2_\mathtt{N}$};
		\node [style=none] (48) at (3, -1.75) {$\scriptstyle j_{\mathtt{C}}m_\mathtt{M}$};
		\node [style=none] (49) at (-3, -1.75) {$\scriptstyle j_{\mathtt{C}}n_\mathtt{N}$};
		\node [style=none] (50) at (0, 4.25) {};
		\node [style=none] (51) at (0, -3.25) {};
	\end{pgfonlayer}
	\begin{pgfonlayer}{edgelayer}
		\draw [style=dashed edge] (11.center) to (1.center);
		\draw [style=dashed edge] (12.center) to (2.center);
		\draw [style=dashed edge] (13.center) to (9.center);
		\draw [style=dashed edge] (14.center) to (7.center);
		\draw (25.center) to (20.center);
		\draw [in=90, out=0] (20.center) to (21.center);
		\draw (22.center) to (21.center);
		\draw [in=0, out=-90] (22.center) to (23.center);
		\draw (23.center) to (24.center);
		\draw (25.center) to (24.center);
		\draw (37.center) to (32.center);
		\draw [in=90, out=180] (32.center) to (33.center);
		\draw (34.center) to (33.center);
		\draw [in=-180, out=-90] (34.center) to (35.center);
		\draw (35.center) to (36.center);
		\draw (37.center) to (36.center);
	\end{pgfonlayer}
\end{tikzpicture}

%% file: arxiv_lqt-v4/figures/transformation.tikz
\begin{tikzpicture}
	\begin{pgfonlayer}{nodelayer}
		\node [style=small square] (0) at (0, 0) {$\T{T},\ \lbrace\T{S}_{k,k'}^{\mathtt{C}}\rbrace_{\mathtt{C}}$};
		\node [style=none] (1) at (3, 0.5) {$\scriptstyle \sys{Q}_{\mathtt{M}}$};
		\node [style=none] (2) at (-3, 0.5) {$\scriptstyle \sys{Q}_{\mathtt{N}}$};
		\node [style=none] (3) at (-3.5, 0) {};
		\node [style=none] (4) at (3.5, 0) {};
		\node [style=none] (5) at (0, 1) {};
		\node [style=none] (6) at (0, -1) {};
	\end{pgfonlayer}
	\begin{pgfonlayer}{edgelayer}
		\draw (3.center) to (0);
		\draw (4.center) to (0);
	\end{pgfonlayer}
\end{tikzpicture}

%% file: arxiv_lqt-v4/figures/quantum_circuit.tikz
\begin{tikzpicture}
	\begin{pgfonlayer}{nodelayer}
		\node [style=transf] (2) at (0, -1.5) {$\T{T}$};
		\node [style=permut] (4) at (0, 2.25) {$\T{S}_{k,k'}^{\mathtt{E}}$};
		\node [style=sigma] (6) at (-4.75, 1.75) {$\sigma_1$};
		\node [style=sigma] (7) at (4.75, 1.75) {$\sigma_2$};
		\node [style=none] (8) at (-4.75, 2.25) {};
		\node [style=none] (9) at (4.75, 2.25) {};
		\node [style=none] (10) at (-4.75, -0.75) {};
		\node [style=none] (11) at (0, -0.75) {};
		\node [style=none] (12) at (4.75, -0.75) {};
		\node [style=none] (13) at (-1.25, 2.25) {};
		\node [style=none] (14) at (1, 2.25) {};
		\node [style=none] (15) at (-4.75, 4) {};
		\node [style=none] (16) at (4.75, 4) {};
		\node [style=none] (17) at (-3, -1.75) {$\scriptstyle \sys{O}_{\mathtt{N}}$};
		\node [style=none] (18) at (3, -1.75) {$\scriptstyle \sys{O}_{\mathtt{M}}$};
		\node [style=none] (19) at (0, -4.25) {$\scriptstyle \sys{O}_{\mathtt{E}}$};
		\node [style=none] (20) at (0, 4.5) {$\scriptstyle \mathtt{E}\odot\mathtt{E}$};
		\node [style=none] (21) at (-3, -0.25) {$\scriptstyle \mathtt{N}\odot\mathtt{N}$};
		\node [style=none] (22) at (3, -0.25) {$\scriptstyle \mathtt{M}\odot\mathtt{M}$};
		\node [style=none] (23) at (-3, 2.75) {$\scriptstyle \mathtt{E}\times\mathtt{N}$};
		\node [style=none] (24) at (3, 2.75) {$\scriptstyle \mathtt{E}\times\mathtt{M}$};
		\node [style=none] (25) at (0, -2.25) {};
		\node [style=none] (26) at (7, -2.25) {};
		\node [style=none] (27) at (-7.25, -2.25) {};
		\node [style=none] (28) at (7, -4.75) {};
		\node [style=none] (30) at (-7.25, -4.75) {};
		\node [style=none] (31) at (-7, 1.75) {};
		\node [style=none] (32) at (7, 1.75) {};
		\node [style=none] (33) at (-5, 1.75) {};
		\node [style=none] (34) at (5, 1.75) {};
		\node [style=none] (35) at (-6.75, 2.25) {$\scriptstyle \left(\mathtt{N}\boxplus\mathtt{E}\right)^{\odot 2}$};
		\node [style=none] (36) at (6.75, 2.25) {$\scriptstyle \left(\mathtt{M}\boxplus\mathtt{E}\right)^{\odot 2}$};
		\node [style=none] (37) at (0, 5.5) {};
		\node [style=none] (38) at (0, -5.75) {};
		\node [style=none] (39) at (-1.75, -3.5) {};
		\node [style=none] (40) at (-1.75, 0.5) {};
		\node [style=none] (41) at (1.75, 0.5) {};
		\node [style=none] (42) at (1.75, -3.5) {};
		\node [style=none] (43) at (0, -1.5) {};
		\node [style=none] (44) at (0, 2.25) {};
		\node [style=none] (45) at (-1.75, 3.5) {};
		\node [style=none] (46) at (1.75, 3.5) {};
		\node [style=none] (47) at (-1.75, 1) {};
		\node [style=none] (48) at (1.75, 1) {};
	\end{pgfonlayer}
	\begin{pgfonlayer}{edgelayer}
		\draw [style=dashed edge] (15.center) to (16.center);
		\draw [style=dashed edge] (30.center) to (28.center);
		\draw [style=dashed edge] (31.center) to (33.center);
		\draw [style=dashed edge] (34.center) to (32.center);
		\draw [style={l_teal}] (40.center)
			 to (39.center)
			 to (42.center)
			 to (41.center)
			 to cycle;
		\draw [style={l_periwinkle}] (47.center)
			 to (45.center)
			 to (46.center)
			 to (48.center)
			 to cycle;
		\draw [style=dashed edge] (13.center) to (8.center);
		\draw [style=dashed edge] (14.center) to (9.center);
		\draw [style=dashed edge] (25.center) to (27.center);
		\draw [style=dashed edge] (11.center) to (10.center);
		\draw [style=dashed edge] (11.center) to (12.center);
		\draw [style=dashed edge] (25.center) to (26.center);
	\end{pgfonlayer}
\end{tikzpicture}

%% file: arxiv_lqt-v4/figures/identity.tikz
\begin{tikzpicture}
	\begin{pgfonlayer}{nodelayer}
		\node [style=none] (2) at (0, 0.5) {$\scriptstyle \sys{Q}_{\mathtt{N}}$};
		\node [style=none] (3) at (-3.5, 0) {};
		\node [style=none] (4) at (3.5, 0) {};
	\end{pgfonlayer}
	\begin{pgfonlayer}{edgelayer}
		\draw (3.center) to (4.center);
	\end{pgfonlayer}
\end{tikzpicture}

%% file: arxiv_lqt-v4/figures/transformation_identity.tikz
\begin{tikzpicture}
	\begin{pgfonlayer}{nodelayer}
		\node [style=small square] (0) at (0, 0) {$\T{I}_{\sys{\mathtt{N}^{\odot 2}}\otimes\sys{O}_{\mathtt{N}}},\ \lbrace\T{I}_{\mathtt{C}\times\mathtt{N}}\rbrace_{\mathtt{C}}$};
		\node [style=none] (1) at (3.75, 0.5) {$\scriptstyle \sys{Q}_{\mathtt{N}}$};
		\node [style=none] (2) at (-3.75, 0.5) {$\scriptstyle \sys{Q}_{\mathtt{N}}$};
		\node [style=none] (3) at (-4.25, 0) {};
		\node [style=none] (4) at (4.25, 0) {};
		\node [style=none] (5) at (0, 1) {};
		\node [style=none] (6) at (0, -1) {};
	\end{pgfonlayer}
	\begin{pgfonlayer}{edgelayer}
		\draw (3.center) to (0);
		\draw (4.center) to (0);
	\end{pgfonlayer}
\end{tikzpicture}

%% file: arxiv_lqt-v4/figures/quantum_circuit_swap.tikz
\begin{tikzpicture}
	\begin{pgfonlayer}{nodelayer}
		\node [style=none] (0) at (0, 7.5) {$\scriptstyle \mathtt{E}\odot\mathtt{E}$};
		\node [style={b_sigma}] (1) at (-12, 3.5) {$\sigma_1$};
		\node [style={b_sigma}] (2) at (11.5, 3.5) {$\sigma_5$};
		\node [style={s_sigma}] (5) at (-6.25, -0.25) {$\sigma_2$};
		\node [style={s_sigma}] (6) at (5.75, -0.25) {$\sigma_4$};
		\node [style=permut] (7) at (0, 4.75) {$\T{S}_{\mathtt{N_1}\mathtt{N_2}}^{\mathtt{E}}$};
		\node [style=permut] (8) at (0, 1.25) {$\sigma_3$};
		\node [style=none] (13) at (-9.25, 5.25) {$\scriptstyle \mathtt{E}\times\left(\mathtt{N}_1\boxplus\mathtt{N}_2\right)$};
		\node [style=none] (14) at (-9.25, 0.25) {$\scriptstyle \left(\mathtt{N}_1\boxplus\mathtt{N}_2\right)^{\odot 2}$};
		\node [style=none] (19) at (-6.25, 1.25) {};
		\node [style=none] (21) at (-0.75, -0.25) {};
		\node [style=none] (22) at (0.75, -1.75) {};
		\node [style=none] (23) at (0.75, -0.25) {};
		\node [style=none] (24) at (-0.75, -1.75) {};
		\node [style=none] (27) at (0.75, -0.25) {};
		\node [style=none] (29) at (-0.75, -3.25) {};
		\node [style=none] (30) at (0.75, -4.75) {};
		\node [style=none] (31) at (0.75, -3.25) {};
		\node [style=none] (32) at (-0.75, -4.75) {};
		\node [style=none] (34) at (0.75, -4.75) {};
		\node [style=none] (41) at (5.75, 1.25) {};
		\node [style=none] (47) at (-6, -1.75) {};
		\node [style=none] (48) at (5.5, -1.75) {};
		\node [style=none] (49) at (-16, -3.25) {};
		\node [style=none] (50) at (-16, -4.75) {};
		\node [style=none] (51) at (-16, -6.75) {};
		\node [style=none] (52) at (15.5, -3.25) {};
		\node [style=none] (53) at (15.5, -4.75) {};
		\node [style=none] (54) at (15.5, -6.75) {};
		\node [style=none] (55) at (-12, 7) {};
		\node [style=none] (56) at (11.5, 7) {};
		\node [style=none] (57) at (-12, 4.75) {};
		\node [style=none] (58) at (11.5, 4.75) {};
		\node [style=none] (59) at (-12, -0.25) {};
		\node [style=none] (60) at (11.5, -0.25) {};
		\node [style=none] (61) at (8.75, 5.25) {$\scriptstyle \mathtt{E}\times\left(\mathtt{N}_2\boxplus\mathtt{N}_1\right)$};
		\node [style=none] (62) at (8.75, 0.25) {$\scriptstyle \left(\mathtt{N}_2\boxplus\mathtt{N}_1\right)^{\odot 2}$};
		\node [style=none] (63) at (-16, 3.5) {};
		\node [style=none] (64) at (15.5, 3.5) {};
		\node [style=none] (65) at (-9.25, -2.75) {$\scriptstyle \sys{O}_{\mathtt{N}_1}$};
		\node [style=none] (66) at (0, -6.25) {$\scriptstyle \sys{O}_{\mathtt{E}}$};
		\node [style=none] (67) at (-3.5, 1.75) {$\scriptstyle \mathtt{N}_2\times\mathtt{N}_1$};
		\node [style=none] (68) at (-3.5, -1.25) {$\scriptstyle \mathtt{N}_1\odot\mathtt{N}_1$};
		\node [style=none] (69) at (-3.5, 0.25) {$\scriptstyle \mathtt{N}_2\odot\mathtt{N}_2$};
		\node [style=none] (70) at (-9.25, -4.25) {$\scriptstyle \sys{O}_{\mathtt{N}_2}$};
		\node [style=none] (71) at (-14.75, 4) {$\scriptstyle \left[\left(\mathtt{N}_1\boxplus\mathtt{N}_2\right)\boxplus\mathtt{E}\right]^{\odot 2}$};
		\node [style=none] (72) at (14.25, 4) {$\scriptstyle \left[\left(\mathtt{N}_2\boxplus\mathtt{N}_1\right)\boxplus\mathtt{E}\right]^{\odot 2}$};
		\node [style=none] (75) at (3.5, 1.75) {$\scriptstyle \mathtt{N}_1\times\mathtt{N}_2$};
		\node [style=none] (78) at (3.5, -1.25) {$\scriptstyle \mathtt{N}_2\odot\mathtt{N}_2$};
		\node [style=none] (80) at (8.75, -2.75) {$\scriptstyle \sys{O}_{\mathtt{N}_2}$};
		\node [style=none] (81) at (3.5, 0.25) {$\scriptstyle \mathtt{N}_1\odot\mathtt{N}_1$};
		\node [style=none] (82) at (0, -7.75) {};
		\node [style=none] (83) at (-2, 6) {};
		\node [style=none] (84) at (2, 6) {};
		\node [style=none] (85) at (-2, 3.5) {};
		\node [style=none] (86) at (2, 3.5) {};
		\node [style=none] (87) at (-7.5, -5.5) {};
		\node [style=none] (88) at (-7.5, 2.75) {};
		\node [style=none] (89) at (7, 2.75) {};
		\node [style=none] (90) at (7, -5.5) {};
		\node [style=none] (91) at (8.75, -4.25) {$\scriptstyle \sys{O}_{\mathtt{N}_1}$};
		\node [style=none] (92) at (0, 8.5) {};
	\end{pgfonlayer}
	\begin{pgfonlayer}{edgelayer}
		\draw [style=dashed edge] (51.center) to (54.center);
		\draw [style=dashed edge] (55.center) to (56.center);
		\draw [style=dashed edge] (63.center) to (1);
		\draw [style=dashed edge] (2) to (64.center);
		\draw [style={l_periwinkle}] (85.center)
			 to (83.center)
			 to (84.center)
			 to (86.center)
			 to cycle;
		\draw [style={l_teal}] (88.center)
			 to (87.center)
			 to (90.center)
			 to (89.center)
			 to cycle;
		\draw [style=dashed edge] (58.center) to (7);
		\draw [style=dashed edge] (7) to (57.center);
		\draw [style=dashed edge] (47.center) to (24.center);
		\draw [style=dashed edge] (22.center) to (48.center);
		\draw [style=dashed edge] (49.center) to (29.center);
		\draw [style=dashed edge] (50.center) to (32.center);
		\draw [style=dashed edge] (34.center) to (53.center);
		\draw [style=dashed edge] (52.center) to (31.center);
		\draw [style=dashed edge] (59.center) to (5);
		\draw [style=dashed edge] (6) to (60.center);
		\draw [style=dashed edge, in=180, out=0] (21.center) to (22.center);
		\draw [style=dashed edge, in=-180, out=0] (29.center) to (34.center);
		\draw [style=dashed edge, in=-180, out=0] (32.center) to (31.center);
		\draw [style=dashed edge, in=-180, out=0] (24.center) to (27.center);
		\draw [style=dashed edge] (19.center) to (8);
		\draw [style=dashed edge] (5) to (21.center);
		\draw [style=dashed edge] (27.center) to (6);
		\draw [style=dashed edge] (8) to (41.center);
	\end{pgfonlayer}
\end{tikzpicture}

%% file: arxiv_lqt-v4/figures/sequential1.tikz
\begin{tikzpicture}
	\begin{pgfonlayer}{nodelayer}
		\node [style=small square] (7) at (0, 0) {$\T{G}\T{F}$};
		\node [style=none] (8) at (3, 0.5) {$\scriptstyle \sys{Q}_{\mathtt{B}}$};
		\node [style=none] (9) at (-3, 0.5) {$\scriptstyle \sys{Q}_{\mathtt{A}}$};
		\node [style=none] (10) at (-3.5, 0) {};
		\node [style=none] (11) at (3.5, 0) {};
		\node [style=none] (12) at (0, 1) {};
		\node [style=none] (13) at (0, -1) {};
	\end{pgfonlayer}
	\begin{pgfonlayer}{edgelayer}
		\draw (10.center) to (7);
		\draw (11.center) to (7);
	\end{pgfonlayer}
\end{tikzpicture}

%% file: arxiv_lqt-v4/figures/sequential_composition.tikz
\begin{tikzpicture}
	\begin{pgfonlayer}{nodelayer}
		\node [style=small square] (0) at (-1.5, 0) {$\T{F}$};
		\node [style=none] (1) at (0, 0.5) {$\scriptstyle \sys{Q}_{\mathtt{B}}$};
		\node [style=none] (2) at (-4, 0.5) {$\scriptstyle \sys{Q}_{\mathtt{A}}$};
		\node [style=none] (3) at (-4.5, 0) {};
		\node [style=small square] (5) at (1.5, 0) {$\T{G}$};
		\node [style=none] (6) at (4, 0.5) {$\scriptstyle \sys{Q}_{\mathtt{C}}$};
		\node [style=none] (9) at (4.5, 0) {};
		\node [style=none] (10) at (-2.75, 1.25) {};
		\node [style=none] (11) at (2.75, 1.25) {};
		\node [style=none] (12) at (-2.75, -1) {};
		\node [style=none] (13) at (2.75, -1) {};
	\end{pgfonlayer}
	\begin{pgfonlayer}{edgelayer}
		\draw (3.center) to (0);
		\draw (9.center) to (5);
		\draw (0) to (5);
		\draw [style={grey_densely_dashed}] (10.center) to (12.center);
		\draw [style={grey_densely_dashed}] (12.center) to (13.center);
		\draw [style={grey_densely_dashed}] (13.center) to (11.center);
		\draw [style={grey_densely_dashed}] (11.center) to (10.center);
	\end{pgfonlayer}
\end{tikzpicture}

%% file: arxiv_lqt-v4/figures/parallel_composition.tikz
\begin{tikzpicture}
	\begin{pgfonlayer}{nodelayer}
		\node [style=small square] (0) at (0, 1) {$\T{T},\ \lbrace\T{S}_{k,k'}^{\mathtt{C}}\rbrace_{\mathtt{C}}$};
		\node [style=none] (1) at (-3.5, -0.5) {$\scriptstyle \sys{Q}_{\mathtt{N}_2}$};
		\node [style=none] (2) at (-4, 1) {};
		\node [style=none] (3) at (4, 1) {};
		\node [style=small square] (10) at (0, -1) {$\tilde{\T{T}},\ \lbrace\tilde{\T{S}}_{z,z'}^{\mathtt{C}}\rbrace_{\mathtt{C}}$};
		\node [style=none] (11) at (3.5, -0.5) {$\scriptstyle \sys{Q}_{\mathtt{M}_2}$};
		\node [style=none] (12) at (-4, -1) {};
		\node [style=none] (13) at (4, -1) {};
		\node [style=none] (14) at (-3.5, 1.5) {$\scriptstyle \sys{Q}_{\mathtt{N}_1}$};
		\node [style=none] (15) at (3.5, 1.5) {$\scriptstyle \sys{Q}_{\mathtt{M}_1}$};
		\node [style=none] (16) at (0, 2.5) {};
		\node [style=none] (17) at (0, -2.5) {};
	\end{pgfonlayer}
	\begin{pgfonlayer}{edgelayer}
		\draw (2.center) to (0);
		\draw (3.center) to (0);
		\draw (12.center) to (10);
		\draw (13.center) to (10);
	\end{pgfonlayer}
\end{tikzpicture}

%% file: arxiv_lqt-v4/figures/quantum_circuit_parallel_composition_alternative.tikz
\begin{tikzpicture}
	\begin{pgfonlayer}{nodelayer}
		\node [style=none] (0) at (-9.25, 3) {};
		\node [style=none] (1) at (-9.25, -5.25) {};
		\node [style=none] (2) at (9.25, -5.25) {};
		\node [style=none] (3) at (9.25, 3) {};
		\node [style=none] (4) at (-4.5, 6.75) {};
		\node [style=none] (5) at (-4.5, 3.75) {};
		\node [style=none] (6) at (4.5, 3.75) {};
		\node [style=none] (7) at (4.5, 6.75) {};
		\node [style=none] (8) at (0, 8.25) {$\scriptstyle \mathtt{E}\odot\mathtt{E}$};
		\node [style={b_sigma}] (9) at (-13.5, 4) {$\sigma_1$};
		\node [style={b_sigma}] (10) at (13.75, 4) {$\sigma_4$};
		\node [style={b_transf}] (11) at (-3, -2.25) {$\T{T}$};
		\node [style={b_transf}] (12) at (1.5, -2.25) {$\tilde{\T{T}}$};
		\node [style={s_sigma}] (13) at (-8, 0) {$\sigma_2$};
		\node [style={s_sigma}] (14) at (8, 0) {$\sigma_3$};
		\node [style=permut] (15) at (0, 5.25) {$\bigotimes_{j_{\mathtt{E}}=1}^{e}\left(\T{S}_{k,k'}^{j_\mathtt{E}}\otimes\tilde{\T{S}}_{z,z'}^{j_\mathtt{E}}\right)$};
		\node [style=permut] (16) at (0, 1.5) {$\lbrace\T{S}_{k,k'}^{\mathtt{C}}\rbrace_{\mathtt{C}}*\lbrace\tilde{\T{S}}_{z,z'}^{\mathtt{C}}\rbrace_{\mathtt{C}}$};
		\node [style=none] (17) at (-8.75, 5.75) {$\scriptstyle \mathtt{E}\times\left(\mathtt{N}_1\boxplus\mathtt{N}_2\right)$};
		\node [style=none] (18) at (-11, 0.5) {$\scriptstyle \left(\mathtt{N}_1\boxplus\mathtt{N}_2\right)^{\odot 2}$};
		\node [style=none] (23) at (-7.75, 1.5) {};
		\node [style=none] (25) at (-1.25, 0) {};
		\node [style=none] (26) at (0.25, -1.5) {};
		\node [style=none] (27) at (0.25, 0) {};
		\node [style=none] (28) at (-1.25, -1.5) {};
		\node [style=none] (29) at (2.75, 0) {};
		\node [style=none] (30) at (4.25, -1.5) {};
		\node [style=none] (31) at (4.25, 0) {};
		\node [style=none] (32) at (2.75, -1.5) {};
		\node [style=none] (33) at (-1.25, -3) {};
		\node [style=none] (34) at (0.25, -4.5) {};
		\node [style=none] (35) at (0.25, -3) {};
		\node [style=none] (36) at (-1.25, -4.5) {};
		\node [style=none] (37) at (2.75, -3) {};
		\node [style=none] (38) at (4.25, -4.5) {};
		\node [style=none] (39) at (4.25, -3) {};
		\node [style=none] (40) at (2.75, -4.5) {};
		\node [style=none] (45) at (7.75, 1.5) {};
		\node [style=none] (51) at (-8, -1.5) {};
		\node [style=none] (52) at (7.75, -1.5) {};
		\node [style=none] (53) at (-17.5, -3) {};
		\node [style=none] (54) at (-17.5, -4.5) {};
		\node [style=none] (55) at (-17.5, -6.5) {};
		\node [style=none] (56) at (17.75, -3) {};
		\node [style=none] (57) at (17.75, -4.5) {};
		\node [style=none] (58) at (17.75, -6.5) {};
		\node [style=none] (59) at (-13.5, 7.75) {};
		\node [style=none] (60) at (13.75, 7.75) {};
		\node [style=none] (61) at (-13.5, 5.25) {};
		\node [style=none] (62) at (13.75, 5.25) {};
		\node [style=none] (63) at (-13.5, 0) {};
		\node [style=none] (64) at (13.75, 0) {};
		\node [style=none] (65) at (8.75, 5.75) {$\scriptstyle \mathtt{E}\times\left(\mathtt{M}_1\boxplus\mathtt{M}_2\right)$};
		\node [style=none] (66) at (11.25, 0.5) {$\scriptstyle \left(\mathtt{M}_1\boxplus\mathtt{M}_2\right)^{\odot 2}$};
		\node [style=none] (67) at (-17.5, 4) {};
		\node [style=none] (68) at (17.75, 4) {};
		\node [style=none] (69) at (-11, -2.5) {$\scriptstyle \sys{O}_{\mathtt{N}_1}$};
		\node [style=none] (70) at (0, -6) {$\scriptstyle \sys{O}_{\mathtt{E}}$};
		\node [style=none] (71) at (-5.75, 2) {$\scriptstyle \mathtt{N}_2\times\mathtt{N}_1$};
		\node [style=none] (72) at (-5.75, -1) {$\scriptstyle \mathtt{N}_1\odot\mathtt{N}_1$};
		\node [style=none] (73) at (-5.75, 0.5) {$\scriptstyle \mathtt{N}_2\odot\mathtt{N}_2$};
		\node [style=none] (74) at (-11, -4) {$\scriptstyle \sys{O}_{\mathtt{N}_2}$};
		\node [style=none] (75) at (-16.25, 4.5) {$\scriptstyle \left[\left(\mathtt{N}_1\boxplus\mathtt{N}_2\right)\boxplus\mathtt{E}\right]^{\odot 2}$};
		\node [style=none] (76) at (16.5, 4.5) {$\scriptstyle \left[\left(\mathtt{M}_1\boxplus\mathtt{M}_2\right)\boxplus\mathtt{E}\right]^{\odot 2}$};
		\node [style=none] (77) at (11.25, -2.5) {$\scriptstyle \sys{O}_{\mathtt{M}_1}$};
		\node [style=none] (78) at (11.25, -4) {$\scriptstyle \sys{O}_{\mathtt{M}_2}$};
		\node [style=none] (79) at (6, 2) {$\scriptstyle \mathtt{M}_2\times\mathtt{M}_1$};
		\node [style=none] (80) at (6, -1) {$\scriptstyle \mathtt{M}_1\odot\mathtt{M}_1$};
		\node [style=none] (81) at (6, 0.5) {$\scriptstyle \mathtt{M}_2\odot\mathtt{M}_2$};
		\node [style=none] (82) at (0, 10) {};
		\node [style=none] (83) at (0, -7.5) {};
	\end{pgfonlayer}
	\begin{pgfonlayer}{edgelayer}
		\draw [style={l_teal}] (1.center)
			 to (0.center)
			 to (3.center)
			 to (2.center)
			 to cycle;
		\draw [style={l_periwinkle}] (5.center)
			 to (4.center)
			 to (7.center)
			 to (6.center)
			 to cycle;
		\draw [style=dashed edge, in=-180, out=0, looseness=0.75] (28.center) to (27.center);
		\draw [style=dashed edge, in=180, out=0] (25.center) to (26.center);
		\draw [style=dashed edge, in=-180, out=0, looseness=0.75] (32.center) to (31.center);
		\draw [style=dashed edge, in=180, out=0] (29.center) to (30.center);
		\draw [style=dashed edge, in=-180, out=0, looseness=0.75] (36.center) to (35.center);
		\draw [style=dashed edge, in=180, out=0] (33.center) to (34.center);
		\draw [style=dashed edge, in=-180, out=0, looseness=0.75] (40.center) to (39.center);
		\draw [style=dashed edge, in=180, out=0] (37.center) to (38.center);
		\draw [style=dashed edge] (27.center) to (29.center);
		\draw [style=dashed edge] (34.center) to (40.center);
		\draw [style=dashed edge] (51.center) to (28.center);
		\draw [style=dashed edge] (30.center) to (52.center);
		\draw [style=dashed edge] (53.center) to (33.center);
		\draw [style=dashed edge] (54.center) to (36.center);
		\draw [style=dashed edge] (38.center) to (57.center);
		\draw [style=dashed edge] (56.center) to (39.center);
		\draw [style=dashed edge] (55.center) to (58.center);
		\draw [style=dashed edge] (59.center) to (60.center);
		\draw [style=dashed edge] (63.center) to (13);
		\draw [style=dashed edge] (64.center) to (14);
		\draw [style=dashed edge] (67.center) to (9);
		\draw [style=dashed edge] (10) to (68.center);
		\draw [style=dashed edge] (61.center) to (15);
		\draw [style=dashed edge] (62.center) to (15);
		\draw [style=dashed edge] (23.center) to (16);
		\draw [style=dashed edge] (16) to (45.center);
		\draw [style=dashed edge] (25.center) to (13);
		\draw [style=dashed edge] (26.center) to (32.center);
		\draw [style=dashed edge] (35.center) to (37.center);
		\draw [style=dashed edge] (31.center) to (14);
	\end{pgfonlayer}
\end{tikzpicture}

%% file: arxiv_lqt-v4/figures/asterisk_product.tikz
\begin{tikzpicture}
	\begin{pgfonlayer}{nodelayer}
		\node [style=small square] (0) at (0, 0) {$\lbrace\T{S}_{k,k'}^{\mathtt{C}}\rbrace_{\mathtt{C}}*\lbrace\tilde{\T{S}}_{z,z'}^{\mathtt{C}}\rbrace_{\mathtt{C}}$};
		\node [style=none] (1) at (4.25, 0.5) {$\scriptstyle \mathtt{M}_2\times\mathtt{M}_1$};
		\node [style=none] (2) at (-4.25, 0.5) {$\scriptstyle \mathtt{N}_2\times\mathtt{N}_1$};
		\node [style=none] (3) at (-4.75, 0) {};
		\node [style=none] (4) at (4.75, 0) {};
		\node [style=none] (5) at (0, 1) {};
		\node [style=none] (6) at (0, -1) {};
	\end{pgfonlayer}
	\begin{pgfonlayer}{edgelayer}
		\draw [style=dashed edge] (3.center) to (0);
		\draw [style=dashed edge] (4.center) to (0);
	\end{pgfonlayer}
\end{tikzpicture}

%% file: arxiv_lqt-v4/figures/asterisk_product_a.tikz
\begin{tikzpicture}
	\begin{pgfonlayer}{nodelayer}
		\node [style=small square] (0) at (-6.25, 0) {$\T{S}_{k,k'}^{\mathtt{N}_2}$};
		\node [style=none] (1) at (-4, 0.5) {$\scriptstyle \mathtt{N}_2\times\mathtt{M}_1$};
		\node [style=none] (2) at (-8.25, 0.5) {$\scriptstyle \mathtt{N}_2\times\mathtt{N}_1$};
		\node [style=none] (3) at (-8.75, 0) {};
		\node [style=small square] (5) at (-2.5, 0) {$\tau$};
		\node [style=small square] (12) at (1.5, 0) {$\tilde{\T{S}}_{z,z'}^{\mathtt{M}_1}$};
		\node [style=small square] (14) at (5.25, 0) {$\tilde\tau$};
		\node [style=none] (15) at (-0.75, 0.5) {$\scriptstyle \mathtt{M}_1\times\mathtt{N}_2$};
		\node [style=none] (16) at (3.75, 0.5) {$\scriptstyle \mathtt{M}_1\times\mathtt{M}_2$};
		\node [style=none] (17) at (7.25, 0) {};
		\node [style=none] (18) at (6.75, 0.5) {$\scriptstyle \mathtt{M}_2\times\mathtt{M}_1$};
		\node [style=none] (19) at (-0.75, 1.25) {};
		\node [style=none] (20) at (-0.75, -1) {};
	\end{pgfonlayer}
	\begin{pgfonlayer}{edgelayer}
		\draw [style=dashed edge] (3.center) to (0);
		\draw [style=dashed edge] (0) to (5);
		\draw [style=dashed edge] (5) to (12);
		\draw [style=dashed edge] (12) to (14);
		\draw [style=dashed edge] (17.center) to (14);
	\end{pgfonlayer}
\end{tikzpicture}

%% file: arxiv_lqt-v4/figures/asterisk_product_b.tikz
\begin{tikzpicture}
	\begin{pgfonlayer}{nodelayer}
		\node [style=small square] (0) at (-5.75, 0) {$\gamma$};
		\node [style=none] (1) at (-4.25, 0.5) {$\scriptstyle \mathtt{N}_1\times\mathtt{N}_2$};
		\node [style=none] (2) at (-7, 0.5) {$\scriptstyle \mathtt{N}_2\times\mathtt{N}_1$};
		\node [style=none] (3) at (-7.5, 0) {};
		\node [style=small square] (4) at (-2.25, 0) {$\tilde{\T{S}}_{z,z'}^{\mathtt{N}_1}$};
		\node [style=small square] (5) at (1.5, 0) {$\tilde\gamma$};
		\node [style=small square] (6) at (5.5, 0) {$\T{S}_{k,k'}^{\mathtt{M}_2}$};
		\node [style=none] (7) at (0, 0.5) {$\scriptstyle \mathtt{N}_1\times\mathtt{M}_2$};
		\node [style=none] (8) at (3.25, 0.5) {$\scriptstyle \mathtt{M}_2\times\mathtt{N}_1$};
		\node [style=none] (9) at (8.25, 0) {};
		\node [style=none] (10) at (7.75, 0.5) {$\scriptstyle \mathtt{M}_2\times\mathtt{M}_1$};
	\end{pgfonlayer}
	\begin{pgfonlayer}{edgelayer}
		\draw [style=dashed edge] (3.center) to (0);
		\draw [style=dashed edge] (0) to (4);
		\draw [style=dashed edge] (4) to (5);
		\draw [style=dashed edge] (5) to (6);
		\draw [style=dashed edge] (9.center) to (6);
	\end{pgfonlayer}
\end{tikzpicture}

%% file: arxiv_lqt-v4/figures/identity_parallel2.tikz
\begin{tikzpicture}
	\begin{pgfonlayer}{nodelayer}
		\node [style=none] (0) at (0, 1.25) {$\scriptstyle \sys{Q}_{\mathtt{N}}$};
		\node [style=none] (1) at (-3.5, 0.75) {};
		\node [style=none] (2) at (3.5, 0.75) {};
		\node [style=none] (3) at (0, -0.5) {$\scriptstyle \sys{Q}_{\mathtt{M}}$};
		\node [style=none] (4) at (-3.5, -1) {};
		\node [style=none] (5) at (3.5, -1) {};
		\node [style=none] (6) at (0, 2) {};
		\node [style=none] (7) at (0, -2) {};
		\node [style=none] (8) at (0, 0) {};
		\node [style=none] (9) at (0, 0) {};
	\end{pgfonlayer}
	\begin{pgfonlayer}{edgelayer}
		\draw (1.center) to (2.center);
		\draw (4.center) to (5.center);
	\end{pgfonlayer}
\end{tikzpicture}

%% file: arxiv_lqt-v4/figures/identity_parallel.tikz
\begin{tikzpicture}
	\begin{pgfonlayer}{nodelayer}
		\node [style=none] (0) at (0, 0.5) {$\scriptstyle \sys{Q}_{\mathtt{N}}\sys{Q}_{\mathtt{M}}$};
		\node [style=none] (1) at (-3.5, -0.25) {};
		\node [style=none] (2) at (3.5, -0.25) {};
	\end{pgfonlayer}
	\begin{pgfonlayer}{edgelayer}
		\draw (1.center) to (2.center);
	\end{pgfonlayer}
\end{tikzpicture}

%% file: arxiv_lqt-v4/figures/pre-associativity_blank2.tikz
\begin{tikzpicture}
	\begin{pgfonlayer}{nodelayer}
		\node [style=none] (0) at (-2.75, 0) {};
		\node [style=none] (1) at (-2.75, 1.75) {};
		\node [style=none] (2) at (2.75, 1.75) {};
		\node [style=none] (3) at (2.75, 0) {};
		\node [style=none] (4) at (2.75, -1.75) {};
		\node [style=none] (5) at (-2.75, -1.75) {};
		\node [style=medium square] (6) at (0, 1.75) {};
		\node [style=none] (7) at (-1.5, 0.5) {};
		\node [style=none] (8) at (1.5, 0.5) {};
		\node [style=none] (9) at (-1.5, -2.25) {};
		\node [style=none] (10) at (1.5, -2.25) {};
		\node [style=none] (11) at (-2, 3.25) {};
		\node [style=none] (12) at (2, 3.25) {};
		\node [style=none] (13) at (-2, -3.25) {};
		\node [style=none] (14) at (2, -3.25) {};
	\end{pgfonlayer}
	\begin{pgfonlayer}{edgelayer}
		\draw (0.center) to (3.center);
		\draw (4.center) to (5.center);
		\draw (1.center) to (6);
		\draw (6) to (2.center);
		\draw [style={grey_densely_dashed}] (7.center) to (9.center);
		\draw [style={grey_densely_dashed}] (9.center) to (10.center);
		\draw [style={grey_densely_dashed}] (10.center) to (8.center);
		\draw [style={grey_densely_dashed}] (8.center) to (7.center);
		\draw [style={grey_densely_dashed}] (11.center) to (13.center);
		\draw [style={grey_densely_dashed}] (13.center) to (14.center);
		\draw [style={grey_densely_dashed}] (14.center) to (12.center);
		\draw [style={grey_densely_dashed}] (12.center) to (11.center);
	\end{pgfonlayer}
\end{tikzpicture}

%% file: arxiv_lqt-v4/figures/pre-associativity_blank.tikz
\begin{tikzpicture}
	\begin{pgfonlayer}{nodelayer}
		\node [style=none] (1) at (-2.75, 0) {};
		\node [style=none] (2) at (-2.75, 1.75) {};
		\node [style=none] (3) at (2.75, 1.75) {};
		\node [style=none] (4) at (2.75, 0) {};
		\node [style=none] (5) at (2.75, -1.75) {};
		\node [style=none] (6) at (-2.75, -1.75) {};
		\node [style=none] (15) at (0, 3.75) {};
		\node [style=none] (16) at (0, -3.75) {};
		\node [style=medium square] (18) at (0, 1.75) {};
		\node [style=none] (19) at (-1.5, 2.75) {};
		\node [style=none] (20) at (1.5, 2.75) {};
		\node [style=none] (21) at (-1.5, -0.5) {};
		\node [style=none] (22) at (1.5, -0.5) {};
		\node [style=none] (23) at (-2, 3.25) {};
		\node [style=none] (24) at (2, 3.25) {};
		\node [style=none] (25) at (-2, -3.25) {};
		\node [style=none] (26) at (2, -3.25) {};
	\end{pgfonlayer}
	\begin{pgfonlayer}{edgelayer}
		\draw (1.center) to (4.center);
		\draw (5.center) to (6.center);
		\draw (2.center) to (18);
		\draw (18) to (3.center);
		\draw [style={grey_densely_dashed}] (19.center) to (21.center);
		\draw [style={grey_densely_dashed}] (21.center) to (22.center);
		\draw [style={grey_densely_dashed}] (22.center) to (20.center);
		\draw [style={grey_densely_dashed}] (20.center) to (19.center);
		\draw [style={grey_densely_dashed}] (23.center) to (25.center);
		\draw [style={grey_densely_dashed}] (25.center) to (26.center);
		\draw [style={grey_densely_dashed}] (26.center) to (24.center);
		\draw [style={grey_densely_dashed}] (24.center) to (23.center);
	\end{pgfonlayer}
\end{tikzpicture}

%% file: arxiv_lqt-v4/figures/pre-associativity2.tikz
\begin{tikzpicture}
	\begin{pgfonlayer}{nodelayer}
		\node [style=small square] (0) at (0, 1.25) {$\T{G}$};
		\node [style=none] (1) at (-1.25, -2) {};
		\node [style=none] (2) at (-3, 1.25) {};
		\node [style=none] (3) at (3, 1.25) {};
		\node [style=none] (4) at (1.25, -2) {};
		\node [style=none] (5) at (1.25, -0.5) {};
		\node [style=none] (6) at (-1.25, -0.5) {};
		\node [style=none] (7) at (-1.75, 0.5) {};
		\node [style=none] (8) at (-1.75, -2.5) {};
		\node [style=none] (9) at (1.75, -2.5) {};
		\node [style=none] (10) at (1.75, 0.5) {};
		\node [style=none] (11) at (-2.75, -0.5) {$\scriptstyle \sys{Q}_{\mathtt{N}\boxplus\mathtt{M}}$};
		\node [style=none] (12) at (2.75, -0.5) {$\scriptstyle \sys{Q}_{\mathtt{N}\boxplus\mathtt{M}}$};
		\node [style=none] (13) at (0, -1.5) {$\scriptstyle \sys{Q}_{\mathtt{M}}$};
		\node [style=none] (14) at (0, 0) {$\scriptstyle \sys{Q}_{\mathtt{N}}$};
		\node [style=none] (17) at (-2, -1) {};
		\node [style=none] (18) at (2, -1) {};
		\node [style=none] (19) at (-3, -1) {};
		\node [style=none] (20) at (3, -1) {};
		\node [style=none] (21) at (-3, 2.5) {};
		\node [style=none] (22) at (3, 2.5) {};
		\node [style=none] (23) at (-3, -2.5) {};
		\node [style=none] (24) at (3, -2.5) {};
		\node [style=none] (25) at (-3.75, 0) {};
		\node [style=none] (26) at (3.75, 0) {};
		\node [style=none] (27) at (-4.5, 0) {};
		\node [style=none] (28) at (4.5, 0) {};
		\node [style=none] (29) at (-4.75, 0.5) {$\scriptstyle \sys{Q}_{\mathtt{A}\boxplus\mathtt{N}\boxplus\mathtt{M}}$};
		\node [style=none] (30) at (4.75, 0.5) {$\scriptstyle \sys{Q}_{\mathtt{B}\boxplus\mathtt{N}\boxplus\mathtt{M}}$};
		\node [style=none] (31) at (-2.25, 1.75) {$\scriptstyle \sys{Q}_{\mathtt{A}}$};
		\node [style=none] (32) at (2.25, 1.75) {$\scriptstyle \sys{Q}_{\mathtt{B}}$};
	\end{pgfonlayer}
	\begin{pgfonlayer}{edgelayer}
		\draw [bend right=15] (7.center) to (8.center);
		\draw [bend right=15] (9.center) to (10.center);
		\draw (3.center) to (0);
		\draw (0) to (2.center);
		\draw (1.center) to (4.center);
		\draw (5.center) to (6.center);
		\draw (19.center) to (17.center);
		\draw (18.center) to (20.center);
		\draw [bend right] (21.center) to (23.center);
		\draw [bend right] (24.center) to (22.center);
		\draw (27.center) to (25.center);
		\draw (26.center) to (28.center);
	\end{pgfonlayer}
\end{tikzpicture}

%% file: arxiv_lqt-v4/figures/pre-associativity.tikz
\begin{tikzpicture}
	\begin{pgfonlayer}{nodelayer}
		\node [style=small square] (0) at (0, 1.25) {$\T{G}$};
		\node [style=none] (1) at (-1.25, -0.5) {};
		\node [style=none] (2) at (-1.25, 1.25) {};
		\node [style=none] (3) at (1.25, 1.25) {};
		\node [style=none] (4) at (1.25, -0.5) {};
		\node [style=none] (5) at (2.75, -2) {};
		\node [style=none] (6) at (-2.75, -2) {};
		\node [style=none] (7) at (-1.75, 2.25) {};
		\node [style=none] (8) at (-1.75, -0.75) {};
		\node [style=none] (9) at (1.75, -0.75) {};
		\node [style=none] (10) at (1.75, 2.25) {};
		\node [style=none] (11) at (-2.75, 1.25) {$\scriptstyle \sys{Q}_{\mathtt{A}\boxplus\mathtt{N}}$};
		\node [style=none] (12) at (2.75, 1.25) {$\scriptstyle \sys{Q}_{\mathtt{B}\boxplus\mathtt{N}}$};
		\node [style=none] (13) at (0, 0) {$\scriptstyle \sys{Q}_{\mathtt{N}}$};
		\node [style=none] (16) at (0, -1.5) {$\scriptstyle \sys{Q}_{\mathtt{M}}$};
		\node [style=none] (17) at (0, 3) {};
		\node [style=none] (18) at (0, -3) {};
		\node [style=none] (19) at (-2, 0.75) {};
		\node [style=none] (20) at (2, 0.75) {};
		\node [style=none] (21) at (-3, 0.75) {};
		\node [style=none] (22) at (3, 0.75) {};
		\node [style=none] (23) at (-3, 2.5) {};
		\node [style=none] (24) at (3, 2.5) {};
		\node [style=none] (25) at (-3, -2.5) {};
		\node [style=none] (26) at (3, -2.5) {};
		\node [style=none] (27) at (-3.75, 0) {};
		\node [style=none] (28) at (3.75, 0) {};
		\node [style=none] (29) at (-4.5, 0) {};
		\node [style=none] (30) at (4.5, 0) {};
		\node [style=none] (31) at (-4.75, 0.5) {$\scriptstyle \sys{Q}_{\mathtt{A}\boxplus\mathtt{N}\boxplus\mathtt{M}}$};
		\node [style=none] (32) at (4.75, 0.5) {$\scriptstyle \sys{Q}_{\mathtt{B}\boxplus\mathtt{N}\boxplus\mathtt{M}}$};
		\node [style=none] (33) at (-1.25, 1.75) {$\scriptstyle \sys{Q}_{\mathtt{A}}$};
		\node [style=none] (34) at (1.25, 1.75) {$\scriptstyle \sys{Q}_{\mathtt{B}}$};
	\end{pgfonlayer}
	\begin{pgfonlayer}{edgelayer}
		\draw [bend right=15] (7.center) to (8.center);
		\draw [bend right=15] (9.center) to (10.center);
		\draw (3.center) to (0);
		\draw (0) to (2.center);
		\draw (1.center) to (4.center);
		\draw (5.center) to (6.center);
		\draw (21.center) to (19.center);
		\draw (20.center) to (22.center);
		\draw [bend right] (23.center) to (25.center);
		\draw [bend right] (26.center) to (24.center);
		\draw (29.center) to (27.center);
		\draw (28.center) to (30.center);
	\end{pgfonlayer}
\end{tikzpicture}

%% file: arxiv_lqt-v4/figures/quantum_circuit_check.tikz
\begin{tikzpicture}
	\begin{pgfonlayer}{nodelayer}
		\node [style=none] (0) at (-9.25, 3) {};
		\node [style=none] (1) at (-9.25, -5.25) {};
		\node [style=none] (2) at (9.25, -5.25) {};
		\node [style=none] (3) at (9.25, 3) {};
		\node [style=none] (4) at (-4.5, 6.75) {};
		\node [style=none] (5) at (-4.5, 3.75) {};
		\node [style=none] (6) at (4.5, 3.75) {};
		\node [style=none] (7) at (4.5, 6.75) {};
		\node [style=none] (8) at (0, 8.25) {$\scriptstyle \mathtt{E}\odot\mathtt{E}$};
		\node [style={b_sigma}] (9) at (-13.5, 4) {$\sigma_1$};
		\node [style={b_sigma}] (10) at (13.75, 4) {$\sigma_4$};
		\node [style={b_transf}] (11) at (0, -2.25) {$\T{T}$};
		\node [style={s_sigma}] (13) at (-8, 0) {$\sigma_2$};
		\node [style={s_sigma}] (14) at (8, 0) {$\sigma_3$};
		\node [style=permut] (15) at (0, 5.25) {$\bigotimes_{j_{\mathtt{E}}=1}^{e}\left(\T{S}_{k,k'}^{j_\mathtt{E}}\otimes\T{I}_{\mathtt{N}\boxplus\mathtt{M}}^{j_\mathtt{E}}\right)$};
		\node [style=permut] (16) at (0, 1.5) {$\T{S}_{k,k'}^{\mathtt{N}\boxplus\mathtt{M}}$};
		\node [style=none] (17) at (-8.75, 5.75) {$\scriptstyle \mathtt{E}\times\left(\mathtt{A}\boxplus\mathtt{N}\boxplus\mathtt{M}\right)$};
		\node [style=none] (18) at (-11, 0.5) {$\scriptstyle \left(\mathtt{A}\boxplus\mathtt{N}\boxplus\mathtt{M}\right)^{\odot 2}$};
		\node [style=none] (23) at (-7.75, 1.5) {};
		\node [style=none] (24) at (-7.75, -0.5) {};
		\node [style=none] (45) at (7.75, 1.5) {};
		\node [style=none] (46) at (7.75, -0.5) {};
		\node [style=none] (51) at (-7.75, -1.5) {};
		\node [style=none] (52) at (7.75, -1.5) {};
		\node [style=none] (53) at (-17.5, -3) {};
		\node [style=none] (54) at (-17.5, -4.5) {};
		\node [style=none] (55) at (-17.5, -6.5) {};
		\node [style=none] (56) at (17.75, -3) {};
		\node [style=none] (57) at (17.75, -4.5) {};
		\node [style=none] (58) at (17.75, -6.5) {};
		\node [style=none] (59) at (-13.5, 7.75) {};
		\node [style=none] (60) at (13.75, 7.75) {};
		\node [style=none] (61) at (-13.5, 5.25) {};
		\node [style=none] (62) at (13.75, 5.25) {};
		\node [style=none] (63) at (-13.5, 0) {};
		\node [style=none] (64) at (13.75, 0) {};
		\node [style=none] (65) at (8.75, 5.75) {$\scriptstyle \mathtt{E}\times\left(\mathtt{B}\boxplus\mathtt{N}\boxplus\mathtt{M}\right)$};
		\node [style=none] (66) at (11.25, 0.5) {$\scriptstyle \left(\mathtt{B}\boxplus\mathtt{N}\boxplus\mathtt{M}\right)^{\odot 2}$};
		\node [style=none] (67) at (-17.5, 4) {};
		\node [style=none] (68) at (17.75, 4) {};
		\node [style=none] (69) at (-11, -2.5) {$\scriptstyle \sys{O}_{\mathtt{A}}$};
		\node [style=none] (70) at (0, -6) {$\scriptstyle \sys{O}_{\mathtt{E}}$};
		\node [style=none] (71) at (-4.5, 2) {$\scriptstyle \left(\mathtt{N}\boxplus\mathtt{M}\right)\times\mathtt{A}$};
		\node [style=none] (72) at (-4.5, -1) {$\scriptstyle \mathtt{A}\odot\mathtt{A}$};
		\node [style=none] (73) at (0, 0) {$\scriptstyle \left(\mathtt{N}\boxplus\mathtt{M}\right)\odot\left(\mathtt{N}\boxplus\mathtt{M}\right)$};
		\node [style=none] (74) at (-11, -4) {$\scriptstyle \sys{O}_{\mathtt{N}\boxplus\mathtt{M}}$};
		\node [style=none] (75) at (-16.25, 4.5) {$\scriptstyle \left[\left(\mathtt{A}\boxplus\mathtt{N}\boxplus\mathtt{M}\right)\boxplus\mathtt{E}\right]^{\odot 2}$};
		\node [style=none] (76) at (16.5, 4.5) {$\scriptstyle \left[\left(\mathtt{B}\boxplus\mathtt{N}\boxplus\mathtt{M}\right)\boxplus\mathtt{E}\right]^{\odot 2}$};
		\node [style=none] (77) at (11.25, -2.5) {$\scriptstyle \sys{O}_{\mathtt{B}}$};
		\node [style=none] (78) at (11.25, -4) {$\scriptstyle \sys{O}_{\mathtt{N}\boxplus\mathtt{M}}$};
		\node [style=none] (79) at (4.5, 2) {$\scriptstyle \left(\mathtt{N}\boxplus\mathtt{M}\right)\times\mathtt{B}$};
		\node [style=none] (80) at (4.5, -1) {$\scriptstyle \mathtt{B}\odot\mathtt{B}$};
		\node [style=none] (82) at (0, 10) {};
		\node [style=none] (83) at (0, -7.5) {};
	\end{pgfonlayer}
	\begin{pgfonlayer}{edgelayer}
		\draw [style={l_teal}] (1.center)
			 to (0.center)
			 to (3.center)
			 to (2.center)
			 to cycle;
		\draw [style={l_periwinkle}] (5.center)
			 to (4.center)
			 to (7.center)
			 to (6.center)
			 to cycle;
		\draw [style=dashed edge] (55.center) to (58.center);
		\draw [style=dashed edge] (59.center) to (60.center);
		\draw [style=dashed edge] (63.center) to (13);
		\draw [style=dashed edge] (64.center) to (14);
		\draw [style=dashed edge] (67.center) to (9);
		\draw [style=dashed edge] (10) to (68.center);
		\draw [style=dashed edge] (61.center) to (15);
		\draw [style=dashed edge] (62.center) to (15);
		\draw [style=dashed edge] (23.center) to (16);
		\draw [style=dashed edge] (16) to (45.center);
		\draw [style=dashed edge] (53.center) to (56.center);
		\draw [style=dashed edge] (57.center) to (54.center);
		\draw [style=dashed edge] (51.center) to (52.center);
		\draw [style=dashed edge] (46.center) to (24.center);
	\end{pgfonlayer}
\end{tikzpicture}

%% file: arxiv_lqt-v4/figures/quantum_circuit_check2.tikz
\begin{tikzpicture}
	\begin{pgfonlayer}{nodelayer}
		\node [style=none] (0) at (-9.25, 3) {};
		\node [style=none] (1) at (-9.25, -10.25) {};
		\node [style=none] (2) at (9.25, -10.25) {};
		\node [style=none] (3) at (9.25, 3) {};
		\node [style=none] (4) at (-5.5, 6.75) {};
		\node [style=none] (5) at (-5.5, 3.75) {};
		\node [style=none] (6) at (5.5, 3.75) {};
		\node [style=none] (7) at (5.5, 6.75) {};
		\node [style=none] (8) at (0, 8.25) {$\scriptstyle \mathtt{E}\odot\mathtt{E}$};
		\node [style={b_sigma}] (9) at (-13.5, 4) {$\sigma_1$};
		\node [style={b_sigma}] (10) at (13.75, 4) {$\sigma_6$};
		\node [style=permut] (15) at (0, 5.25) {$\bigotimes_{j_{\mathtt{E}}=1}^{e}\left[\left(\T{S}_{k,k'}^{j_\mathtt{E}}\otimes\T{I}_{\mathtt{N}}^{j_\mathtt{E}}\right)\otimes\T{I}_{\mathtt{M}}^{j_\mathtt{E}}\right]$};
		\node [style=permut] (16) at (0, 1) {$\bigotimes_{j_{\mathtt{M}}=1}^{m}\left(\T{S}_{k,k'}^{j_\mathtt{M}}\otimes\T{I}_{\mathtt{N}}^{j_\mathtt{M}}\right)$};
		\node [style=none] (17) at (-8.75, 5.75) {$\scriptstyle \mathtt{E}\times\left(\mathtt{A}\boxplus\mathtt{N}\boxplus\mathtt{M}\right)$};
		\node [style=none] (18) at (-11, 0.5) {$\scriptstyle \left(\mathtt{A}\boxplus\mathtt{N}\boxplus\mathtt{M}\right)^{\odot 2}$};
		\node [style=none] (23) at (-7.75, 1) {};
		\node [style=none] (24) at (-7.75, -1.5) {};
		\node [style=none] (45) at (7.75, 1) {};
		\node [style=none] (46) at (7.75, -1.5) {};
		\node [style=none] (55) at (-17.5, -11.5) {};
		\node [style=none] (58) at (17.75, -11.5) {};
		\node [style=none] (59) at (-13.5, 7.75) {};
		\node [style=none] (60) at (13.75, 7.75) {};
		\node [style=none] (61) at (-13.5, 5.25) {};
		\node [style=none] (62) at (13.75, 5.25) {};
		\node [style=none] (63) at (-13.5, 0) {};
		\node [style=none] (64) at (13.75, 0) {};
		\node [style=none] (65) at (8.75, 5.75) {$\scriptstyle \mathtt{E}\times\left(\mathtt{B}\boxplus\mathtt{N}\boxplus\mathtt{M}\right)$};
		\node [style=none] (66) at (11.25, 0.5) {$\scriptstyle \left(\mathtt{B}\boxplus\mathtt{N}\boxplus\mathtt{M}\right)^{\odot 2}$};
		\node [style=none] (67) at (-17.5, 4) {};
		\node [style=none] (68) at (17.75, 4) {};
		\node [style=none] (70) at (0, -11) {$\scriptstyle \sys{O}_{\mathtt{E}}$};
		\node [style=none] (71) at (-5.5, 1.5) {$\scriptstyle \mathtt{M}\times\left(\mathtt{A}\boxplus\mathtt{N}\right)$};
		\node [style=none] (73) at (0, -1) {$\scriptstyle \mathtt{M}\odot\mathtt{M}$};
		\node [style=none] (75) at (-16.25, 4.5) {$\scriptstyle \left[\left(\mathtt{A}\boxplus\mathtt{N}\boxplus\mathtt{M}\right)\boxplus\mathtt{E}\right]^{\odot 2}$};
		\node [style=none] (76) at (16.5, 4.5) {$\scriptstyle \left[\left(\mathtt{B}\boxplus\mathtt{N}\boxplus\mathtt{M}\right)\boxplus\mathtt{E}\right]^{\odot 2}$};
		\node [style=none] (79) at (5.5, 1.5) {$\scriptstyle \mathtt{M}\times\left(\mathtt{B}\boxplus\mathtt{N}\right)$};
		\node [style=none] (82) at (0, 10) {};
		\node [style=none] (83) at (0, -12.75) {};
		\node [style={b_sigma}] (127) at (-8, -2.5) {$\sigma_2$};
		\node [style={b_transf}] (128) at (0, -7.25) {$\T{T}$};
		\node [style={s_sigma}] (129) at (-4.5, -5) {$\sigma_3$};
		\node [style={s_sigma}] (130) at (4.5, -5) {$\sigma_4$};
		\node [style=permut] (131) at (0, -3.5) {$\T{S}_{k,k'}^{\mathtt{N}}$};
		\node [style=none] (133) at (-4.25, -3.5) {};
		\node [style=none] (134) at (-4.25, -5.5) {};
		\node [style=none] (143) at (4.25, -3.5) {};
		\node [style=none] (144) at (4.25, -5.5) {};
		\node [style=none] (145) at (-4.25, -6.5) {};
		\node [style=none] (146) at (4.25, -6.5) {};
		\node [style=none] (147) at (-17.5, -8) {};
		\node [style=none] (148) at (-17.5, -9.5) {};
		\node [style=none] (149) at (17.75, -8) {};
		\node [style=none] (150) at (17.75, -9.5) {};
		\node [style=none] (151) at (-8, -5) {};
		\node [style=none] (152) at (8, -5) {};
		\node [style=none] (154) at (-11, -7.5) {$\scriptstyle \sys{O}_{\mathtt{A}}$};
		\node [style=none] (158) at (-11, -9) {$\scriptstyle \sys{O}_{\mathtt{N}\boxplus\mathtt{M}}$};
		\node [style=none] (159) at (11.25, -7.5) {$\scriptstyle \sys{O}_{\mathtt{B}}$};
		\node [style=none] (160) at (11.25, -9) {$\scriptstyle \sys{O}_{\mathtt{N}\boxplus\mathtt{M}}$};
		\node [style={b_sigma}] (164) at (8, -2.5) {$\sigma_5$};
		\node [style=none] (167) at (0, -5) {$\scriptstyle \mathtt{N}\odot\mathtt{N}$};
		\node [style=none] (168) at (2.5, -6) {$\scriptstyle \mathtt{B}\odot\mathtt{B}$};
		\node [style=none] (169) at (-2.5, -6) {$\scriptstyle \mathtt{A}\odot\mathtt{A}$};
		\node [style=none] (170) at (-7.75, 0) {};
		\node [style=none] (171) at (7.75, 0) {};
		\node [style=none] (172) at (-6.25, -4.5) {$\scriptstyle \left(\mathtt{A}\boxplus\mathtt{N}\right)^{\odot2}$};
		\node [style=none] (173) at (6.25, -4.5) {$\scriptstyle \left(\mathtt{B}\boxplus\mathtt{N}\right)^{\odot2}$};
		\node [style=none] (174) at (-2.75, -3) {$\scriptstyle \mathtt{N}\times\mathtt{A}$};
		\node [style=none] (175) at (2.75, -3) {$\scriptstyle \mathtt{N}\times\mathtt{B}$};
	\end{pgfonlayer}
	\begin{pgfonlayer}{edgelayer}
		\draw [style={l_teal}] (1.center)
			 to (0.center)
			 to (3.center)
			 to (2.center)
			 to cycle;
		\draw [style={l_periwinkle}] (5.center)
			 to (4.center)
			 to (7.center)
			 to (6.center)
			 to cycle;
		\draw [style=dashed edge] (55.center) to (58.center);
		\draw [style=dashed edge] (59.center) to (60.center);
		\draw [style=dashed edge] (67.center) to (9);
		\draw [style=dashed edge] (10) to (68.center);
		\draw [style=dashed edge] (61.center) to (15);
		\draw [style=dashed edge] (62.center) to (15);
		\draw [style=dashed edge] (23.center) to (16);
		\draw [style=dashed edge] (16) to (45.center);
		\draw [style=dashed edge] (133.center) to (131);
		\draw [style=dashed edge] (131) to (143.center);
		\draw [style=dashed edge] (63.center) to (170.center);
		\draw [style=dashed edge] (171.center) to (64.center);
		\draw [style=dashed edge] (24.center) to (46.center);
		\draw [style=dashed edge] (151.center) to (129);
		\draw [style=dashed edge] (130) to (152.center);
		\draw [style=dashed edge] (144.center) to (134.center);
		\draw [style=dashed edge] (145.center) to (146.center);
		\draw [style=dashed edge] (147.center) to (149.center);
		\draw [style=dashed edge] (150.center) to (148.center);
	\end{pgfonlayer}
\end{tikzpicture}

%% file: arxiv_lqt-v4/figures/pre-associativity_blank_var_1.tikz
\begin{tikzpicture}
	\begin{pgfonlayer}{nodelayer}
		\node [style=none] (2) at (-2.75, 0) {};
		\node [style=none] (3) at (2.75, 0) {};
		\node [style=none] (5) at (2.75, -1.75) {};
		\node [style=none] (6) at (-2.75, -1.75) {};
		\node [style=none] (15) at (0, 3.25) {};
		\node [style=none] (16) at (0, -3.25) {};
		\node [style=medium square] (18) at (0, 0) {};
		\node [style=none] (19) at (-1.5, 2.25) {};
		\node [style=none] (20) at (1.5, 2.25) {};
		\node [style=none] (21) at (-1.5, -1) {};
		\node [style=none] (22) at (1.5, -1) {};
		\node [style=none] (23) at (-2, 2.75) {};
		\node [style=none] (24) at (2, 2.75) {};
		\node [style=none] (25) at (-2, -2.75) {};
		\node [style=none] (26) at (2, -2.75) {};
		\node [style=none] (27) at (-2.75, 1.75) {};
		\node [style=none] (28) at (2.75, 1.75) {};
	\end{pgfonlayer}
	\begin{pgfonlayer}{edgelayer}
		\draw (5.center) to (6.center);
		\draw (2.center) to (18);
		\draw (18) to (3.center);
		\draw [style={grey_densely_dashed}] (19.center) to (21.center);
		\draw [style={grey_densely_dashed}] (21.center) to (22.center);
		\draw [style={grey_densely_dashed}] (22.center) to (20.center);
		\draw [style={grey_densely_dashed}] (20.center) to (19.center);
		\draw [style={grey_densely_dashed}] (23.center) to (25.center);
		\draw [style={grey_densely_dashed}] (25.center) to (26.center);
		\draw [style={grey_densely_dashed}] (26.center) to (24.center);
		\draw [style={grey_densely_dashed}] (24.center) to (23.center);
		\draw (27.center) to (28.center);
	\end{pgfonlayer}
\end{tikzpicture}

%% file: arxiv_lqt-v4/figures/pre-associativity_blank2_var_1.tikz
\begin{tikzpicture}
	\begin{pgfonlayer}{nodelayer}
		\node [style=none] (1) at (-2.75, 0) {};
		\node [style=none] (2) at (2.75, 0) {};
		\node [style=none] (4) at (2.75, -1.75) {};
		\node [style=none] (5) at (-2.75, -1.75) {};
		\node [style=medium square] (6) at (0, 0) {};
		\node [style=none] (7) at (-1.5, 1) {};
		\node [style=none] (8) at (1.5, 1) {};
		\node [style=none] (9) at (-1.5, -2.25) {};
		\node [style=none] (10) at (1.5, -2.25) {};
		\node [style=none] (11) at (-2, 2.75) {};
		\node [style=none] (12) at (2, 2.75) {};
		\node [style=none] (13) at (-2, -2.75) {};
		\node [style=none] (14) at (2, -2.75) {};
		\node [style=none] (15) at (-2.75, 1.75) {};
		\node [style=none] (16) at (2.75, 1.75) {};
	\end{pgfonlayer}
	\begin{pgfonlayer}{edgelayer}
		\draw (4.center) to (5.center);
		\draw (1.center) to (6);
		\draw (6) to (2.center);
		\draw [style={grey_densely_dashed}] (7.center) to (9.center);
		\draw [style={grey_densely_dashed}] (9.center) to (10.center);
		\draw [style={grey_densely_dashed}] (10.center) to (8.center);
		\draw [style={grey_densely_dashed}] (8.center) to (7.center);
		\draw [style={grey_densely_dashed}] (11.center) to (13.center);
		\draw [style={grey_densely_dashed}] (13.center) to (14.center);
		\draw [style={grey_densely_dashed}] (14.center) to (12.center);
		\draw [style={grey_densely_dashed}] (12.center) to (11.center);
		\draw (15.center) to (16.center);
	\end{pgfonlayer}
\end{tikzpicture}

%% file: arxiv_lqt-v4/figures/pre-associativity_blank_var_2.tikz
\begin{tikzpicture}
	\begin{pgfonlayer}{nodelayer}
		\node [style=none] (1) at (-2.75, 0) {};
		\node [style=none] (2) at (-2.75, -1.75) {};
		\node [style=none] (3) at (2.75, -1.75) {};
		\node [style=none] (4) at (2.75, 0) {};
		\node [style=none] (15) at (0, 3.75) {};
		\node [style=none] (16) at (0, -3.75) {};
		\node [style=medium square] (18) at (0, -1.75) {};
		\node [style=none] (19) at (-1.5, 2.25) {};
		\node [style=none] (20) at (1.5, 2.25) {};
		\node [style=none] (21) at (-1.5, -0.5) {};
		\node [style=none] (22) at (1.5, -0.5) {};
		\node [style=none] (23) at (-2, 3.25) {};
		\node [style=none] (24) at (2, 3.25) {};
		\node [style=none] (25) at (-2, -3.25) {};
		\node [style=none] (26) at (2, -3.25) {};
		\node [style=none] (27) at (2.75, 1.75) {};
		\node [style=none] (28) at (-2.75, 1.75) {};
	\end{pgfonlayer}
	\begin{pgfonlayer}{edgelayer}
		\draw (1.center) to (4.center);
		\draw (2.center) to (18);
		\draw (18) to (3.center);
		\draw [style={grey_densely_dashed}] (19.center) to (21.center);
		\draw [style={grey_densely_dashed}] (21.center) to (22.center);
		\draw [style={grey_densely_dashed}] (22.center) to (20.center);
		\draw [style={grey_densely_dashed}] (20.center) to (19.center);
		\draw [style={grey_densely_dashed}] (23.center) to (25.center);
		\draw [style={grey_densely_dashed}] (25.center) to (26.center);
		\draw [style={grey_densely_dashed}] (26.center) to (24.center);
		\draw [style={grey_densely_dashed}] (24.center) to (23.center);
		\draw (27.center) to (28.center);
	\end{pgfonlayer}
\end{tikzpicture}

%% file: arxiv_lqt-v4/figures/pre-associativity_blank2_var_2.tikz
\begin{tikzpicture}
	\begin{pgfonlayer}{nodelayer}
		\node [style=none] (0) at (-2.75, 0) {};
		\node [style=none] (1) at (-2.75, -1.75) {};
		\node [style=none] (2) at (2.75, -1.75) {};
		\node [style=none] (3) at (2.75, 0) {};
		\node [style=medium square] (6) at (0, -1.75) {};
		\node [style=none] (7) at (-1.5, 0.5) {};
		\node [style=none] (8) at (1.5, 0.5) {};
		\node [style=none] (9) at (-1.5, -2.75) {};
		\node [style=none] (10) at (1.5, -2.75) {};
		\node [style=none] (11) at (-2, 3.25) {};
		\node [style=none] (12) at (2, 3.25) {};
		\node [style=none] (13) at (-2, -3.25) {};
		\node [style=none] (14) at (2, -3.25) {};
		\node [style=none] (15) at (2.75, 1.75) {};
		\node [style=none] (16) at (-2.75, 1.75) {};
	\end{pgfonlayer}
	\begin{pgfonlayer}{edgelayer}
		\draw (0.center) to (3.center);
		\draw (1.center) to (6);
		\draw (6) to (2.center);
		\draw [style={grey_densely_dashed}] (7.center) to (9.center);
		\draw [style={grey_densely_dashed}] (9.center) to (10.center);
		\draw [style={grey_densely_dashed}] (10.center) to (8.center);
		\draw [style={grey_densely_dashed}] (8.center) to (7.center);
		\draw [style={grey_densely_dashed}] (11.center) to (13.center);
		\draw [style={grey_densely_dashed}] (13.center) to (14.center);
		\draw [style={grey_densely_dashed}] (14.center) to (12.center);
		\draw [style={grey_densely_dashed}] (12.center) to (11.center);
		\draw (15.center) to (16.center);
	\end{pgfonlayer}
\end{tikzpicture}

%% file: arxiv_lqt-v4/figures/prepare_and_measure_lqt.tikz
\begin{tikzpicture}
	\begin{pgfonlayer}{nodelayer}
		\node [style=none] (0) at (-1, 2.5) {};
		\node [style=none] (3) at (1, 2.5) {};
		\node [style=none] (7) at (-3, 2.5) {};
		\node [style=none] (11) at (3, 2.5) {};
		\node [style=tall rectangle] (15) at (0, 0) {$\mathcal{S}$};
		\node [style=none] (16) at (6, 3) {};
		\node [style=none] (17) at (6.5, 2.5) {};
		\node [style=none] (18) at (6.5, 2.5) {};
		\node [style=none] (19) at (6, 2) {};
		\node [style=none] (20) at (4.5, 2) {};
		\node [style=none] (21) at (4.5, 3) {};
		\node [style=none] (22) at (5.5, 2.5) {$a_1$};
		\node [style=none] (23) at (4.5, 2.5) {};
		\node [style=none] (24) at (3, 2.5) {};
		\node [style=none] (26) at (-5.75, 3) {};
		\node [style=none] (27) at (-6.25, 2.5) {};
		\node [style=none] (28) at (-6.25, 2.5) {};
		\node [style=none] (29) at (-5.75, 2) {};
		\node [style=none] (30) at (-4.5, 2) {};
		\node [style=none] (31) at (-4.5, 3) {};
		\node [style=none] (32) at (-5.25, 2.5) {$\Sigma_1$};
		\node [style=none] (33) at (-3, 2.5) {};
		\node [style=none] (34) at (-4.5, 2.5) {};
		\node [style=none] (35) at (-3, 3) {$\scriptstyle \mathrm{Q}_{\mathtt{S}_1}$};
		\node [style=none] (42) at (0, 3.75) {};
		\node [style=none] (43) at (0, -3.75) {};
		\node [style=none] (44) at (-1, 0.5) {};
		\node [style=none] (45) at (1, 0.5) {};
		\node [style=none] (46) at (-3, 0.5) {};
		\node [style=none] (47) at (3, 0.5) {};
		\node [style=none] (48) at (6, 1) {};
		\node [style=none] (49) at (6.5, 0.5) {};
		\node [style=none] (50) at (6.5, 0.5) {};
		\node [style=none] (51) at (6, 0) {};
		\node [style=none] (52) at (4.5, 0) {};
		\node [style=none] (53) at (4.5, 1) {};
		\node [style=none] (54) at (5.5, 0.5) {$a_2$};
		\node [style=none] (55) at (4.5, 0.5) {};
		\node [style=none] (56) at (3, 0.5) {};
		\node [style=none] (58) at (-5.75, 1) {};
		\node [style=none] (59) at (-6.25, 0.5) {};
		\node [style=none] (60) at (-6.25, 0.5) {};
		\node [style=none] (61) at (-5.75, 0) {};
		\node [style=none] (62) at (-4.5, 0) {};
		\node [style=none] (63) at (-4.5, 1) {};
		\node [style=none] (64) at (-5.25, 0.5) {$\Sigma_2$};
		\node [style=none] (65) at (-3, 0.5) {};
		\node [style=none] (66) at (-4.5, 0.5) {};
		\node [style=none] (68) at (-1, -2.5) {};
		\node [style=none] (69) at (1, -2.5) {};
		\node [style=none] (70) at (-3, -2.5) {};
		\node [style=none] (71) at (3, -2.5) {};
		\node [style=none] (72) at (6, -2) {};
		\node [style=none] (73) at (6.5, -2.5) {};
		\node [style=none] (74) at (6.5, -2.5) {};
		\node [style=none] (75) at (6, -3) {};
		\node [style=none] (76) at (4.5, -3) {};
		\node [style=none] (77) at (4.5, -2) {};
		\node [style=none] (78) at (5.5, -2.5) {$a_n$};
		\node [style=none] (79) at (4.5, -2.5) {};
		\node [style=none] (80) at (3, -2.5) {};
		\node [style=none] (82) at (-5.75, -2) {};
		\node [style=none] (83) at (-6.25, -2.5) {};
		\node [style=none] (84) at (-6.25, -2.5) {};
		\node [style=none] (85) at (-5.75, -3) {};
		\node [style=none] (86) at (-4.5, -3) {};
		\node [style=none] (87) at (-4.5, -2) {};
		\node [style=none] (88) at (-5.25, -2.5) {$\Sigma_m$};
		\node [style=none] (89) at (-3, -2.5) {};
		\node [style=none] (90) at (-4.5, -2.5) {};
		\node [style=none] (95) at (-3, -0.5) {\vdots};
		\node [style=none] (97) at (3, -0.5) {\vdots};
		\node [style=none] (98) at (-3, 1) {$\scriptstyle \mathrm{Q}_{\mathtt{S}_2}$};
		\node [style=none] (99) at (-3, -2) {$\scriptstyle \mathrm{Q}_{\mathtt{S}_m}$};
		\node [style=none] (100) at (3, 3) {$\scriptstyle \mathrm{Q}_{\mathtt{S}'_1}$};
		\node [style=none] (101) at (3, 1) {$\scriptstyle \mathrm{Q}_{\mathtt{S}'_2}$};
		\node [style=none] (102) at (3, -2) {$\scriptstyle \mathrm{Q}_{\mathtt{S}'_n}$};
	\end{pgfonlayer}
	\begin{pgfonlayer}{edgelayer}
		\draw (7.center) to (0.center);
		\draw (11.center) to (3.center);
		\draw (21.center) to (16.center);
		\draw [in=90, out=0] (16.center) to (17.center);
		\draw (18.center) to (17.center);
		\draw [in=0, out=-90] (18.center) to (19.center);
		\draw (19.center) to (20.center);
		\draw (21.center) to (20.center);
		\draw (24.center) to (23.center);
		\draw (31.center) to (26.center);
		\draw [in=90, out=180] (26.center) to (27.center);
		\draw (28.center) to (27.center);
		\draw [in=-180, out=-90] (28.center) to (29.center);
		\draw (29.center) to (30.center);
		\draw (31.center) to (30.center);
		\draw (34.center) to (33.center);
		\draw (46.center) to (44.center);
		\draw (47.center) to (45.center);
		\draw (53.center) to (48.center);
		\draw [in=90, out=0] (48.center) to (49.center);
		\draw (50.center) to (49.center);
		\draw [in=0, out=-90] (50.center) to (51.center);
		\draw (51.center) to (52.center);
		\draw (53.center) to (52.center);
		\draw (56.center) to (55.center);
		\draw (63.center) to (58.center);
		\draw [in=90, out=180] (58.center) to (59.center);
		\draw (60.center) to (59.center);
		\draw [in=-180, out=-90] (60.center) to (61.center);
		\draw (61.center) to (62.center);
		\draw (63.center) to (62.center);
		\draw (66.center) to (65.center);
		\draw (70.center) to (68.center);
		\draw (71.center) to (69.center);
		\draw (77.center) to (72.center);
		\draw [in=90, out=0] (72.center) to (73.center);
		\draw (74.center) to (73.center);
		\draw [in=0, out=-90] (74.center) to (75.center);
		\draw (75.center) to (76.center);
		\draw (77.center) to (76.center);
		\draw (80.center) to (79.center);
		\draw (87.center) to (82.center);
		\draw [in=90, out=180] (82.center) to (83.center);
		\draw (84.center) to (83.center);
		\draw [in=-180, out=-90] (84.center) to (85.center);
		\draw (85.center) to (86.center);
		\draw (87.center) to (86.center);
		\draw (90.center) to (89.center);
	\end{pgfonlayer}
\end{tikzpicture}

%% file: arxiv_lqt-v4/figures/prepare_and_measure_qt.tikz
\begin{tikzpicture}
	\begin{pgfonlayer}{nodelayer}
		\node [style=tall rectangle] (4) at (0, 0) {$\tilde{\mathcal{S}}$};
		\node [style=none] (5) at (6, 3) {};
		\node [style=none] (6) at (6.5, 2.5) {};
		\node [style=none] (7) at (6.5, 2.5) {};
		\node [style=none] (8) at (6, 2) {};
		\node [style=none] (9) at (4.5, 2) {};
		\node [style=none] (10) at (4.5, 3) {};
		\node [style=none] (11) at (5.5, 2.5) {Tr};
		\node [style=none] (12) at (4.5, 2.5) {};
		\node [style=none] (13) at (0.75, 2.5) {};
		\node [style=none] (15) at (-5.75, 3) {};
		\node [style=none] (16) at (-6.25, 2.5) {};
		\node [style=none] (17) at (-6.25, 2.5) {};
		\node [style=none] (18) at (-5.75, 2) {};
		\node [style=none] (19) at (-4.5, 2) {};
		\node [style=none] (20) at (-4.5, 3) {};
		\node [style=none] (21) at (-5.25, 2.5) {$\xi$};
		\node [style=none] (22) at (-0.75, 2.5) {};
		\node [style=none] (23) at (-4.5, 2.5) {};
		\node [style=none] (31) at (6, 1) {};
		\node [style=none] (32) at (6.5, 0.5) {};
		\node [style=none] (33) at (6.5, 0.5) {};
		\node [style=none] (34) at (6, 0) {};
		\node [style=none] (35) at (4.5, 0) {};
		\node [style=none] (36) at (4.5, 1) {};
		\node [style=none] (37) at (5.5, 0.5) {$a_1$};
		\node [style=none] (38) at (4.5, 0.5) {};
		\node [style=none] (39) at (0.75, 0.5) {};
		\node [style=none] (41) at (-5.75, 1) {};
		\node [style=none] (42) at (-6.25, 0.5) {};
		\node [style=none] (43) at (-6.25, 0.5) {};
		\node [style=none] (44) at (-5.75, 0) {};
		\node [style=none] (45) at (-4.5, 0) {};
		\node [style=none] (46) at (-4.5, 1) {};
		\node [style=none] (47) at (-5.25, 0.5) {$\Sigma_1$};
		\node [style=none] (48) at (-0.75, 0.5) {};
		\node [style=none] (49) at (-4.5, 0.5) {};
		\node [style=none] (55) at (6, -2) {};
		\node [style=none] (56) at (6.5, -2.5) {};
		\node [style=none] (57) at (6.5, -2.5) {};
		\node [style=none] (58) at (6, -3) {};
		\node [style=none] (59) at (4.5, -3) {};
		\node [style=none] (60) at (4.5, -2) {};
		\node [style=none] (61) at (5.5, -2.5) {$a_n$};
		\node [style=none] (62) at (4.5, -2.5) {};
		\node [style=none] (63) at (0.75, -2.5) {};
		\node [style=none] (65) at (-5.75, -2) {};
		\node [style=none] (66) at (-6.25, -2.5) {};
		\node [style=none] (67) at (-6.25, -2.5) {};
		\node [style=none] (68) at (-5.75, -3) {};
		\node [style=none] (69) at (-4.5, -3) {};
		\node [style=none] (70) at (-4.5, -2) {};
		\node [style=none] (71) at (-5.25, -2.5) {$\Sigma_m$};
		\node [style=none] (72) at (-0.75, -2.5) {};
		\node [style=none] (73) at (-4.5, -2.5) {};
		\node [style=none] (80) at (-3, 3) {$\scriptstyle \tilde{\mathrm{L}}'$};
		\node [style=none] (81) at (-3, -0.5) {\vdots};
		\node [style=none] (82) at (3, -0.5) {\vdots};
		\node [style=none] (83) at (-2.75, 1) {$\scriptstyle \mathtt{S}_1^{\odot 2}\otimes\mathrm{O}_{\mathtt{S}_1}$};
		\node [style=none] (84) at (-2.75, -2) {$\scriptstyle \mathtt{S}_m^{\odot 2}\otimes\mathrm{O}_{\mathtt{S}_m}$};
		\node [style=none] (85) at (3, 3) {$\scriptstyle \tilde{\mathrm{L}}$};
		\node [style=none] (86) at (3, 1) {$\scriptstyle \mathtt{S}_1'^{\odot 2}\otimes\mathrm{O}_{\mathtt{S}_1'}$};
		\node [style=none] (87) at (3, -2) {$\scriptstyle \mathtt{S}_n'^{\odot 2}\otimes\mathrm{O}_{\mathtt{S}_n'}$};
	\end{pgfonlayer}
	\begin{pgfonlayer}{edgelayer}
		\draw (10.center) to (5.center);
		\draw [in=90, out=0] (5.center) to (6.center);
		\draw (7.center) to (6.center);
		\draw [in=0, out=-90] (7.center) to (8.center);
		\draw (8.center) to (9.center);
		\draw (10.center) to (9.center);
		\draw [style=dashed edge] (13.center) to (12.center);
		\draw (20.center) to (15.center);
		\draw [in=90, out=180] (15.center) to (16.center);
		\draw (17.center) to (16.center);
		\draw [in=-180, out=-90] (17.center) to (18.center);
		\draw (18.center) to (19.center);
		\draw (20.center) to (19.center);
		\draw [style=dashed edge] (23.center) to (22.center);
		\draw (36.center) to (31.center);
		\draw [in=90, out=0] (31.center) to (32.center);
		\draw (33.center) to (32.center);
		\draw [in=0, out=-90] (33.center) to (34.center);
		\draw (34.center) to (35.center);
		\draw (36.center) to (35.center);
		\draw [style=dashed edge] (39.center) to (38.center);
		\draw (46.center) to (41.center);
		\draw [in=90, out=180] (41.center) to (42.center);
		\draw (43.center) to (42.center);
		\draw [in=-180, out=-90] (43.center) to (44.center);
		\draw (44.center) to (45.center);
		\draw (46.center) to (45.center);
		\draw [style=dashed edge] (49.center) to (48.center);
		\draw (60.center) to (55.center);
		\draw [in=90, out=0] (55.center) to (56.center);
		\draw (57.center) to (56.center);
		\draw [in=0, out=-90] (57.center) to (58.center);
		\draw (58.center) to (59.center);
		\draw (60.center) to (59.center);
		\draw [style=dashed edge] (63.center) to (62.center);
		\draw (70.center) to (65.center);
		\draw [in=90, out=180] (65.center) to (66.center);
		\draw (67.center) to (66.center);
		\draw [in=-180, out=-90] (67.center) to (68.center);
		\draw (68.center) to (69.center);
		\draw (70.center) to (69.center);
		\draw [style=dashed edge] (73.center) to (72.center);
	\end{pgfonlayer}
\end{tikzpicture}

%% file: arxiv_lqt-v4/figures/prepare_and_measure_qt_final.tikz
\begin{tikzpicture}
	\begin{pgfonlayer}{nodelayer}
		\node [style=tall rectangle] (4) at (0, 0) {$\mathcal{S}$};
		\node [style=none] (5) at (5.75, 3.25) {};
		\node [style=none] (6) at (6.25, 2.5) {};
		\node [style=none] (7) at (6.25, 2.5) {};
		\node [style=none] (8) at (5.75, 1.75) {};
		\node [style=none] (9) at (3.75, 1.75) {};
		\node [style=none] (10) at (3.75, 3.25) {};
		\node [style=none] (11) at (5, 2.5) {$\xi_{\tilde{\mathrm{L}}_{1}'}a_1$};
		\node [style=none] (12) at (3.75, 2.5) {};
		\node [style=none] (13) at (0.75, 2.5) {};
		\node [style=none] (14) at (-6, 3.25) {};
		\node [style=none] (15) at (-6.5, 2.5) {};
		\node [style=none] (16) at (-6.5, 2.5) {};
		\node [style=none] (17) at (-6, 1.75) {};
		\node [style=none] (18) at (-3.75, 1.75) {};
		\node [style=none] (19) at (-3.75, 3.25) {};
		\node [style=none] (20) at (-5, 2.5) {$\text{Tr}_{\tilde{\mathrm{L}}_{1}}\!{\Sigma_1}$};
		\node [style=none] (21) at (-0.75, 2.5) {};
		\node [style=none] (22) at (-3.75, 2.5) {};
		\node [style=none] (23) at (-2.5, 3) {$\scriptstyle \tilde{\mathrm{Q}}_{\mathtt{S}_1}$};
		\node [style=none] (24) at (0, 3.75) {};
		\node [style=none] (25) at (0, -3.75) {};
		\node [style=none] (30) at (5.75, 1.25) {};
		\node [style=none] (31) at (6.25, 0.5) {};
		\node [style=none] (32) at (6.25, 0.5) {};
		\node [style=none] (33) at (5.75, -0.25) {};
		\node [style=none] (34) at (3.75, -0.25) {};
		\node [style=none] (35) at (3.75, 1.25) {};
		\node [style=none] (36) at (5, 0.5) {$\xi_{\tilde{\mathrm{L}}_{2}'}a_2$};
		\node [style=none] (37) at (3.75, 0.5) {};
		\node [style=none] (38) at (0.75, 0.5) {};
		\node [style=none] (52) at (5.75, -1.75) {};
		\node [style=none] (53) at (6.25, -2.5) {};
		\node [style=none] (54) at (6.25, -2.5) {};
		\node [style=none] (55) at (5.75, -3.25) {};
		\node [style=none] (56) at (3.75, -3.25) {};
		\node [style=none] (57) at (3.75, -1.75) {};
		\node [style=none] (58) at (5, -2.5) {$\xi_{\tilde{\mathrm{L}}_{n}'}a_n$};
		\node [style=none] (59) at (3.75, -2.5) {};
		\node [style=none] (60) at (0.75, -2.5) {};
		\node [style=none] (73) at (-2.5, -0.5) {\vdots};
		\node [style=none] (74) at (2.5, -0.5) {\vdots};
		\node [style=none] (77) at (2.5, 3) {$\scriptstyle \tilde{\mathrm{Q}}_{\mathtt{S}_1'}$};
		\node [style=none] (78) at (2.5, 1) {$\scriptstyle \tilde{\mathrm{Q}}_{\mathtt{S}_2'}$};
		\node [style=none] (79) at (2.5, -2) {$\scriptstyle \tilde{\mathrm{Q}}_{\mathtt{S}_n'}$};
		\node [style=none] (82) at (-6, 1.25) {};
		\node [style=none] (83) at (-6.5, 0.5) {};
		\node [style=none] (84) at (-6.5, 0.5) {};
		\node [style=none] (85) at (-6, -0.25) {};
		\node [style=none] (86) at (-3.75, -0.25) {};
		\node [style=none] (87) at (-3.75, 1.25) {};
		\node [style=none] (88) at (-5, 0.5) {$\text{Tr}_{\tilde{\mathrm{L}}_{2}}\!{\Sigma_2}$};
		\node [style=none] (89) at (-0.75, 0.5) {};
		\node [style=none] (90) at (-3.75, 0.5) {};
		\node [style=none] (91) at (-2.5, 1) {$\scriptstyle \tilde{\mathrm{Q}}_{\mathtt{S}_2}$};
		\node [style=none] (94) at (-6, -1.75) {};
		\node [style=none] (95) at (-6.5, -2.5) {};
		\node [style=none] (96) at (-6.5, -2.5) {};
		\node [style=none] (97) at (-6, -3.25) {};
		\node [style=none] (98) at (-3.75, -3.25) {};
		\node [style=none] (99) at (-3.75, -1.75) {};
		\node [style=none] (100) at (-5, -2.5) {$\text{Tr}_{\tilde{\mathrm{L}}_{m}}\!\!{\Sigma_m}$};
		\node [style=none] (101) at (-0.75, -2.5) {};
		\node [style=none] (102) at (-3.75, -2.5) {};
		\node [style=none] (103) at (-2.5, -2) {$\scriptstyle \tilde{\mathrm{Q}}_{\mathtt{S}_m}$};
	\end{pgfonlayer}
	\begin{pgfonlayer}{edgelayer}
		\draw (10.center) to (5.center);
		\draw [in=90, out=0] (5.center) to (6.center);
		\draw (7.center) to (6.center);
		\draw [in=0, out=-90] (7.center) to (8.center);
		\draw (8.center) to (9.center);
		\draw (10.center) to (9.center);
		\draw [style=dashed edge] (13.center) to (12.center);
		\draw (19.center) to (14.center);
		\draw [in=90, out=180] (14.center) to (15.center);
		\draw (16.center) to (15.center);
		\draw [in=-180, out=-90] (16.center) to (17.center);
		\draw (17.center) to (18.center);
		\draw (19.center) to (18.center);
		\draw [style=dashed edge] (22.center) to (21.center);
		\draw (35.center) to (30.center);
		\draw [in=90, out=0] (30.center) to (31.center);
		\draw (32.center) to (31.center);
		\draw [in=0, out=-90] (32.center) to (33.center);
		\draw (33.center) to (34.center);
		\draw (35.center) to (34.center);
		\draw [style=dashed edge] (38.center) to (37.center);
		\draw (57.center) to (52.center);
		\draw [in=90, out=0] (52.center) to (53.center);
		\draw (54.center) to (53.center);
		\draw [in=0, out=-90] (54.center) to (55.center);
		\draw (55.center) to (56.center);
		\draw (57.center) to (56.center);
		\draw [style=dashed edge] (60.center) to (59.center);
		\draw (87.center) to (82.center);
		\draw [in=90, out=180] (82.center) to (83.center);
		\draw (84.center) to (83.center);
		\draw [in=-180, out=-90] (84.center) to (85.center);
		\draw (85.center) to (86.center);
		\draw (87.center) to (86.center);
		\draw [style=dashed edge] (90.center) to (89.center);
		\draw (99.center) to (94.center);
		\draw [in=90, out=180] (94.center) to (95.center);
		\draw (96.center) to (95.center);
		\draw [in=-180, out=-90] (96.center) to (97.center);
		\draw (97.center) to (98.center);
		\draw (99.center) to (98.center);
		\draw [style=dashed edge] (102.center) to (101.center);
	\end{pgfonlayer}
\end{tikzpicture}